\documentclass[aps,prd, twoside,superscriptaddress,showpacs, preprintnumbers, showpacs, nofootinbib,twocolumn]{revtex4-1}

\pdfoutput=1
\usepackage{amssymb}
\usepackage{amsthm}
\usepackage{graphicx,amsmath}
\usepackage{color}
\newcommand{\beq}{\begin{eqnarray}}
\newcommand{\eeq}{\end{eqnarray}}

\newcommand{\bmp}{\noindent\begin{minipage}{16cm}}
\newcommand{\emp}{\end{minipage}\vskip 7mm} 

\newcommand{\lcdm}{$\Lambda$CDM\ }

\usepackage{dcolumn}
\usepackage{bm}
\usepackage{bbm}
\usepackage{subfigure}
\usepackage{slashed} 
\usepackage{mathrsfs}

\usepackage{epsfig}

\usepackage{hyperref}

\newcommand{\bea}{\begin{eqnarray}}
\newcommand{\eea}{\end{eqnarray}}

\newcommand{\ba}{\begin{eqnarray}}
\newcommand{\ea}{\end{eqnarray}}

\newcommand{\be}{\begin{eqnarray}}
\newcommand{\ee}{\end{eqnarray}}

\begin{document}

\title{Closing in on Resonantly Produced Sterile Neutrino Dark Matter}

%

\author{John F.~Cherry}
\email{jfcherry@vt.edu}
\author{Shunsaku Horiuchi}
\email{horiuchi@vt.edu}
\affiliation{Center for Neutrino Physics, Department of Physics, Virginia Tech, 
Blacksburg, VA 24061, USA}
\affiliation{Neutrino Engineering Institute, New Mexico Consortium, Los Alamos, New Mexico 87545, USA}


\begin{abstract}
We perform an exhaustive scan of the allowed resonant production regime for sterile neutrino dark matter in order to improve constraints for dark matter structures which arise from the non-thermal sterile neutrino energy spectra.  Small-scale structure constraints are particularly sensitive to large lepton asymmetries/small mixing angles which result in relatively warmer sterile neutrino momentum distributions.   We revisit Milky Way galaxy subhalo count constraints and combine them with recent searches for X-ray emission from sterile neutrino decays. Together they rule out models outside the mass range $ 7.0\, {\rm keV} \leq  m_{\nu_s} \leq 36\, {\rm keV}$ and lepton asymmetries smaller than $15\times 10^{-6}$ per unit entropy density at $95\%$ CI or greater.  We also find that while a portion of the parameter space remains unconstrained, {the combination of subhalo counts and X-ray data indicate the candidate $3.55\, \rm keV$ X-ray line signal potentially originating from a $7.1\,\rm keV$ sterile neutrino decay to be disfavored at $93\%$ CI.} 
 \end{abstract}

\preprint{}

\maketitle

\section{Introduction}

One of the most trying aspects of the search for the nature and origin of the dark matter is that the dark matter itself has proved to be difficult to directly interact with, except through the force of gravity.  It may come as no surprise now, that we concern ourselves with {\it sterile} neutrino dark matter, which by definition has no direct interaction with Standard Model (SM) particles at all!  But the sterile neutrino's saving grace is the flavor mixing that it shares with SM neutrino species, allowing them to form a natural dark matter candidate.  Frequent collisions between SM neutrinos drive production of sterile neutrinos in the early Universe through quantum decoherence.  With the added constraint that the production of sterile neutrinos result in the correct dark matter abundance this is known as the Dodelson-Widrow (DW) mechanism~\cite{Dodelson:1994aa}.  Alternately, the presence of net lepton asymmetry in the primordial plasma may drive the production of sterile neutrino dark matter via in-medium neutrino mixing angle enhancement known as {\it resonance}.  This resonant production of neutrinos was first described by Shi \& Fuller~\cite{Shi:1999aa} and allows for sterile neutrinos with mixing angles much smaller than those allowed for DW to obtain the necessary relic abundance to account for the dark matter.

The resonantly produced sterile neutrinos will typically have non-thermal, distorted momentum distributions.  Of critical importance, these non-thermal momentum spectra have a host of dynamically salutary effects on the structure of sterile neutrino dark matter halos (e.g., \cite{Bozek:2015bdo}).  Many of the dark matter structure anomalies such as the cusp-versus-core problem~\cite{Moore:1994yx,Flores:1994gz,Navarro:1996gj,Gentile:2004aa,Gilmore:2007aa,Frusciante:2012aa}, the too-big-to-fail problem~\cite{BoylanKolchin:2011de,Walker:2013aa}, and the missing satellites problem~\cite{Klypin:1999uc,Moore:1999nt,Kauffmann:1993gv} are all eased if the dark matter has a warm, non-thermal spectral distribution.  Resonant production of sterile neutrino dark matter satisfies the spectral distortion requirements of these structure anomalies quite readily~\cite{Lovell:2012aa,Anderhalden:2013aa,Abazajian:2014aa,Horiuchi:2016aa,Harada:2014lma,Bose:2016irl,Lovell:2016ab}.

The probes of the resonantly produced sterile neutrino dark matter parameter space are rapidly closing in on the allowed region from all sides (see, e.g., review articles~\cite{Kusenko:2009aa,Boyarsky:2009ab,Boyarsky:2012rt,Adhikari:2017aa}).  Counts of Milky Way satellite galaxies place strong constraints on the low-mass portion of the parameter space, X-ray decay line searches efficiently bound the large mixing angle and high mass regimes, and constraints on the net lepton asymmetry of the Universe bound the minimum allowable mixing angles.  Taken together these form a fully bounded box with good prospects for either observing or definitively excluding the presence of sterile neutrino dark matter in the future.

Recent works on Milky Way satellite counts have focused on limited or coarse scans of the parameter space to estimate the constraints which can be inferred from structure formation~\cite{Horiuchi:2016aa,Schneider:2016aa}.  In this paper we endeavor to place such constraints on firmer footing.  We use the calculated matter power spectra and thermal energy distributions of resonantly produced sterile neutrino dark matter on a finer grid, as well as conduct a likelihood analysis of predicted vs.~observed combined Milky Way and Andromeda (M31) subhalo counts.  Further, we also investigate the similarities and discrepancies between two separate theoretical descriptions of galaxy structure formation in order to gauge the model sensitivity of subhalo count predictions.  Lastly, we use our calculation to investigate the joint goodness of fit with X-ray observations for a $m_{\nu_s} = 7.1 \, \rm keV$ sterile neutrino which is a candidate for explaining the anomalous $3.55\, \rm keV$ X-ray line signal \cite{Bulbul:2014aa,Boyarsky:2014aa,Urban:2015aa,Boyarsky:2015aa,Iakubovskyi:2016aa,Franse:2016aa} (but, see also null claims by Refs.~\cite{Anderson:2015aa,Carlson:2015aa,Horiuchi:2014aa,Tamura:2015aa,Jeltema:2016aa,Ruchayskiy:2016aa,Sekiya:2016aa}). 

{In section II we review our methodology for constructing our resonantly produced sterile neutrino momentum distributions, their resultant matter power spectrum, and small scale structure formation.  In section III we compute predicted subhalo counts for the Milky Way and M31 galaxies.  In section IV we derive constraints for sterile neutrino dark matter based on observations, and discuss the implications of combining them with other constraints. We close with conclusions in section V. 

\section{Modeling the Matter Power Spectrum}\label{sec:Struct}

\subsection{Sterile neutrino dark matter production}

In the simplest sense, production of sterile neutrinos in the environment of the early Universe can be thought of in terms of the behavior of a quantum mechanical damped oscillator~\cite{stodolsky:1987QK,Barbieri:1991aa,Enqvist:1991aa}.  Though sterile neutrino flavor states posses no direct coupling to the plasma, SM neutrinos are simultaneously oscillating into the sterile flavor state and scattering with the plasma frequently enough to be in thermal equilibrium above a temperature of $T\sim 3\, \rm MeV$.  The constant scattering events experienced by the SM neutrinos interrupts their coherent evolution, leading to a damping of the oscillation between SM and sterile flavor states.  In a {\it static} Universe, this would lead to the eventual equilibration of the SM and sterile neutrino flavor state populations on a relaxation time scale~\cite{stodolsky:1987QK},
\begin{equation}
\tau = \frac{V_T^2}{D^2+V_z^2+V_T^2}D\, ,
\label{trelax}
\end{equation}
where for a neutrino of energy $E_\nu$, the damping rate is $D  = \Gamma_{\rm all\ interactions}/2 \sim g_\star G_F T^5$, $V_T = \delta m^2 / 2E_\nu \, \sin 2\theta_V$, and $V_z = \delta m^2 / 2E_\nu \, \cos 2\theta_V + V_L$, the SM-sterile mass squared splitting is $\delta m^2$, and the neutrino mixing angle is $\theta_V$.   In this sense $V_L$ is an effective mass term which neutrinos acquire through coherent forward scattering interactions with the plasma, which can be quite complicated in general~\cite{Venumadhav:2015aa}.

In the case of an expanding Universe, with $V_L \equiv 0$, this damped oscillatory behavior can lead to partial population of the sterile neutrino states, producing just the right amount of sterile neutrino dark matter through the DW mechanism~\cite{Dodelson:1994aa}.  In this case one must solve the co-moving Boltzmann equation,
\begin{equation}
\left( \frac{d}{dt} - HE_\nu\frac{d}{dE_\nu}\right)f_{\rm s}(E_\nu,t) = \frac{1}{\tau(E_\nu,t)}f_{\rm SM}(E_\nu,t)\, ,
\label{Boltz}
\end{equation}
where $f_{\rm s}$ and $f_{\rm SM}$ are the sterile and SM neutrino distribution functions, respectively.  If the relaxation rate, $1/\tau$, is significantly less than the Hubble rate, $H$, for the entire history of the Universe, the sterile neutrino population will \lq\lq freeze-in\rq\rq\ without fully thermalizing with the plasma.  This leads to the non-resonant production of sterile neutrinos which may be the source of the dark matter. 

Should the Universe posses a sufficiently large lepton asymmetry, for which we define $L_6 = 10^6 (n_{\nu_e} - n_{\bar\nu_e})/s$ with entropy density $s$, then neutrino flavor states can acquire large effective mass differences, $V_L$, through their interactions with the plasma and experience a resonance when $V_z = 0$.  This produces a temporary and strongly energy dependent reduction of the equilibration time between the sterile and SM neutrino populations.  This was pointed out by Shi \& Fuller~\cite{Shi:1999aa}, who also noted that the energy dependence of the resonant condition produced strongly distorted spectral energy distributions in the resultant sterile neutrino population.  

While helpful, this mechanism is not without limitations.  The maximum lepton asymmetry allowed in the early Universe is bounded by observation of the products of Big Bang Nucleosynthesis (BBN), particularly by the primordial He abundance~\cite{Dolgov:2002bb,Abazajian:2002lr,Serpico:2005aa}.  If the lepton asymmetry in the primordial plasma is sufficiently large the SM neutrino spectral energy distributions will become degenerate, leading to alterations in the rates for $\nu$-baryon interactions and changes in the proton fraction of the plasma during BBN.  Because the observed values of the SM neutrino mixing angles are large~\cite{Mangano:2012aa,Castorina:2012aa}, current observations place an upper limit on the lepton asymmetry of $L_6 = 2500$~\cite{Serpico:2005aa,Mangano:2012aa,Castorina:2012aa,Steigman:2012sp}. 

To solve the general problem of the production of sterile neutrino dark matter we have performed an exhaustive parameter space scan using the publicly available {\texttt sterile-dm} code~\cite{Venumadhav:2015aa}.  This code assumes the Boltzmann equation approach outlined above which requires that neutrino transport and oscillation be collisionally dominated.  In particular, the authors of {\texttt sterile-dm} emphasized the treatment of neutrino opacities and transport during the QCD phase transition epoch, which is notoriously difficult to model.  As the resonant production of sterile neutrinos frequently takes place above a temperature of $100\, \rm MeV$, this careful treatment of the neutrino transport makes {\texttt sterile-dm} a significant improvement on available software~\cite{Abazajian:2001ab}.   

In Fig.~\ref{fig:lepasym} we show our results of running {\texttt sterile-dm} over the allowed sterile neutrino parameter space in terms of the necessary lepton asymmetry, $L_6$, required to produce the correct relic density of dark matter, $\Omega_{DM}h^2 = 0.1188$~\cite{Planck-Collaboration:2016aa}.  These results are obtained on a $100\times100$ grid within the paramter space, which allows the smooth features within the results to be observed.  We do not consider $m_{\nu_s} \leq 1.7\, \rm keV$ since they are robustly bound by phase-space constraints~\cite{Tremaine:1979aa,Gorbunov:2008aa,Boyarsky:2009aa}. 

\begin{figure}[t] 
\begin{center}
 \includegraphics[width=.48\textwidth]{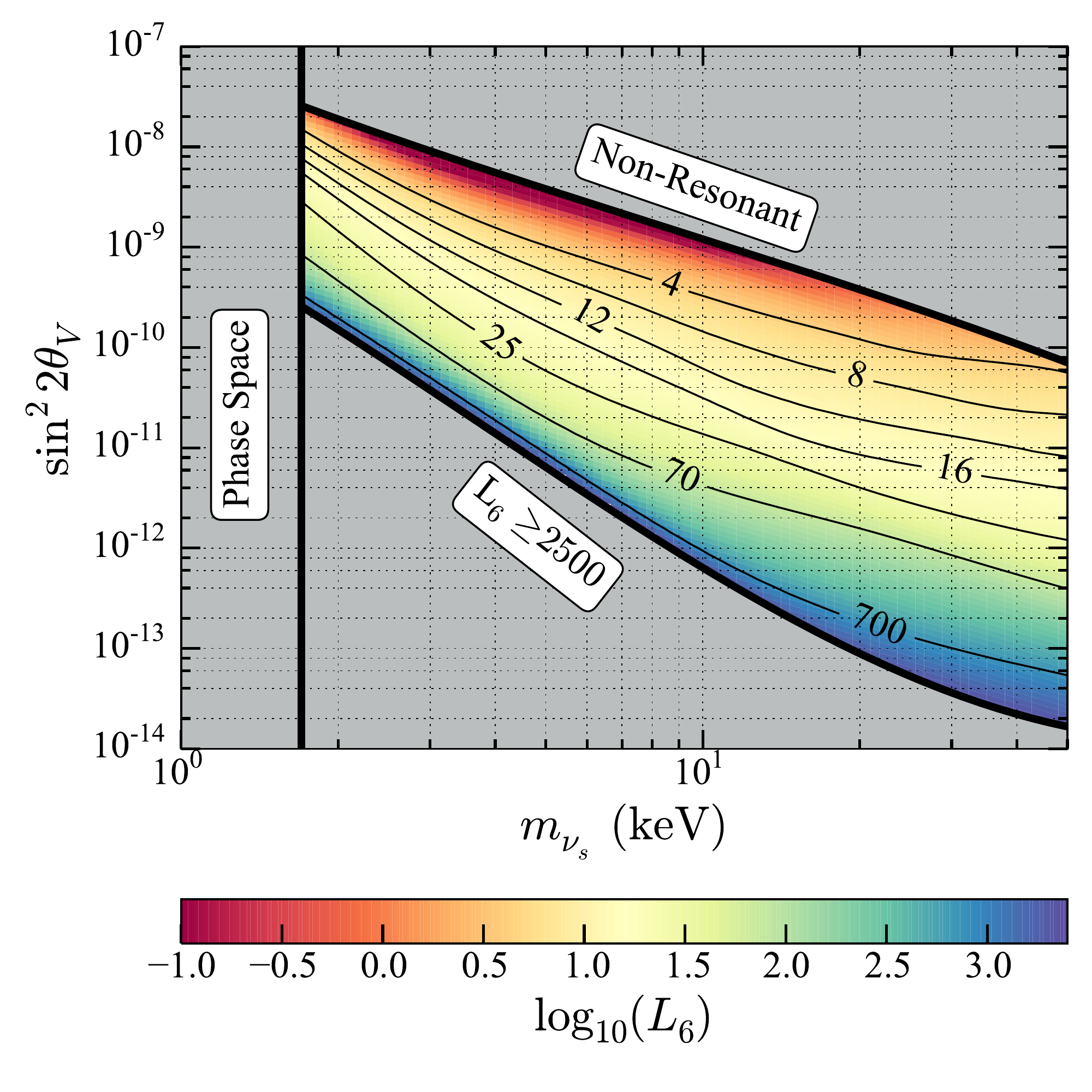}

\caption{ The lepton asymmetry required to produce the correct relic abundance of SDM.  The color scale is logarithmic, while contours are labeled with the absolute value of $L_6$.  Mixing angles above the non-resonant production line over-produce DM, while mixing angles below the $L_6 \geq 2500$ line violate BBN bounds~\cite{Serpico:2005aa}.  The left hand side is capped by the phase space bound of $m_{\nu_s}\geq 1.7\,\rm keV$~\cite{Boyarsky:2009aa}.}
\label{fig:lepasym}
\end{center}
\end{figure}

Notably, the BBN constraint of $L_6 \geq 2500$ produced by the {\texttt sterile-dm} code does not fully agree with the results of~\cite{Boyarsky:2009ab}, who report a lower range of allowable $\sin^2 2\theta_V$ for masses below $m_{\nu_s} \leq 10\,\rm keV$.  We have taken care to check that our results agree exactly with those reported in the original {\texttt sterile-dm} paper by Venumadhav et al.~\cite{Venumadhav:2015aa} for $m_s = 7.11\,\rm keV$.  This disagreement is not unexpected given {\texttt sterile-dm}'s updates to neutrino transport opacities during the QCD epoch, highlighting the importance of the SM plasma and its effect on the production of sterile neutrino dark matter.

\subsection{Spectral Distortions}

\begin{figure}[t] 
\begin{center}
 \includegraphics[width=.48\textwidth]{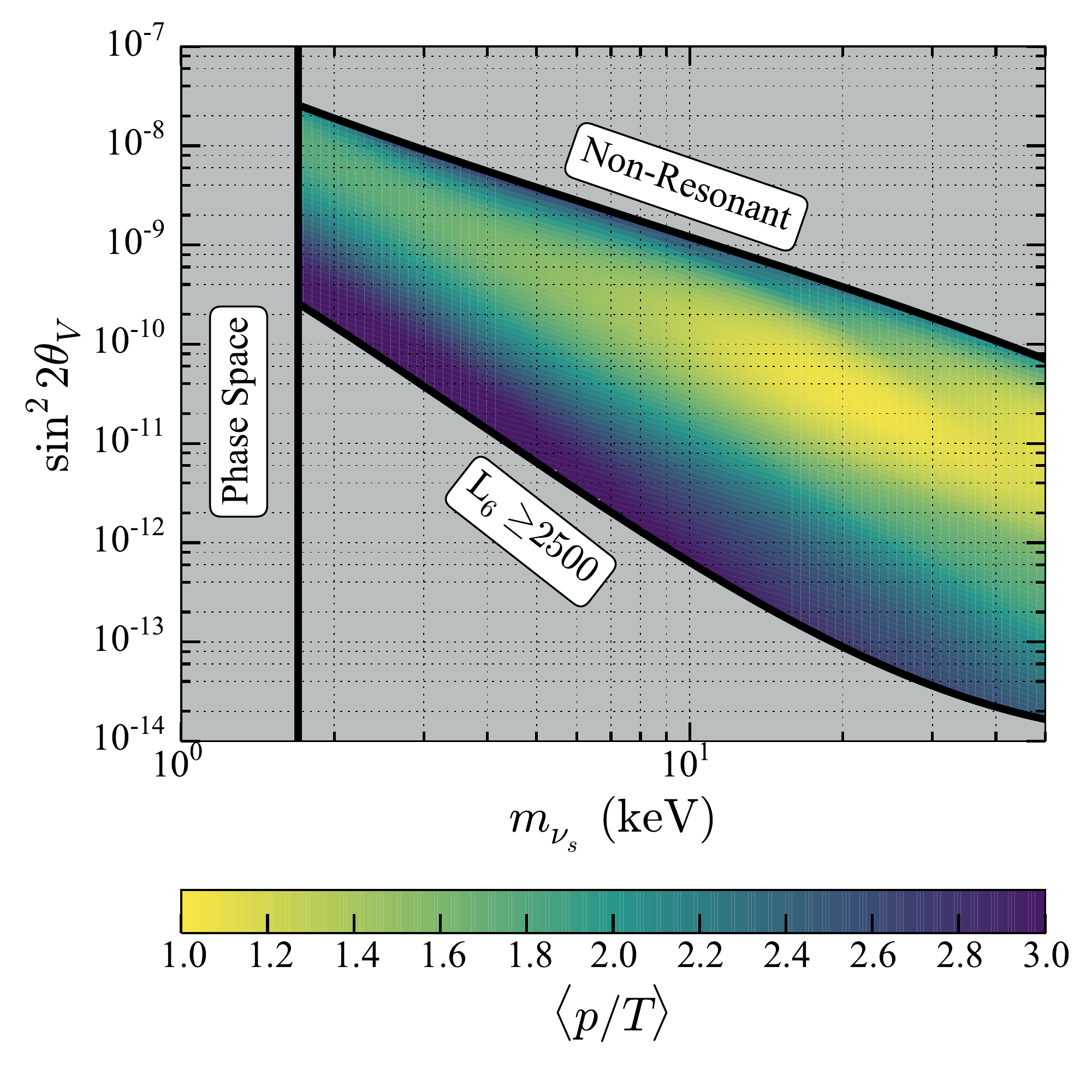}
\caption{The average momentum of the sum of $\nu_s$ and $\bar\nu_s$ populations.  For reference, a non-degenerate Fermi gas has average momentum $\langle p/T\rangle = 3.15$.}
\label{fig:pmin}
\end{center}
\end{figure}

We show the ensemble averaged momentum of the $\nu_s + \bar\nu_s$ distributions produced by the {\texttt sterile-dm} calculation in Fig.~\ref{fig:pmin}.  The results are in good agreement with expectations in which resonant production yields significantly colder sterile neutrino spectra than the non-resonant production mechanism, provided that the peak of sterile neutrino production is before or during the QCD phase transition~\cite{Abazajian:2001ab,Venumadhav:2015aa}.  For scenarios which require very large lepton asymmetry, sterile neutrino production is delayed to temperatures well below $T\sim 170\, \rm MeV$, producing sterile neutrino spectra which asymptotically tend to warmer thermal distributions.

Importantly, the resonant production of sterile neutrinos does not treat $\nu_s$ and $\bar\nu_s$ equally.  Because the sign of $V_L$ is opposite for particles and anti-particles, either of $\nu_s$ or $\bar\nu_s$ will experience resonant production, but the condition $V_z = 0$ can never be satisfied simultaneously for both.  As a result $\nu_s$ and $\bar\nu_s$ populations have dissimilar energy spectra and relic abundances, one species will be relatively hot and significantly underproduced by non-resonant scattering processes.

\subsection{Matter Power Spectrum}

In order to gauge the observable effects of the resonantly produced, non-thermal spectra of sterile neutrinos we have used the software {\texttt CLASS}~\cite{Lesgourgues:2011aa} to process the $\nu_s$ and $\bar\nu_s$ populations produced by {\texttt sterile-dm}.  {\texttt CLASS} is able to solve for the evolution of linear density perturbations by taking in user specified phase-space distributions for multiple species of dark matter.  From this solution, {\texttt CLASS} generates 3D matter power spectra consistent with the cosmological parameters which can be specified, and in this work have been drawn directly from the {\it Planck} best fit cosmological values~\cite{Planck-Collaboration:2016aa}. 

Because sterile neutrino dark matter is generally warmer than the cold dark matter (\lq\lq CDM\rq\rq ) of \lcdm (the standard cosmological model of $\Lambda$ + CDM), the amount of power present on small scales will be reduced by free streaming.  This effect is due directly to the additional momentum carried by the warm sterile neutrinos streaming out of density perturbations insufficiently massive to keep them gravitationally bound.  This outward flow of warm particles damps the amplitude of dark matter density fluctuations resulting in less power (and fewer density perturbations) on small length scales~\cite{Silk:1968aa}.  Precisely how much the matter power spectrum is damped relative to \lcdm on the scale of a given co-moving inverse length $k$ may depend sensitively on the spectral distribution of momentum within the relic sterile neutrino population.  

To illustrate these effects over the sterile neutrino parameter space, we define $k_{\rm cut}$ to be the inverse length scale at which the 3D matter power spectrum predicted by {\texttt CLASS} using the sterile neutrino spectra generated by {\texttt sterile-dm} is suppressed by a factor of $1/e$ relative to the 3D matter power spectrum of CDM.  We show the results of our calculation in Fig.~\ref{fig:kcut}.  The results are a match for the predictions of linear perturbation theory.  Lighter sterile neutrino masses result in matter power suppression on larger length scales (smaller $k_{\rm cut}$) owing to the longer free-streaming length of such neutrinos.  Further, considering fixed sterile neutrino masses shows that the ensemble averaged momentum $\langle p/T\rangle$ is a good proxy for matter power suppression.  Colder sterile neutrino spectra are unable to free-stream over long distances and only suppress the power on small length scales (larger $k_{\rm cut}$).

The potential consequence of this matter power suppression are important tools for indirectly probing the sterile neutrino dark matter parameter space.  A lack of power on small length scales will reduce the number of low mass dark matter halos below some threshold which implies a concomitant reduction in the number of dwarf galaxies in the Universe.  Counts of the observed number of dwarf galaxies orbiting the Milky Way can be compared to the results of N-body simulations and sterile neutrino dark matter models which {\it under predict} the observed number Milky Way satellites can be excluded~\cite{Polisensky:2011aa,Horiuchi:2014aa}.  Further, fewer dark matter halos would be capable of accreting dense clouds of Hydrogen gas during the epoch of re-ionization.  This may lead to changes in the distinct absorption patterns of photons as a function of redshift in the Lyman-$\alpha$ forest~\cite{Abazajian:2006aa,Viel:2006aa,Seljak:2006aa,Boyarsky:2009ac,Viel:2013aa,Garzilli:2015aa}.

\begin{figure}[t] 
\begin{center}
 \includegraphics[width=.48\textwidth]{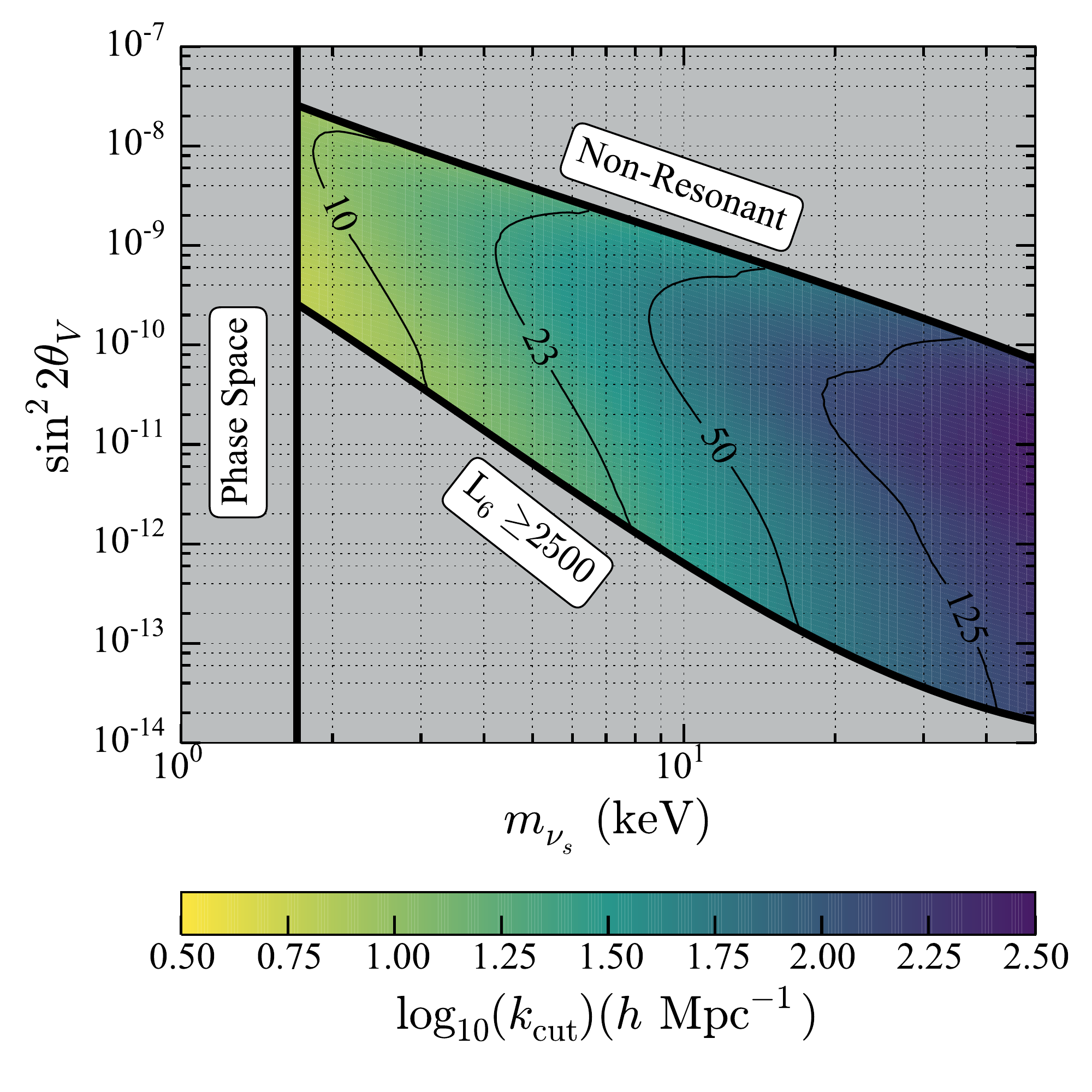}
\caption{The cutoff value of $k_{\rm cut}$ beyond which the matter power spectrum predicted by a given SDM model is suppressed by more than a factor of $1/e$ compared to the \lcdm prediction.  The color scale is logarithmic, while contours are labeled with the absolute value of $k_{\rm cut}$. }
\label{fig:kcut}
\end{center}
\end{figure}

\section{Subhalo Counts}\label{sec:SubH}

One of the most widely known anomalies related to the nature of dark matter is the {\it missing\ satellites} problem~\cite{Klypin:1999uc,Moore:1999nt,Kauffmann:1993gv} wherein the number of satellite galaxies observed orbiting the Milky Way is significantly lower than expected from the hierarchical structure formation of \lcdm cosmology.  Sterile neutrino dark matter offers a tempting solution to this discrepancy, as the suppression of the matter power spectrum at small scales prevents the formation of the dark matter subhalos which play host to these missing Milky Way satellites.  However, numerous baryonic effects such as reionization~\cite{Bullock:2000aa}, supernova feedback \cite{Governato:2012fa}, or tidal stripping can easily rob one of these dark matter subhalos of its observable baryons~\cite{Font:2011aa,Lovell:2016aa}.

While the efficacy of baryonic feedback mechanisms in explaining the missing satellites problem is a topic of ongoing research, the dearth of observable subhalos around the Milky Way can still be used to constrain the sterile neutrino dark matter parameter space.  Because there are a few dozen observed Milky Way subhalos which have been catalogued classically and by  surveys such as the SDSS and DES, we can place a statistically meaningful lower bound on the number of subhalos which are bound to the Milky Way galaxy.  While baryonic effects may prevent extant subhalos from being observed, they cannot create satellite galaxies in the absence of a host dark matter halo.  In this sense we can say conservatively and regardless of the effects of baryons on the satellite galaxy counts, the sterile neutrino dark matter models we are considering must plausibly predict as many subhalos as are observed.

In order to provide some measure of the uncertainty that subhalo population models {\it themselves} introduce into the constraints which can be placed on sterile neutrino dark matter we consider the predictions of two separate models of structure formation which attempt to incorporate the effects of Warm Dark Matter (\lq\lq WDM\rq\rq ), Model 1~\cite{Schneider:2015aa} and Model 2~\cite{Dunstan:2011aa}.  Both of these models are based on the extended Press-Schechter approach~\cite{Press:1974aa,Bond:1991aa} for semi-analytically relating the primordial dark matter power spectrum to the mass functions of halos and subhalos, yet take different approaches to accommodating WDM within this formalism, as well as normalizing their results to different suites of numerical N-body simulations.  Model 1 is constructed with an aim to provide a generic formula for including the effects of matter power suppression with arbitrary shape and scale, which is implemented through the use of a sharp-k filter~\cite{Schneider:2015aa}.  Model 2 attempts to include the effects of WDM within the Press-Schechter formalism using two-point statistics along with a host of individual adaptations for treating WDM's consequences for various subcomponents of the model, such as biasing of the smooth dark matter component~\cite{Dunstan:2011aa}.  Further, these models are conditioned on numerical models of structure formation which have been created with different cosmological parameters.  Model 2 employs the best fit parameters from WMAP7~\cite{Komatsu:2011aa}, while Model 1 is based on the Planck 2013 best fit~\cite{Planck-Collaboration:2014aa}.  The updated Planck fit favors slightly more dark matter and slightly more clustering as compared to the WMAP7 fit.  In the \lcdm limit, these differences are expected to produce a preference for more subhalos in Model 1, which we will discuss shortly.

\begin{figure}[t] 
\begin{center}
 \includegraphics[width=.48\textwidth]{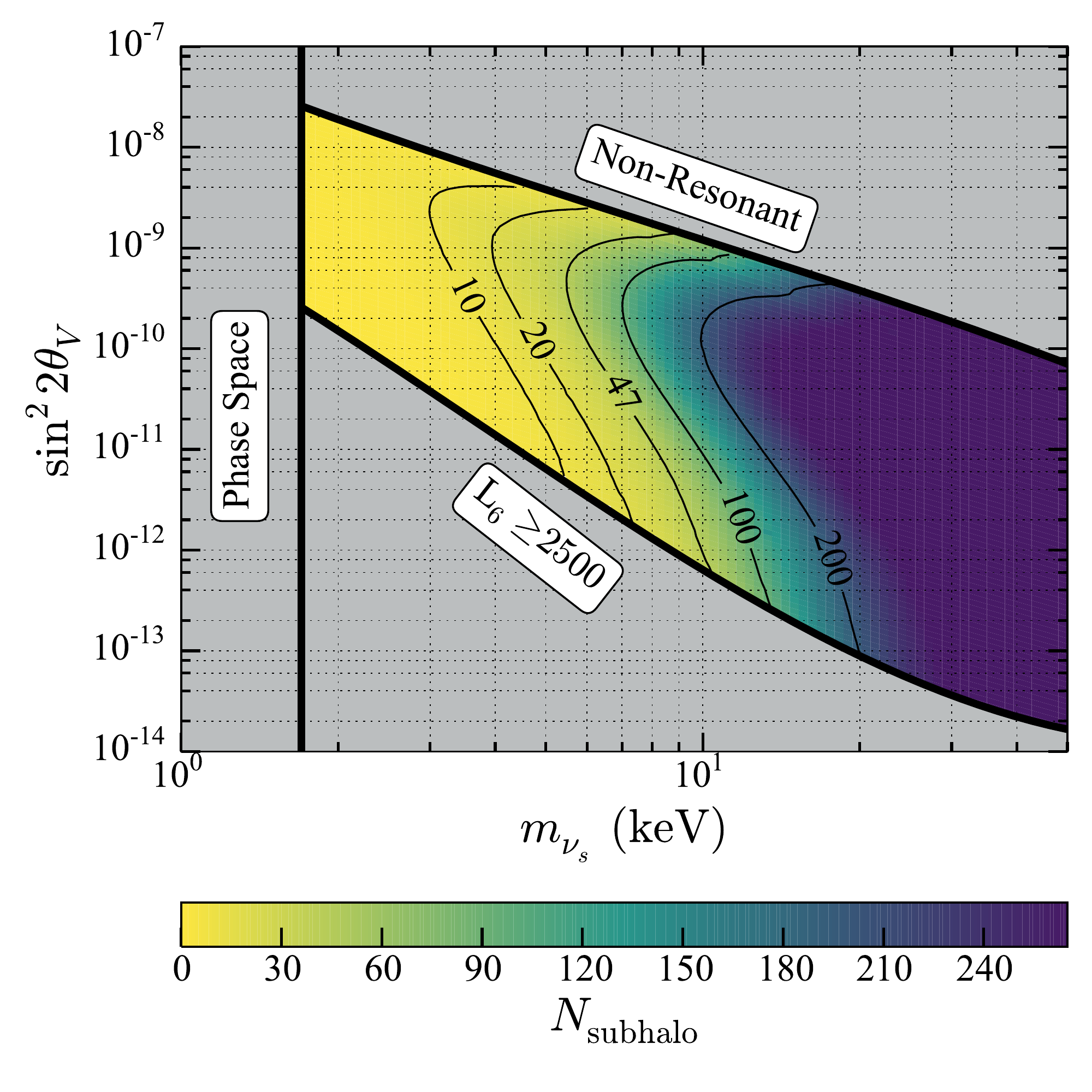}
\caption{The number of detectable subhalos surrounding the Milky Way galaxy for Model 1~\cite{Schneider:2015aa}.  The isocontour of 47 subhalos is the $95\%$ CI limit on the all-sky corrected SDSS subhalo counts.}
\label{fig:subh}
\end{center}
\end{figure}

\begin{figure}[t] 
\begin{center}
 \includegraphics[width=.48\textwidth]{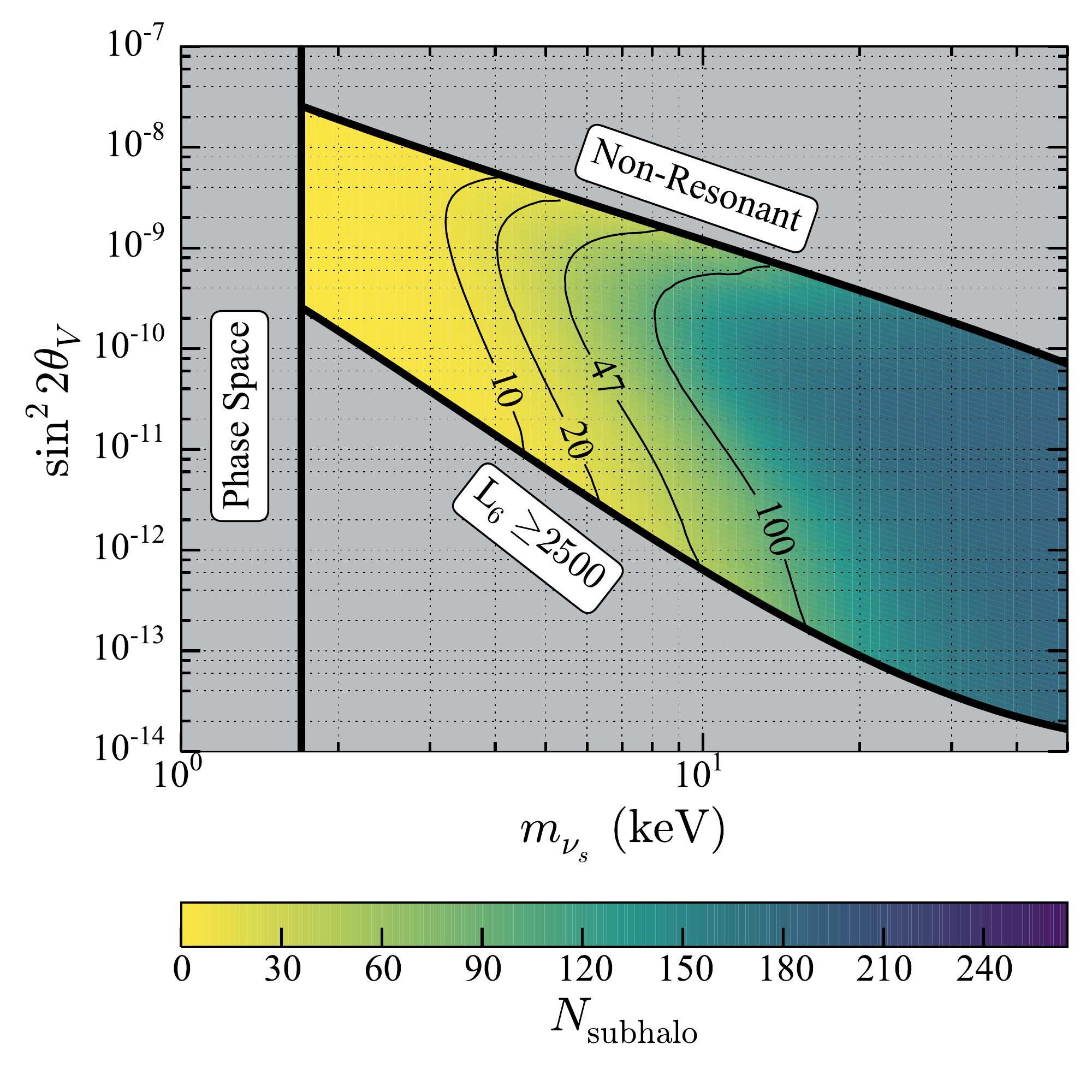}
\caption{The number of detectable subhalos surrounding the Milky Way galaxy for Model 2~\cite{Dunstan:2011aa}.  The isocontour of 47 subhalos is the $95\%$ CI limit on the all-sky corrected SDSS subhalo counts.}
\label{fig:Kevsubh}
\end{center}
\end{figure}

We show in Fig.~\ref{fig:subh} the results of Model 1's predictions for the subhalo counts using identical input parameters to Ref.~\cite{Schneider:2016aa}.  To conservatively estimate the number of subhalos when integrating the subhalo mass distribution function we take the upper bound on the mass of the Milky Way Galaxy, $3\times 10^{12}\,M_\odot/h$, and the lower bound comparable to the subhalo mass of known Milky Way and M31 satellite galaxies, $10^8\, M_\odot/h$ (see below).  Model 1 is a duplicate of the calculation of~\cite{Schneider:2015aa,Schneider:2016aa} and, unsurprisingly, the results of our calculation of the number of Milky Way satellites agree. 

In Fig.~\ref{fig:Kevsubh} we show the results of our subhalo number calculation for Model 2.  Again we use identical limits of integration for the subhalo mass function of $10^8\, M_\odot/h$ and $3\times 10^{12}\,M_\odot/h$.  This helps define the quantity which we will refer to as $N_{\rm subhalo}$, which is the number of observable satellite galaxies within this mass range that orbit a Milky Way sized host galaxy.

There is appreciable disagreement between these two models in the large $k_{\rm cut}$ limit, when the matter power suppression is limited and dark matter is effectively CDM-like.  Model 1 predicts $N_{\rm subhalo} = 265$ in this limit while Model 2 predicts $N_{\rm subhalo} = 185$.  While this $40\%$ disagreement is quite large, we find that it is attributable to a disagreement the overall slope of the subhalo mass function, $dN/dm$, which biases Model 1 toward a greater preponderance of low mass halos near the lower bound of our mass cut, $10^8\, M_\odot/h$.  This is a result which is not unexpected, given the distinct initial assumptions for cosmological parameters in numerical simulations and the subsequent mapping of those results into each analytic framework.

The agreement between models is considerably better when restricting the subhalo population to its more massive constituents through matter power spectrum suppression.  In the regime where $k_{\rm cut} < 40 \,{h \rm Mpc^{-1}}$, both Model 1 and Model 2 obtain a good agreement on the Milky Way subhalo count to within (a few)$\%$.  This is a fortuitous result because the current limits on subhalo counts from direct observation of galactic satellites fall within the regime where both models under consideration agree, as we shall see in the next section.  For the sake of being maximally conservative, we will compare observational counts of Milky Way satellites to which ever model predicts {\it more} satellites for a given point in the sterile neutrino parameter space.

To increase the sample size we are considering, we also model the expected number of subhalos of M31.  Based on current estimates, M31 is roughly equivalent to the Milky Way in total dark matter halo mass (perhaps slightly more massive)~\cite{Wang:2015aa}.  We thus set the M31 halo mass to $3\times 10^{12}\,M_\odot/h$, similar to the Milky Way projection.  Based on the results of~\cite{Brooks:2014aa} we infer the mass within the half light radius of the M31 satellites based on their absolute magnitude, $M_V$, and from there extrapolate the full halo mass~\cite{Bullock:2001aa}.  This results in a nearly identical subhalo mass range for the M31 satellites to that found for the Milky Way satellite galaxies in~\cite{Strigari:2008aa}.  From this we also conclude that a conservative lower bound on M31 subhalo masses would be $10^8\, M_\odot/h$, making it a twin for the Milky Way in terms of our model predictions of subhalo populations.

To determine the relevant subhalo mass range, we use the results of~\cite{Brooks:2014aa} and extend the mass to light ratio scaling relation of dwarf spheroidal satellites to typical V band magnitudes for candidates in the DES sample~\cite{DES-Collaboration:2016aa}.  We take a parametric fit to the data in~\cite{Brooks:2014aa} and find,
\begin{equation}
{\rm log}_{10} (M/L_V) = 0.26 \times M_V + 4.3\, .
\end{equation}
This gives a typical inferred halo mass of (a few)$\times 10^8 M_\odot/h$ to (a few)$\times 10^9 M_\odot/h$ for the DES candidates.  This suggests we do not need to lower the subhalo lower mass bound below $10^8\, M_\odot/h$ in our model predictions to account for the additional observed satellites in the DES sample.

\section{Sterile Neutrino Constraints}\label{sec:Disc}

\subsection{Revisiting subhalo count constraints}

The formation of galaxies is through a process of hierarchical mergers of smaller progenitor galaxies.  Although the population distributions of host halo to subhalos are strongly correlated, the structure formation process remains fundamentally {\it stochastic} in nature.  The timing and number of satellites presently orbiting or merging with the Milky Way varies randomly because the primordial distribution of dark matter subhalos is similarly a random variation about the primordial dark matter power spectrum.  We therefore use a binned maximum likelihood method~\cite{PDG:2013xk} with the likelihood defined by Poisson statistics (in contrast with previous study of Ref.~\cite{Schneider:2016aa}). The negative log of the likelihood function for observed counts of subhalos orbiting the host galaxies is,
\be
\mathcal{L}\left( \{ x_i\},\{\mu_i\}\right) = \sum_i \left( \mu_i - x_i + x_i \ln \frac{x_i}{\mu_i} \right) \, ,
\label{LLratio}
\ee
where the observed number of sub halos in the data, $\{ x_i\}$, is to be found from the expected number predicted by the model, $\{\mu_i\}$, with $i = 1,...,N$ for $N$ hosts which have satellite count data (for us this is simply $i=1,2$ for the Milky Way and M31, respectively).  An important property of $\mathcal{L}$ is that in the limit of \lq\lq large\rq\rq $\{ x_i\}$ the likelihood ratio probability distribution asymptotes to the $\chi^2$ probability distribution, with the relation $\Delta\mathcal{L} \simeq \chi^2/2$, a result known as Wilks' Theorem~\cite{Wilks:1938uq}.

We define our exclusion criterion as a $95\%$ confidence interval (\lq\lq CI\rq\rq ) overabundance of observed satellites for a one sided Likelihood ratio distribution function, i.e., we exclude models  considering only fluctuations {\it above} the model expectation.  Note that to obtain the likelihood function for the one sided distribution, we set the contribution to the sum in Equation~\ref{LLratio} for galaxy $i$ equal to $0$ if $\mu_i > x_i$.  In the large sub halo count limit, with 2 degrees of freedom (for the two free parameters $\theta_V$ and $m_s$), this gives the asymptotic value $\Delta\mathcal{L} = \Delta \chi^2/2 =  2.3$ from the observed sub halo population to set our lower bound on acceptable predictions.   

The most pessimistic constraint we can construct is to assume the distribution of Milky Way satellite galaxies is {\it maximally} anisotropic, i.e., that $100\%$ of the non-classical satellite galaxies orbiting the Milky Way are found within the SDSS survey field of view.  This gives a total of 26 Milky Way (11 classical and 15 found by SDSS) and 35 M31 satellites~\cite{McConnachie:2012aa}.  Combining these counts we can exclude models with matter power spectra which produce fewer than 24 Milky Way satellite galaxies.  

More realistically, we can take the SDSS limited field of view into account and correct for all sky coverage within the SDSS data set.  Given the $29\%$ sky coverage~\cite{SDSS-Collaboration:2016aa} of the SDSS survey, we find the an isotropic distribution of Milky Way satellites would produce a total of 63 satellite subhalos.   We exclude at $95\%$ CI models which predict less than $47$ subhalos, c.f., the $N_{\rm subhalo} = 47$ contour on Figs.~\ref{fig:subh} and ~\ref{fig:Kevsubh}.  

Interestingly, DES has found a number of new faint satellite candidates~\cite{Koposov:2015aa,Bechtol:2015aa,Drlica-Wagner:2015aa,DES-Collaboration:2016aa}.  However, a number of issues must be addressed before these candidates satellites can be confidently included.  The DES sample shows strong evidence that these objects cluster around the more massive Milky Way satellite galaxies, making an estimate of the all sky number of satellites within the DES observable magnitude range difficult.  Nonetheless the effort has been undertaken by the collaboration to estimate the total number of Milky Way satellites observable by DES, finding $N_{\rm subhalo} \approx 100$~\cite{Drlica-Wagner:2015aa}.  Using this result we project a $95\%$ CI constraint for models which predict fewer than 80 Milky Way satellites. It should nevertheless be cautioned that a number of the DES candidate satellites have not yet had their stellar kinematics measured to infer the masses of their dark matter halos.  Rather than attempting to make a concrete exclusion using the DES satellite candidates, we will make a projection of the possible strength of the combined surveys with the DES data set to constrain sterile neutrino dark matter models.  This allows us to place our projected constraint for the combined candidates counts within the sterile neutrino parameter space, with the caveat the result is dependent on validation of our halo mass estimates with proper stellar kinematics measurements.

We show our results for the combined parameter space exclusion for resonantly produced sterile neutrino dark matter in Fig.~\ref{fig:constraint} on the $\sin^2 2\theta_V\ {\rm vs.}\ m_{\nu_s}$ plane.  Our $95\%$ CI Milky Way satellite count limits for the SDSS subhalo counts are shown in dark blue, with light blue denoting the moderately improved exclusion obtained by applying the all sky correction to the SDSS catalog.  By allowing for stochastic variability of the local subhalo population, we obtain more conservative constraints than Ref.~\cite{Schneider:2016aa}, which did not quantify the goodness of fit for model predictions and imposed an arbitrary subhalo count cutoff.  For both SDSS constraints, we find that the subhalo count predictions of Model 2 are slightly larger and hence more conservative, while for the DES constraint Model 1 produces the more conservative constraint. 

Shown in red is our hypothetical projection for the limit which may be placed by the combined satellite sample.  We emphasize this limit is beholden to kinematic studies of the stellar populations of the DES satellites which will set the host dark matter halo masses and establish the minimum halo mass cutoff needed for DES.  A large impact on the constraint will arise from exclusion of satellite candidates as mis-identified stellar clusters, which lowers the inferred all-sky satellite count below $100$ and driving the combined constraint asymptotically toward the SDSS constraint.  

\begin{figure}[t] 
\begin{center}
\includegraphics[width=.48\textwidth]{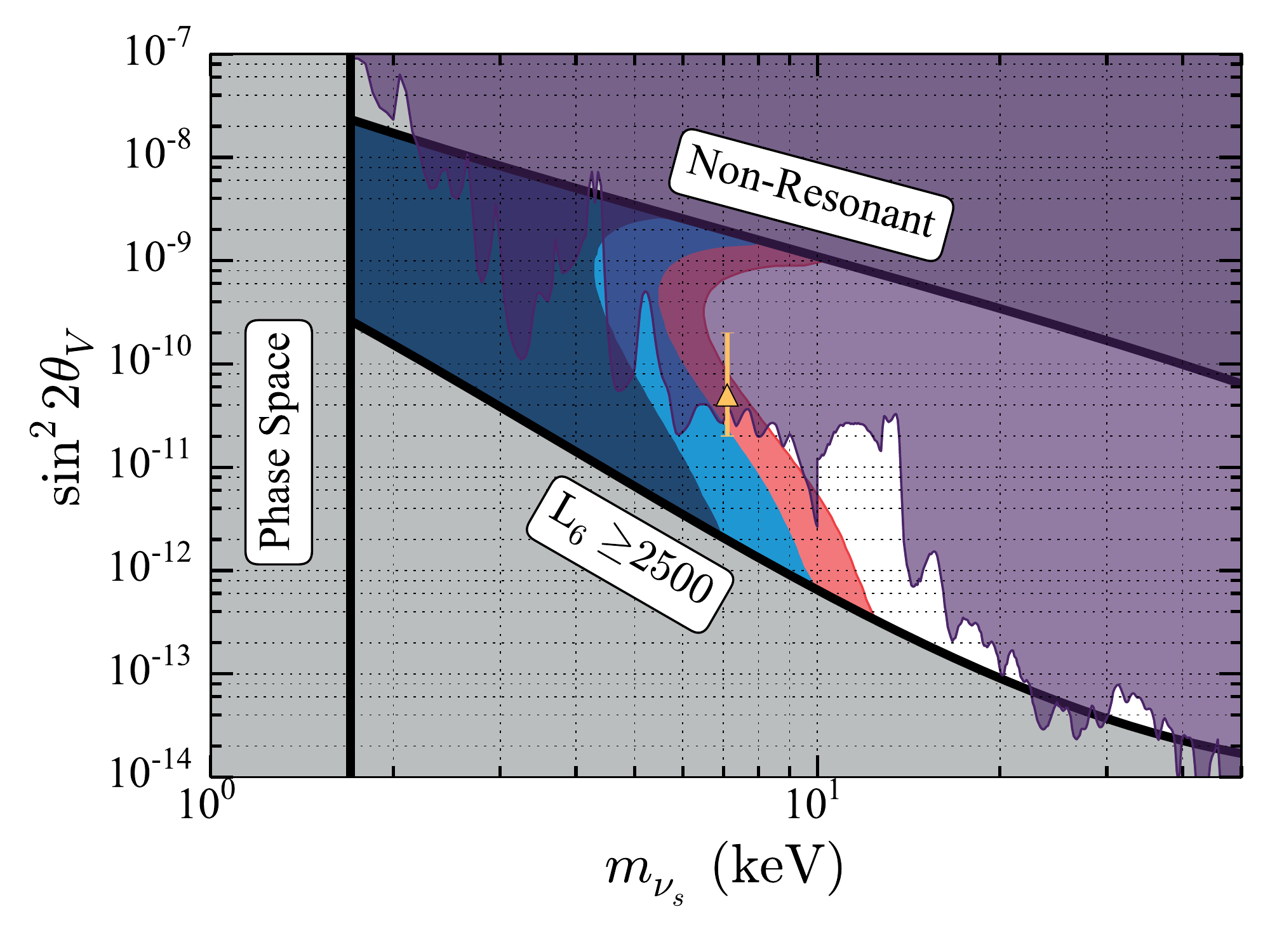}
\caption{Combined constraints for sterile neutrino dark matter.  The dark blue region indicates the SSDS + M31 $95\%$ CI exclusion, while the light blue represents the same exclusion using the all-sky corrected SDSS counts.  The red region indicates our projection for the $95\%$ CI exclusion for the combined surveys.  The purple contour represents the combined $95\%$ CI X-ray exclusion for {\it Chandra} observation of M31 and NuSTAR observations of the galactic center~\cite{Horiuchi:2014aa,Perez:2016aa}.  The orange point indicates the claimed $7.1\, \rm keV$ sterile neutrino decay line candidate~\cite{Bulbul:2014aa,Boyarsky:2014aa}.  }
\label{fig:constraint}
\end{center}
\end{figure}

\subsection{Summary of additional constraints}\label{sec:Disc:others}

So far, we have considered constraints on sterile neutrino dark matter arising from the lepton asymmetry during production, phase-space constraints, and a re-analysis of subhalo constraints. Here we summarize two additional widely considered constraints in the literature: X-ray constraints, which is highly complementary to our subhalo constraints, and Lyman-$\alpha$ forest constraints, which relies on the suppression of power on small scales. 

\textbf{X-ray lines from decays}: Massive sterile neutrinos are not indefinitely stable particles.  First order corrections allow massive neutrinos to branch into a loop containing charged leptons and bosons which can then radiate before the loop closes, leading to the decay of the neutrino into a lighter mass eigen state.  This produces a narrow line emitted photon with $E = m_{\nu_s}/2$.  The decay rate for this process~\cite{Pal:1982aa} is given by,
\begin{equation}
\Gamma = 1.37\times10^{-15}\left(\frac{\sin^2 2\theta_V}{10^{-7}}\right)\left(\frac{m_{\nu_s}}{1\,\rm keV}\right)^5 \, \rm s^{-1}\, .
\end{equation}
Searching for this radiative decay emission provides an important probe for the presence of sterile neutrino dark matter and since its theoretical inception \cite{Abazajian:2001aa,Dolgov:2002aa} has been investigated by many authors, e.g., \cite{Watson:2006aa,Yuksel:2008aa} (see recent review \cite{Adhikari:2017aa} for a full discussion). Limits improve with larger mixing angle and mass, which both increase X-ray fluxes, making X-ray limits highly complementary to subhalo count limits. The $95\%$ CI X-ray exclusion limits of Refs.~\cite{Horiuchi:2014aa,Perez:2016aa} are shown in purple in Fig.~\ref{fig:constraint}.  

\textbf{Lyman-$\alpha$ forest}: the suppression of small-scale matter power will manifest in the Lyman-$\alpha$ forest flux power spectrum, and intriguingly the recently reported limits are stronger than those from subhalo counts \cite{Viel:2013aa,Baur:2016aa}. However, it has recently been pointed out that the gas dynamics of the Inter-Galactic Medium (IGM) can have a dramatic and confounding effect on the absorption spectrum of the Lyman-$\alpha$ forest for redshifts $z<6$~\cite{Kulkarni:2015aa}.  Those authors found that pressure smoothing of the gas in the IGM deviated widely from the predictions of linear theory, and that the gas density power spectrum small scale cut-off for \lcdm was completely erased from the Lyman-$\alpha$ signal.  The conclusion of Ref.~\cite{Kulkarni:2015aa} was that the absorption features of the Lyman-$\alpha$ forest can potentially be dominated by the pressure smoothing scale in the IGM, and until such time as observations are able to characterize which scale there is no known way to disentangle the Lyman-$\alpha$ absorption from IGM pressure smoothing from the absorption due to the matter power spectrum on small length scales. In light of this result, we feel that it is premature to consider the exclusion of the reference models used by~\cite{Schneider:2016aa} to be robust.  As the real-world scale of the pressure smoothing of the IGM remains unknown, we elect not to draw any inferences for resonantly produced sterile neutrino dark matter based on the Lyman-$\alpha$ absorption data.

\subsection{Combined Constraints}\label{sec:Disc:combine}

\begin{figure}[t] 
\begin{center}
 \includegraphics[width=.48\textwidth]{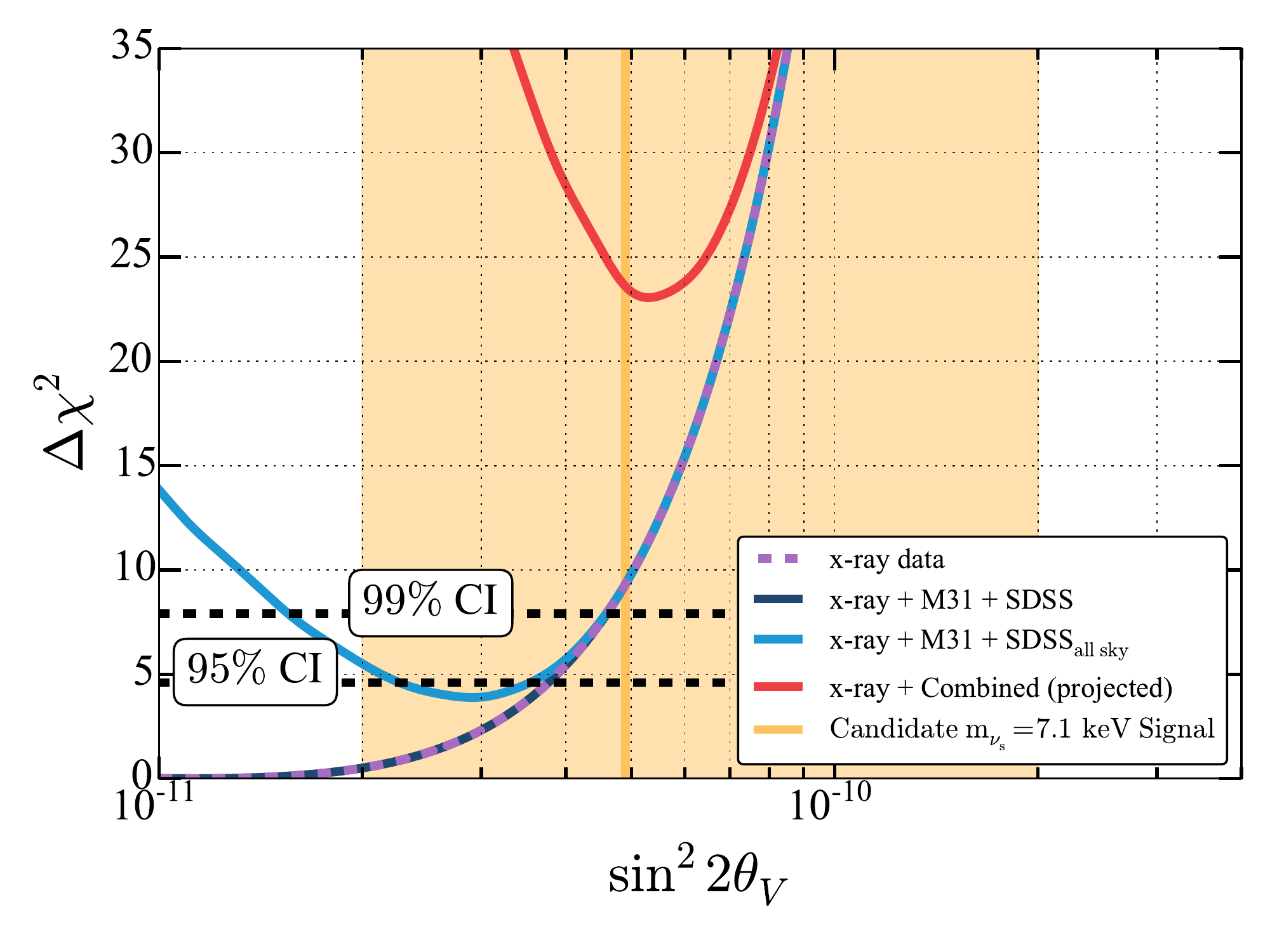}
\caption{The combined goodness of fit for the claimed $7.1\, \rm keV$ sterile neutrino dark matter candidate (the solid orange line indicates the best fit value, with light orange showing the error bars~\cite{Boyarsky:2014aa}).  The purple contours represent the $95\%$ CI Chandra M31 X-ray observation~\cite{Horiuchi:2014aa}.  Dark blue indicates the SSDS + M31 subhalo counts + X-ray data, and light blue represents the SDSS all sky corrected cluster counts + X-ray data.  The red line indicates our projection for the combined satellite survey counts along with X-ray data.}
\label{fig:constraint1D}
\end{center}
\end{figure}

In Fig.~\ref{fig:constraint}, we show both our $95\%$ CI subhalo count limits and the $95\%$ CI X-ray limits of Refs.~\cite{Horiuchi:2014aa,Perez:2016aa}.  We find that by combining SDSS subhalo counts with X-ray data we can exclude sterile neutrino masses $m_{\nu_s} \leq 7.0 \,\rm keV$ with a significance of at least $95\%$ CI, and rule out net lepton asymmetries less than $L_6 \leq 15$. There are only isolated islands of parameter space left for sterile neutrinos produced via oscillations to be the sole dark matter particle. A larger island remains between 10--20 keV mass, and a smaller one at 30--35 keV; the latter is already excluded in $\nu$MSM models \cite{Asaka:2005an,Boyarsky:2006jm,Asaka:2006nq,Canetti:2012vf,Canetti:2012kh} due to lepton asymmetry bounds, leaving the lighter mass range which will further be reduced by accumulated X-ray statistics by NuSTAR \cite{Perez:2016aa} and on-going and future satellite galaxy searches. 

We demonstrate the significance of our subhalo constraints and the power of combining them with X-ray limits by considering the tantalizing but controversial emission line at $3.55\, \rm keV$ reported from various studies observing galaxies and galaxy clusters~\cite{Bulbul:2014aa,Boyarsky:2014aa,Urban:2015aa,Boyarsky:2015aa,Iakubovskyi:2016aa,Franse:2016aa}; other studies find null results that are in conflict with these claims~\cite{Anderson:2015aa,Carlson:2015aa,Horiuchi:2014aa,Tamura:2015aa,Jeltema:2016aa,Ruchayskiy:2016aa,Sekiya:2016aa}. Interpreted as sterile neutrino decay, the require parameters are shown by the orange triangle with error bars in Fig.~\ref{fig:constraint}. In Fig.~\ref{fig:constraint1D} we show the combined goodness of fit for the sterile neutrino interpretation.  We combine the X-ray and subhalo measurement significances in an un-binned fit assuming {${\mathcal L}\approx \Delta \chi^2/2$}.   The dashed black lines indicate the $\Delta \chi^2$ statistical significances for a 1 sided $\chi^2$ distribution function with 2 degrees of freedom.  Presently, the error bars on the value of $\sin^2 2\theta_V$ place the observation of the $3.55\,\rm keV$ X-ray line in moderate, $95\%$ CI, tension with the sterile neutrino dark matter decay explanation.  

The projected goodness of fit from the DES satellite counts + X-ray data are not intended to be conclusive, but they do illustrate an important point about the complementarity of the subhalo count constraints and X-ray emission constraints in the remainder of the resonant regime.  Simply put, incremental increases in subhalo counts result in qualitative jumps in the impacts of sterile neutrino mixing angle constraints.  The majority of the remaining parameter space is dominated by resonant production with large lepton asymmetry.  This favors sterile neutrino populations with larger $\langle p/T\rangle$, which in turn effectively suppress the formation of small scale structure and the number of observable Milk Way satellite galaxies.  As a result, improving the satellite count constraints inexorably closes the parameter space off in the direction of the X-ray limits.  Additionally, improving the X-ray limits above $20\,\rm keV$ photon energy ($m_{\nu_s}>10\,\rm keV$) will close the last portion of the parameter space as yet unconstrained by satellite counts~\cite{Perez:2016aa}.

\section{Conclusions}

We have performed a thorough scan of oscillation-based production of sterile neutrino dark matter with the aim of improving constraints for small-scale structures which arise from the non-thermal sterile neutrino energy spectra. By implementing the updated production calculation {\texttt sterile-dm}~\cite{Venumadhav:2015aa} and considering multiple subhalo count prediction prescriptions, we show that the Milky Way $+$ Andromeda galaxy subhalo counts combined with recent X-ray data leave only small portions of allowed sterile neutrino parameter space: $ 7.0\, {\rm keV} \leq  m_{\nu_s} \leq 36\, {\rm keV}$ and lepton asymmetries $L_6 \geq 15$ at $95\%$ CI or greater. 

We find that the general trends in resonantly produced sterile neutrino dark matter structure formation are robust. Although the asymmetry of the production of $\nu_s$ versus $\bar\nu_s$ in the resonant regime makes prediction of the matter power spectrum challenging we find the ensemble average $\langle p/T\rangle$ remains a reasonable predictor for the matter power spectrum cutoff. However, we find mild disagreement of our results using {\texttt sterile-dm} for computing location of the BBN limit on $L_6$ as compared to those of~\cite{Boyarsky:2009ab}, demonstrating that understanding neutrino transport in the plasma of the early Universe remains one of the most important open topics in the study of sterile neutrino dark matter.  Ref.~\cite{Schneider:2016aa} points out the effect is a weakening of the structure constraints by some $\sim 1\,\rm keV$.  Transport has always been an important part of the resonant production mechanism and we have seen here that even transport below the QCD phase transition temperature can appreciably impact the result.

We do not use our predictions for the dark matter power spectrum to draw conclusions from the Lyman-$\alpha$ forest absorption.  While gas pressure smoothing in the IGM renders interpretation of the Lyman-$\alpha$ spectral features ambiguous, the authors of~\cite{Kulkarni:2015aa} report they will soon be able to tease out smoothing length scale from observation of high redshift quasar pairs.  This will hopefully allow the sterile neutrino modifications to the matter power spectrum contribution of the Lyman-$\alpha$ forest to be isolated and confidently measured.

Finally, we find the $m_{\nu_s} = 7.1 \, \rm keV$ sterile neutrino candidate for explaining the anomalous $3.55\, \rm keV$ X-ray line signal is in moderate, $93\%$ CI at minimum, tension with the combined observation of Milky Way subhalos in the SDSS catalog and non-observation of M31 X-ray emission by {\it Chandra}.  The projected goodness of fit from the inclusion of the DES satellites, while not an actual constraint, illustrates the power of subhalo counts to probe the remaining portions of the allowed sterile neutrino dark matter parameter space.  More stellar kinematic studies, already ongoing~\cite{Koposov:2015ab,Simon:2015aa,Walker:2015aa,Walker:2016aa}, will be needed so that the constraints from observed faint dwarf galaxy counts can be placed definitively within the sterile neutrino dark matter picture. Combined with future X-ray data and new identification strategies \cite{Speckhard:2015eva,Powell:2016zbo}, the nature of the anomalous 3.55 keV signal may be tested in the near future by multiple independent methods. 

\begin{acknowledgments}

We thank Kevork Abazajian, John Beacom, Kenny Ng, and Jose O\~norbe for useful discussions and careful reading of the manuscript. SH acknowledges support from the Institute for Nuclear Theory program ``Neutrino Astrophysics and Fundamental Properties'' 15-2a where this work was initiated. 
\end{acknowledgments}

\bibliography{allref.bib}

\begin{thebibliography}{104}%
\makeatletter
\providecommand \@ifxundefined [1]{%
 \@ifx{#1\undefined}
}%
\providecommand \@ifnum [1]{%
 \ifnum #1\expandafter \@firstoftwo
 \else \expandafter \@secondoftwo
 \fi
}%
\providecommand \@ifx [1]{%
 \ifx #1\expandafter \@firstoftwo
 \else \expandafter \@secondoftwo
 \fi
}%
\providecommand \natexlab [1]{#1}%
\providecommand \enquote  [1]{``#1''}%
\providecommand \bibnamefont  [1]{#1}%
\providecommand \bibfnamefont [1]{#1}%
\providecommand \citenamefont [1]{#1}%
\providecommand \href@noop [0]{\@secondoftwo}%
\providecommand \href [0]{\begingroup \@sanitize@url \@href}%
\providecommand \@href[1]{\@@startlink{#1}\@@href}%
\providecommand \@@href[1]{\endgroup#1\@@endlink}%
\providecommand \@sanitize@url [0]{\catcode `\\12\catcode `\$12\catcode
  `\&12\catcode `\#12\catcode `\^12\catcode `\_12\catcode `\%12\relax}%
\providecommand \@@startlink[1]{}%
\providecommand \@@endlink[0]{}%
\providecommand \url  [0]{\begingroup\@sanitize@url \@url }%
\providecommand \@url [1]{\endgroup\@href {#1}{\urlprefix }}%
\providecommand \urlprefix  [0]{URL }%
\providecommand \Eprint [0]{\href }%
\providecommand \doibase [0]{http://dx.doi.org/}%
\providecommand \selectlanguage [0]{\@gobble}%
\providecommand \bibinfo  [0]{\@secondoftwo}%
\providecommand \bibfield  [0]{\@secondoftwo}%
\providecommand \translation [1]{[#1]}%
\providecommand \BibitemOpen [0]{}%
\providecommand \bibitemStop [0]{}%
\providecommand \bibitemNoStop [0]{.\EOS\space}%
\providecommand \EOS [0]{\spacefactor3000\relax}%
\providecommand \BibitemShut  [1]{\csname bibitem#1\endcsname}%
\let\auto@bib@innerbib\@empty
\bibitem [{\citenamefont {{Dodelson}}\ and\ \citenamefont
  {{Widrow}}(1994)}]{Dodelson:1994aa}%
  \BibitemOpen
  \bibfield  {author} {\bibinfo {author} {\bibfnamefont {S.}~\bibnamefont
  {{Dodelson}}}\ and\ \bibinfo {author} {\bibfnamefont {L.~M.}\ \bibnamefont
  {{Widrow}}},\ }\href {\doibase 10.1103/PhysRevLett.72.17} {\bibfield
  {journal} {\bibinfo  {journal} {Physical Review Letters}\ }\textbf {\bibinfo
  {volume} {72}},\ \bibinfo {pages} {17} (\bibinfo {year} {1994})},\ \Eprint
  {http://arxiv.org/abs/hep-ph/9303287} {hep-ph/9303287} \BibitemShut {NoStop}%
\bibitem [{\citenamefont {{Shi}}\ and\ \citenamefont
  {{Fuller}}(1999)}]{Shi:1999aa}%
  \BibitemOpen
  \bibfield  {author} {\bibinfo {author} {\bibfnamefont {X.}~\bibnamefont
  {{Shi}}}\ and\ \bibinfo {author} {\bibfnamefont {G.~M.}\ \bibnamefont
  {{Fuller}}},\ }\href {\doibase 10.1103/PhysRevLett.83.3120} {\bibfield
  {journal} {\bibinfo  {journal} {Physical Review Letters}\ }\textbf {\bibinfo
  {volume} {83}},\ \bibinfo {pages} {3120} (\bibinfo {year} {1999})},\ \Eprint
  {http://arxiv.org/abs/astro-ph/9904041} {astro-ph/9904041} \BibitemShut
  {NoStop}%
\bibitem [{\citenamefont {Bozek}\ \emph {et~al.}(2016)\citenamefont {Bozek},
  \citenamefont {Boylan-Kolchin}, \citenamefont {Horiuchi}, \citenamefont
  {Garrison-Kimmel}, \citenamefont {Abazajian},\ and\ \citenamefont
  {Bullock}}]{Bozek:2015bdo}%
  \BibitemOpen
  \bibfield  {author} {\bibinfo {author} {\bibfnamefont {B.}~\bibnamefont
  {Bozek}}, \bibinfo {author} {\bibfnamefont {M.}~\bibnamefont
  {Boylan-Kolchin}}, \bibinfo {author} {\bibfnamefont {S.}~\bibnamefont
  {Horiuchi}}, \bibinfo {author} {\bibfnamefont {S.}~\bibnamefont
  {Garrison-Kimmel}}, \bibinfo {author} {\bibfnamefont {K.}~\bibnamefont
  {Abazajian}}, \ and\ \bibinfo {author} {\bibfnamefont {J.~S.}\ \bibnamefont
  {Bullock}},\ }\href {\doibase 10.1093/mnras/stw688} {\bibfield  {journal}
  {\bibinfo  {journal} {Mon. Not. Roy. Astron. Soc.}\ }\textbf {\bibinfo
  {volume} {459}},\ \bibinfo {pages} {1489} (\bibinfo {year} {2016})},\ \Eprint
  {http://arxiv.org/abs/1512.04544} {arXiv:1512.04544 [astro-ph.CO]}
  \BibitemShut {NoStop}%
\bibitem [{\citenamefont {Moore}(1994)}]{Moore:1994yx}%
  \BibitemOpen
  \bibfield  {author} {\bibinfo {author} {\bibfnamefont {B.}~\bibnamefont
  {Moore}},\ }\href {\doibase 10.1038/370629a0} {\bibfield  {journal} {\bibinfo
   {journal} {Nature}\ }\textbf {\bibinfo {volume} {370}},\ \bibinfo {pages}
  {629} (\bibinfo {year} {1994})}\BibitemShut {NoStop}%
\bibitem [{\citenamefont {Flores}\ and\ \citenamefont
  {Primack}(1994)}]{Flores:1994gz}%
  \BibitemOpen
  \bibfield  {author} {\bibinfo {author} {\bibfnamefont {R.~A.}\ \bibnamefont
  {Flores}}\ and\ \bibinfo {author} {\bibfnamefont {J.~R.}\ \bibnamefont
  {Primack}},\ }\href {\doibase 10.1086/187350} {\bibfield  {journal} {\bibinfo
   {journal} {Astrophys.J.}\ }\textbf {\bibinfo {volume} {427}},\ \bibinfo
  {pages} {L1} (\bibinfo {year} {1994})},\ \Eprint
  {http://arxiv.org/abs/astro-ph/9402004} {arXiv:astro-ph/9402004 [astro-ph]}
  \BibitemShut {NoStop}%
\bibitem [{\citenamefont {Navarro}\ \emph {et~al.}(1997)\citenamefont
  {Navarro}, \citenamefont {Frenk},\ and\ \citenamefont
  {White}}]{Navarro:1996gj}%
  \BibitemOpen
  \bibfield  {author} {\bibinfo {author} {\bibfnamefont {J.~F.}\ \bibnamefont
  {Navarro}}, \bibinfo {author} {\bibfnamefont {C.~S.}\ \bibnamefont {Frenk}},
  \ and\ \bibinfo {author} {\bibfnamefont {S.~D.}\ \bibnamefont {White}},\
  }\href {\doibase 10.1086/304888} {\bibfield  {journal} {\bibinfo  {journal}
  {Astrophys.J.}\ }\textbf {\bibinfo {volume} {490}},\ \bibinfo {pages} {493}
  (\bibinfo {year} {1997})},\ \Eprint {http://arxiv.org/abs/astro-ph/9611107}
  {arXiv:astro-ph/9611107 [astro-ph]} \BibitemShut {NoStop}%
\bibitem [{\citenamefont {{Gentile}}\ \emph {et~al.}(2004)\citenamefont
  {{Gentile}}, \citenamefont {{Salucci}}, \citenamefont {{Klein}},
  \citenamefont {{Vergani}},\ and\ \citenamefont
  {{Kalberla}}}]{Gentile:2004aa}%
  \BibitemOpen
  \bibfield  {author} {\bibinfo {author} {\bibfnamefont {G.}~\bibnamefont
  {{Gentile}}}, \bibinfo {author} {\bibfnamefont {P.}~\bibnamefont
  {{Salucci}}}, \bibinfo {author} {\bibfnamefont {U.}~\bibnamefont {{Klein}}},
  \bibinfo {author} {\bibfnamefont {D.}~\bibnamefont {{Vergani}}}, \ and\
  \bibinfo {author} {\bibfnamefont {P.}~\bibnamefont {{Kalberla}}},\ }\href
  {\doibase 10.1111/j.1365-2966.2004.07836.x} {\bibfield  {journal} {\bibinfo
  {journal} {MNRAS}\ }\textbf {\bibinfo {volume} {351}},\ \bibinfo {pages}
  {903} (\bibinfo {year} {2004})},\ \Eprint
  {http://arxiv.org/abs/astro-ph/0403154} {astro-ph/0403154} \BibitemShut
  {NoStop}%
\bibitem [{\citenamefont {{Gilmore}}\ \emph {et~al.}(2007)\citenamefont
  {{Gilmore}}, \citenamefont {{Wilkinson}}, \citenamefont {{Kleyna}},
  \citenamefont {{Koch}}, \citenamefont {{Evans}}, \citenamefont {{Wyse}},\
  and\ \citenamefont {{Grebel}}}]{Gilmore:2007aa}%
  \BibitemOpen
  \bibfield  {author} {\bibinfo {author} {\bibfnamefont {G.}~\bibnamefont
  {{Gilmore}}}, \bibinfo {author} {\bibfnamefont {M.}~\bibnamefont
  {{Wilkinson}}}, \bibinfo {author} {\bibfnamefont {J.}~\bibnamefont
  {{Kleyna}}}, \bibinfo {author} {\bibfnamefont {A.}~\bibnamefont {{Koch}}},
  \bibinfo {author} {\bibfnamefont {W.}~\bibnamefont {{Evans}}}, \bibinfo
  {author} {\bibfnamefont {R.~F.~G.}\ \bibnamefont {{Wyse}}}, \ and\ \bibinfo
  {author} {\bibfnamefont {E.~K.}\ \bibnamefont {{Grebel}}},\ }\href {\doibase
  10.1016/j.nuclphysbps.2007.08.143} {\bibfield  {journal} {\bibinfo  {journal}
  {Nuclear Physics B Proceedings Supplements}\ }\textbf {\bibinfo {volume}
  {173}},\ \bibinfo {pages} {15} (\bibinfo {year} {2007})},\ \Eprint
  {http://arxiv.org/abs/astro-ph/0608528} {astro-ph/0608528} \BibitemShut
  {NoStop}%
\bibitem [{\citenamefont {{Frusciante}}\ \emph {et~al.}(2012)\citenamefont
  {{Frusciante}}, \citenamefont {{Salucci}}, \citenamefont {{Vernieri}},
  \citenamefont {{Cannon}},\ and\ \citenamefont {{Elson}}}]{Frusciante:2012aa}%
  \BibitemOpen
  \bibfield  {author} {\bibinfo {author} {\bibfnamefont {N.}~\bibnamefont
  {{Frusciante}}}, \bibinfo {author} {\bibfnamefont {P.}~\bibnamefont
  {{Salucci}}}, \bibinfo {author} {\bibfnamefont {D.}~\bibnamefont
  {{Vernieri}}}, \bibinfo {author} {\bibfnamefont {J.~M.}\ \bibnamefont
  {{Cannon}}}, \ and\ \bibinfo {author} {\bibfnamefont {E.~C.}\ \bibnamefont
  {{Elson}}},\ }\href {\doibase 10.1111/j.1365-2966.2012.21495.x} {\bibfield
  {journal} {\bibinfo  {journal} {MNRAS}\ }\textbf {\bibinfo {volume} {426}},\
  \bibinfo {pages} {751} (\bibinfo {year} {2012})},\ \Eprint
  {http://arxiv.org/abs/1206.0314} {arXiv:1206.0314 [astro-ph.CO]} \BibitemShut
  {NoStop}%
\bibitem [{\citenamefont {Boylan-Kolchin}\ \emph {et~al.}(2011)\citenamefont
  {Boylan-Kolchin}, \citenamefont {Bullock},\ and\ \citenamefont
  {Kaplinghat}}]{BoylanKolchin:2011de}%
  \BibitemOpen
  \bibfield  {author} {\bibinfo {author} {\bibfnamefont {M.}~\bibnamefont
  {Boylan-Kolchin}}, \bibinfo {author} {\bibfnamefont {J.~S.}\ \bibnamefont
  {Bullock}}, \ and\ \bibinfo {author} {\bibfnamefont {M.}~\bibnamefont
  {Kaplinghat}},\ }\href@noop {} {\bibfield  {journal} {\bibinfo  {journal}
  {Mon.Not.Roy.Astron.Soc.}\ }\textbf {\bibinfo {volume} {415}},\ \bibinfo
  {pages} {L40} (\bibinfo {year} {2011})},\ \Eprint
  {http://arxiv.org/abs/1103.0007} {arXiv:1103.0007 [astro-ph.CO]} \BibitemShut
  {NoStop}%
\bibitem [{\citenamefont {{Walker}}(2013)}]{Walker:2013aa}%
  \BibitemOpen
  \bibfield  {author} {\bibinfo {author} {\bibfnamefont {M.}~\bibnamefont
  {{Walker}}},\ }\enquote {\bibinfo {title} {{Dark Matter in the Galactic Dwarf
  Spheroidal Satellites}},}\ in\ \href {\doibase 10.1007/978-94-007-5612-0_20}
  {\emph {\bibinfo {booktitle} {Planets, Stars and Stellar Systems.~Volume 5:
  Galactic Structure and Stellar Populations}}},\ \bibinfo {editor} {edited by\
  \bibinfo {editor} {\bibfnamefont {T.~D.}\ \bibnamefont {{Oswalt}}}\ and\
  \bibinfo {editor} {\bibfnamefont {G.}~\bibnamefont {{Gilmore}}}}\ (\bibinfo
  {year} {2013})\ p.\ \bibinfo {pages} {1039},\ \Eprint
  {http://arxiv.org/abs/1205.0311} {arXiv:1205.0311 [astro-ph.CO]} \BibitemShut
  {NoStop}%
\bibitem [{\citenamefont {Klypin}\ \emph {et~al.}(1999)\citenamefont {Klypin},
  \citenamefont {Kravtsov}, \citenamefont {Valenzuela},\ and\ \citenamefont
  {Prada}}]{Klypin:1999uc}%
  \BibitemOpen
  \bibfield  {author} {\bibinfo {author} {\bibfnamefont {A.~A.}\ \bibnamefont
  {Klypin}}, \bibinfo {author} {\bibfnamefont {A.~V.}\ \bibnamefont
  {Kravtsov}}, \bibinfo {author} {\bibfnamefont {O.}~\bibnamefont
  {Valenzuela}}, \ and\ \bibinfo {author} {\bibfnamefont {F.}~\bibnamefont
  {Prada}},\ }\href {\doibase 10.1086/307643} {\bibfield  {journal} {\bibinfo
  {journal} {Astrophys.J.}\ }\textbf {\bibinfo {volume} {522}},\ \bibinfo
  {pages} {82} (\bibinfo {year} {1999})},\ \Eprint
  {http://arxiv.org/abs/astro-ph/9901240} {arXiv:astro-ph/9901240 [astro-ph]}
  \BibitemShut {NoStop}%
\bibitem [{\citenamefont {Moore}\ \emph {et~al.}(1999)\citenamefont {Moore},
  \citenamefont {Ghigna}, \citenamefont {Governato}, \citenamefont {Lake},
  \citenamefont {Quinn} \emph {et~al.}}]{Moore:1999nt}%
  \BibitemOpen
  \bibfield  {author} {\bibinfo {author} {\bibfnamefont {B.}~\bibnamefont
  {Moore}}, \bibinfo {author} {\bibfnamefont {S.}~\bibnamefont {Ghigna}},
  \bibinfo {author} {\bibfnamefont {F.}~\bibnamefont {Governato}}, \bibinfo
  {author} {\bibfnamefont {G.}~\bibnamefont {Lake}}, \bibinfo {author}
  {\bibfnamefont {T.~R.}\ \bibnamefont {Quinn}},  \emph {et~al.},\ }\href
  {\doibase 10.1086/312287} {\bibfield  {journal} {\bibinfo  {journal}
  {Astrophys.J.}\ }\textbf {\bibinfo {volume} {524}},\ \bibinfo {pages} {L19}
  (\bibinfo {year} {1999})},\ \Eprint {http://arxiv.org/abs/astro-ph/9907411}
  {arXiv:astro-ph/9907411 [astro-ph]} \BibitemShut {NoStop}%
\bibitem [{\citenamefont {Kauffmann}\ \emph {et~al.}(1993)\citenamefont
  {Kauffmann}, \citenamefont {White},\ and\ \citenamefont
  {Guiderdoni}}]{Kauffmann:1993gv}%
  \BibitemOpen
  \bibfield  {author} {\bibinfo {author} {\bibfnamefont {G.}~\bibnamefont
  {Kauffmann}}, \bibinfo {author} {\bibfnamefont {S.~D.}\ \bibnamefont
  {White}}, \ and\ \bibinfo {author} {\bibfnamefont {B.}~\bibnamefont
  {Guiderdoni}},\ }\href@noop {} {\bibfield  {journal} {\bibinfo  {journal}
  {Mon.Not.Roy.Astron.Soc.}\ }\textbf {\bibinfo {volume} {264}},\ \bibinfo
  {pages} {201} (\bibinfo {year} {1993})}\BibitemShut {NoStop}%
\bibitem [{\citenamefont {{Lovell}}\ \emph {et~al.}(2012)\citenamefont
  {{Lovell}}, \citenamefont {{Eke}}, \citenamefont {{Frenk}}, \citenamefont
  {{Gao}}, \citenamefont {{Jenkins}}, \citenamefont {{Theuns}}, \citenamefont
  {{Wang}}, \citenamefont {{White}}, \citenamefont {{Boyarsky}},\ and\
  \citenamefont {{Ruchayskiy}}}]{Lovell:2012aa}%
  \BibitemOpen
  \bibfield  {author} {\bibinfo {author} {\bibfnamefont {M.~R.}\ \bibnamefont
  {{Lovell}}}, \bibinfo {author} {\bibfnamefont {V.}~\bibnamefont {{Eke}}},
  \bibinfo {author} {\bibfnamefont {C.~S.}\ \bibnamefont {{Frenk}}}, \bibinfo
  {author} {\bibfnamefont {L.}~\bibnamefont {{Gao}}}, \bibinfo {author}
  {\bibfnamefont {A.}~\bibnamefont {{Jenkins}}}, \bibinfo {author}
  {\bibfnamefont {T.}~\bibnamefont {{Theuns}}}, \bibinfo {author}
  {\bibfnamefont {J.}~\bibnamefont {{Wang}}}, \bibinfo {author} {\bibfnamefont
  {S.~D.~M.}\ \bibnamefont {{White}}}, \bibinfo {author} {\bibfnamefont
  {A.}~\bibnamefont {{Boyarsky}}}, \ and\ \bibinfo {author} {\bibfnamefont
  {O.}~\bibnamefont {{Ruchayskiy}}},\ }\href {\doibase
  10.1111/j.1365-2966.2011.20200.x} {\bibfield  {journal} {\bibinfo  {journal}
  {MNRAS}\ }\textbf {\bibinfo {volume} {420}},\ \bibinfo {pages} {2318}
  (\bibinfo {year} {2012})},\ \Eprint {http://arxiv.org/abs/1104.2929}
  {arXiv:1104.2929} \BibitemShut {NoStop}%
\bibitem [{\citenamefont {{Anderhalden}}\ \emph {et~al.}(2013)\citenamefont
  {{Anderhalden}}, \citenamefont {{Schneider}}, \citenamefont {{Macci{\`o}}},
  \citenamefont {{Diemand}},\ and\ \citenamefont
  {{Bertone}}}]{Anderhalden:2013aa}%
  \BibitemOpen
  \bibfield  {author} {\bibinfo {author} {\bibfnamefont {D.}~\bibnamefont
  {{Anderhalden}}}, \bibinfo {author} {\bibfnamefont {A.}~\bibnamefont
  {{Schneider}}}, \bibinfo {author} {\bibfnamefont {A.~V.}\ \bibnamefont
  {{Macci{\`o}}}}, \bibinfo {author} {\bibfnamefont {J.}~\bibnamefont
  {{Diemand}}}, \ and\ \bibinfo {author} {\bibfnamefont {G.}~\bibnamefont
  {{Bertone}}},\ }\href {\doibase 10.1088/1475-7516/2013/03/014} {\bibfield
  {journal} {\bibinfo  {journal} {J. Cosmol. Astropart. Phys.}\ }\textbf
  {\bibinfo {volume} {3}},\ \bibinfo {eid} {014} (\bibinfo {year} {2013})},\
  \Eprint {http://arxiv.org/abs/1212.2967} {arXiv:1212.2967} \BibitemShut
  {NoStop}%
\bibitem [{\citenamefont {{Abazajian}}\ \emph {et~al.}(2014)\citenamefont
  {{Abazajian}}, \citenamefont {{Canac}}, \citenamefont {{Horiuchi}},\ and\
  \citenamefont {{Kaplinghat}}}]{Abazajian:2014aa}%
  \BibitemOpen
  \bibfield  {author} {\bibinfo {author} {\bibfnamefont {K.~N.}\ \bibnamefont
  {{Abazajian}}}, \bibinfo {author} {\bibfnamefont {N.}~\bibnamefont
  {{Canac}}}, \bibinfo {author} {\bibfnamefont {S.}~\bibnamefont {{Horiuchi}}},
  \ and\ \bibinfo {author} {\bibfnamefont {M.}~\bibnamefont {{Kaplinghat}}},\
  }\href {\doibase 10.1103/PhysRevD.90.023526} {\bibfield  {journal} {\bibinfo
  {journal} {\prd}\ }\textbf {\bibinfo {volume} {90}},\ \bibinfo {eid} {023526}
  (\bibinfo {year} {2014})},\ \Eprint {http://arxiv.org/abs/1402.4090}
  {arXiv:1402.4090 [astro-ph.HE]} \BibitemShut {NoStop}%
\bibitem [{\citenamefont {{Horiuchi}}\ \emph {et~al.}(2016)\citenamefont
  {{Horiuchi}}, \citenamefont {{Bozek}}, \citenamefont {{Abazajian}},
  \citenamefont {{Boylan-Kolchin}}, \citenamefont {{Bullock}}, \citenamefont
  {{Garrison-Kimmel}},\ and\ \citenamefont {{Onorbe}}}]{Horiuchi:2016aa}%
  \BibitemOpen
  \bibfield  {author} {\bibinfo {author} {\bibfnamefont {S.}~\bibnamefont
  {{Horiuchi}}}, \bibinfo {author} {\bibfnamefont {B.}~\bibnamefont {{Bozek}}},
  \bibinfo {author} {\bibfnamefont {K.~N.}\ \bibnamefont {{Abazajian}}},
  \bibinfo {author} {\bibfnamefont {M.}~\bibnamefont {{Boylan-Kolchin}}},
  \bibinfo {author} {\bibfnamefont {J.~S.}\ \bibnamefont {{Bullock}}}, \bibinfo
  {author} {\bibfnamefont {S.}~\bibnamefont {{Garrison-Kimmel}}}, \ and\
  \bibinfo {author} {\bibfnamefont {J.}~\bibnamefont {{Onorbe}}},\ }\href
  {\doibase 10.1093/mnras/stv2922} {\bibfield  {journal} {\bibinfo  {journal}
  {MNRAS}\ }\textbf {\bibinfo {volume} {456}},\ \bibinfo {pages} {4346}
  (\bibinfo {year} {2016})},\ \Eprint {http://arxiv.org/abs/1512.04548}
  {arXiv:1512.04548} \BibitemShut {NoStop}%
\bibitem [{\citenamefont {Harada}\ and\ \citenamefont
  {Kamada}(2016)}]{Harada:2014lma}%
  \BibitemOpen
  \bibfield  {author} {\bibinfo {author} {\bibfnamefont {A.}~\bibnamefont
  {Harada}}\ and\ \bibinfo {author} {\bibfnamefont {A.}~\bibnamefont
  {Kamada}},\ }\href {\doibase 10.1088/1475-7516/2016/01/031} {\bibfield
  {journal} {\bibinfo  {journal} {JCAP}\ }\textbf {\bibinfo {volume} {1601}},\
  \bibinfo {pages} {031} (\bibinfo {year} {2016})},\ \Eprint
  {http://arxiv.org/abs/1412.1592} {arXiv:1412.1592 [astro-ph.CO]} \BibitemShut
  {NoStop}%
\bibitem [{\citenamefont {Bose}\ \emph {et~al.}(2016)\citenamefont {Bose},
  \citenamefont {Hellwing}, \citenamefont {Frenk}, \citenamefont {Jenkins},
  \citenamefont {Lovell}, \citenamefont {Helly}, \citenamefont {Li},\ and\
  \citenamefont {Gao}}]{Bose:2016irl}%
  \BibitemOpen
  \bibfield  {author} {\bibinfo {author} {\bibfnamefont {S.}~\bibnamefont
  {Bose}}, \bibinfo {author} {\bibfnamefont {W.~A.}\ \bibnamefont {Hellwing}},
  \bibinfo {author} {\bibfnamefont {C.~S.}\ \bibnamefont {Frenk}}, \bibinfo
  {author} {\bibfnamefont {A.}~\bibnamefont {Jenkins}}, \bibinfo {author}
  {\bibfnamefont {M.~R.}\ \bibnamefont {Lovell}}, \bibinfo {author}
  {\bibfnamefont {J.~C.}\ \bibnamefont {Helly}}, \bibinfo {author}
  {\bibfnamefont {B.}~\bibnamefont {Li}}, \ and\ \bibinfo {author}
  {\bibfnamefont {L.}~\bibnamefont {Gao}},\ }\href {\doibase
  10.1093/mnras/stw2686} {\  (\bibinfo {year} {2016}),\
  10.1093/mnras/stw2686},\ \Eprint {http://arxiv.org/abs/1604.07409}
  {arXiv:1604.07409 [astro-ph.CO]} \BibitemShut {NoStop}%
\bibitem [{\citenamefont {{Lovell}}\ \emph
  {et~al.}(2016{\natexlab{a}})\citenamefont {{Lovell}}, \citenamefont {{Bose}},
  \citenamefont {{Boyarsky}}, \citenamefont {{Crain}}, \citenamefont {{Frenk}},
  \citenamefont {{Hellwing}}, \citenamefont {{Ludlow}}, \citenamefont
  {{Navarro}}, \citenamefont {{Ruchayskiy}}, \citenamefont {{Sawala}},
  \citenamefont {{Schaller}}, \citenamefont {{Schaye}},\ and\ \citenamefont
  {{Theuns}}}]{Lovell:2016ab}%
  \BibitemOpen
  \bibfield  {author} {\bibinfo {author} {\bibfnamefont {M.~R.}\ \bibnamefont
  {{Lovell}}}, \bibinfo {author} {\bibfnamefont {S.}~\bibnamefont {{Bose}}},
  \bibinfo {author} {\bibfnamefont {A.}~\bibnamefont {{Boyarsky}}}, \bibinfo
  {author} {\bibfnamefont {R.~A.}\ \bibnamefont {{Crain}}}, \bibinfo {author}
  {\bibfnamefont {C.~S.}\ \bibnamefont {{Frenk}}}, \bibinfo {author}
  {\bibfnamefont {W.~A.}\ \bibnamefont {{Hellwing}}}, \bibinfo {author}
  {\bibfnamefont {A.~D.}\ \bibnamefont {{Ludlow}}}, \bibinfo {author}
  {\bibfnamefont {J.~F.}\ \bibnamefont {{Navarro}}}, \bibinfo {author}
  {\bibfnamefont {O.}~\bibnamefont {{Ruchayskiy}}}, \bibinfo {author}
  {\bibfnamefont {T.}~\bibnamefont {{Sawala}}}, \bibinfo {author}
  {\bibfnamefont {M.}~\bibnamefont {{Schaller}}}, \bibinfo {author}
  {\bibfnamefont {J.}~\bibnamefont {{Schaye}}}, \ and\ \bibinfo {author}
  {\bibfnamefont {T.}~\bibnamefont {{Theuns}}},\ }\href@noop {} {\bibfield
  {journal} {\bibinfo  {journal} {ArXiv e-prints}\ } (\bibinfo {year}
  {2016}{\natexlab{a}})},\ \Eprint {http://arxiv.org/abs/1611.00010}
  {arXiv:1611.00010} \BibitemShut {NoStop}%
\bibitem [{\citenamefont {{Kusenko}}(2009)}]{Kusenko:2009aa}%
  \BibitemOpen
  \bibfield  {author} {\bibinfo {author} {\bibfnamefont {A.}~\bibnamefont
  {{Kusenko}}},\ }\href@noop {} {\bibfield  {journal} {\bibinfo  {journal}
  {Phys. Rep.}\ }\textbf {\bibinfo {volume} {481}},\ \bibinfo {pages} {1}
  (\bibinfo {year} {2009})},\ \Eprint {http://arxiv.org/abs/0906.2968v3}
  {arXiv:0906.2968v3 [hep-ph]} \BibitemShut {NoStop}%
\bibitem [{\citenamefont {{Boyarsky}}\ \emph
  {et~al.}(2009{\natexlab{a}})\citenamefont {{Boyarsky}}, \citenamefont
  {{Ruchayskiy}},\ and\ \citenamefont {{Shaposhnikov}}}]{Boyarsky:2009ab}%
  \BibitemOpen
  \bibfield  {author} {\bibinfo {author} {\bibfnamefont {A.}~\bibnamefont
  {{Boyarsky}}}, \bibinfo {author} {\bibfnamefont {O.}~\bibnamefont
  {{Ruchayskiy}}}, \ and\ \bibinfo {author} {\bibfnamefont {M.}~\bibnamefont
  {{Shaposhnikov}}},\ }\href {\doibase 10.1146/annurev.nucl.010909.083654}
  {\bibfield  {journal} {\bibinfo  {journal} {Annual Review of Nuclear and
  Particle Science}\ }\textbf {\bibinfo {volume} {59}},\ \bibinfo {pages} {191}
  (\bibinfo {year} {2009}{\natexlab{a}})},\ \Eprint
  {http://arxiv.org/abs/0901.0011} {arXiv:0901.0011 [hep-ph]} \BibitemShut
  {NoStop}%
\bibitem [{\citenamefont {Boyarsky}\ \emph {et~al.}(2012)\citenamefont
  {Boyarsky}, \citenamefont {Iakubovskyi},\ and\ \citenamefont
  {Ruchayskiy}}]{Boyarsky:2012rt}%
  \BibitemOpen
  \bibfield  {author} {\bibinfo {author} {\bibfnamefont {A.}~\bibnamefont
  {Boyarsky}}, \bibinfo {author} {\bibfnamefont {D.}~\bibnamefont
  {Iakubovskyi}}, \ and\ \bibinfo {author} {\bibfnamefont {O.}~\bibnamefont
  {Ruchayskiy}},\ }\href {\doibase 10.1016/j.dark.2012.11.001} {\bibfield
  {journal} {\bibinfo  {journal} {Phys. Dark Univ.}\ }\textbf {\bibinfo
  {volume} {1}},\ \bibinfo {pages} {136} (\bibinfo {year} {2012})},\ \Eprint
  {http://arxiv.org/abs/1306.4954} {arXiv:1306.4954 [astro-ph.CO]} \BibitemShut
  {NoStop}%
\bibitem [{\citenamefont {{Adhikari}}\ \emph {et~al.}(2017)\citenamefont
  {{Adhikari}}, \citenamefont {{Agostini}}, \citenamefont {{Ky}}, \citenamefont
  {{Araki}}, \citenamefont {{Archidiacono}}, \citenamefont {{Bahr}},
  \citenamefont {{Baur}}, \citenamefont {{Behrens}}, \citenamefont
  {{Bezrukov}}, \citenamefont {{Bhupal Dev}}, \citenamefont {{Borah}},
  \citenamefont {{Boyarsky}}, \citenamefont {{de Gouvea}}, \citenamefont
  {{Pires}}, \citenamefont {{de Vega}}, \citenamefont {{Dias}}, \citenamefont
  {{Di Bari}}, \citenamefont {{Djurcic}}, \citenamefont {{Dolde}},
  \citenamefont {{Dorrer}}, \citenamefont {{Durero}}, \citenamefont
  {{Dragoun}}, \citenamefont {{Drewes}}, \citenamefont {{Drexlin}},
  \citenamefont {{D{\"u}llmann}}, \citenamefont {{Eberhardt}}, \citenamefont
  {{Eliseev}}, \citenamefont {{Enss}}, \citenamefont {{Evans}}, \citenamefont
  {{Faessler}}, \citenamefont {{Filianin}}, \citenamefont {{Fischer}},
  \citenamefont {{Fleischmann}}, \citenamefont {{Formaggio}}, \citenamefont
  {{Franse}}, \citenamefont {{Fraenkle}}, \citenamefont {{Frenk}},
  \citenamefont {{Fuller}}, \citenamefont {{Gastaldo}}, \citenamefont
  {{Garzilli}}, \citenamefont {{Giunti}}, \citenamefont {{Gl{\"u}ck}},
  \citenamefont {{Goodman}}, \citenamefont {{Gonzalez-Garcia}}, \citenamefont
  {{Gorbunov}}, \citenamefont {{Hamann}}, \citenamefont {{Hannen}},
  \citenamefont {{Hannestad}}, \citenamefont {{Hansen}}, \citenamefont
  {{Hassel}}, \citenamefont {{Heeck}}, \citenamefont {{Hofmann}}, \citenamefont
  {{Houdy}}, \citenamefont {{Huber}}, \citenamefont {{Iakubovskyi}},
  \citenamefont {{Ianni}}, \citenamefont {{Ibarra}}, \citenamefont
  {{Jacobsson}}, \citenamefont {{Jeltema}}, \citenamefont {{Jochum}},
  \citenamefont {{Kempf}}, \citenamefont {{Kieck}}, \citenamefont
  {{Korzeczek}}, \citenamefont {{Kornoukhov}}, \citenamefont {{Lachenmaier}},
  \citenamefont {{Laine}}, \citenamefont {{Langacker}}, \citenamefont
  {{Lasserre}}, \citenamefont {{Lesgourgues}}, \citenamefont {{Lhuillier}},
  \citenamefont {{Li}}, \citenamefont {{Liao}}, \citenamefont {{Long}},
  \citenamefont {{Maltoni}}, \citenamefont {{Mangano}}, \citenamefont
  {{Mavromatos}}, \citenamefont {{Menci}}, \citenamefont {{Merle}},
  \citenamefont {{Mertens}}, \citenamefont {{Mirizzi}}, \citenamefont
  {{Monreal}}, \citenamefont {{Nozik}}, \citenamefont {{Neronov}},
  \citenamefont {{Niro}}, \citenamefont {{Novikov}}, \citenamefont
  {{Oberauer}}, \citenamefont {{Otten}}, \citenamefont
  {{Palanque-Delabrouille}}, \citenamefont {{Pallavicini}}, \citenamefont
  {{Pantuev}}, \citenamefont {{Papastergis}}, \citenamefont {{Parke}},
  \citenamefont {{Pascoli}}, \citenamefont {{Pastor}}, \citenamefont
  {{Patwardhan}}, \citenamefont {{Pilaftsis}}, \citenamefont {{Radford}},
  \citenamefont {{Ranitzsch}}, \citenamefont {{Rest}}, \citenamefont
  {{Robinson}}, \citenamefont {{Rodrigues da Silva}}, \citenamefont
  {{Ruchayskiy}}, \citenamefont {{Sanchez}}, \citenamefont {{Sasaki}},
  \citenamefont {{Saviano}}, \citenamefont {{Schneider}}, \citenamefont
  {{Schneider}}, \citenamefont {{Schwetz}}, \citenamefont {{Sch{\"o}nert}},
  \citenamefont {{Scholl}}, \citenamefont {{Shankar}}, \citenamefont
  {{Shrock}}, \citenamefont {{Steinbrink}}, \citenamefont {{Strigari}},
  \citenamefont {{Suekane}}, \citenamefont {{Suerfu}}, \citenamefont
  {{Takahashi}}, \citenamefont {{Van}}, \citenamefont {{Tkachev}},
  \citenamefont {{Totzauer}}, \citenamefont {{Tsai}}, \citenamefont {{Tully}},
  \citenamefont {{Valerius}}, \citenamefont {{Valle}}, \citenamefont {{Venos}},
  \citenamefont {{Viel}}, \citenamefont {{Vivier}}, \citenamefont {{Wang}},
  \citenamefont {{Weinheimer}}, \citenamefont {{Wendt}}, \citenamefont
  {{Winslow}}, \citenamefont {{Wolf}}, \citenamefont {{Wurm}}, \citenamefont
  {{Xing}}, \citenamefont {{Zhou}},\ and\ \citenamefont
  {{Zuber}}}]{Adhikari:2017aa}%
  \BibitemOpen
  \bibfield  {author} {\bibinfo {author} {\bibfnamefont {R.}~\bibnamefont
  {{Adhikari}}}, \bibinfo {author} {\bibfnamefont {M.}~\bibnamefont
  {{Agostini}}}, \bibinfo {author} {\bibfnamefont {N.~A.}\ \bibnamefont
  {{Ky}}}, \bibinfo {author} {\bibfnamefont {T.}~\bibnamefont {{Araki}}},
  \bibinfo {author} {\bibfnamefont {M.}~\bibnamefont {{Archidiacono}}},
  \bibinfo {author} {\bibfnamefont {M.}~\bibnamefont {{Bahr}}}, \bibinfo
  {author} {\bibfnamefont {J.}~\bibnamefont {{Baur}}}, \bibinfo {author}
  {\bibfnamefont {J.}~\bibnamefont {{Behrens}}}, \bibinfo {author}
  {\bibfnamefont {F.}~\bibnamefont {{Bezrukov}}}, \bibinfo {author}
  {\bibfnamefont {P.~S.}\ \bibnamefont {{Bhupal Dev}}}, \bibinfo {author}
  {\bibfnamefont {D.}~\bibnamefont {{Borah}}}, \bibinfo {author} {\bibfnamefont
  {A.}~\bibnamefont {{Boyarsky}}}, \bibinfo {author} {\bibfnamefont
  {A.}~\bibnamefont {{de Gouvea}}}, \bibinfo {author} {\bibfnamefont
  {C.~A.~d.~S.}\ \bibnamefont {{Pires}}}, \bibinfo {author} {\bibfnamefont
  {H.~J.}\ \bibnamefont {{de Vega}}}, \bibinfo {author} {\bibfnamefont {A.~G.}\
  \bibnamefont {{Dias}}}, \bibinfo {author} {\bibfnamefont {P.}~\bibnamefont
  {{Di Bari}}}, \bibinfo {author} {\bibfnamefont {Z.}~\bibnamefont
  {{Djurcic}}}, \bibinfo {author} {\bibfnamefont {K.}~\bibnamefont {{Dolde}}},
  \bibinfo {author} {\bibfnamefont {H.}~\bibnamefont {{Dorrer}}}, \bibinfo
  {author} {\bibfnamefont {M.}~\bibnamefont {{Durero}}}, \bibinfo {author}
  {\bibfnamefont {O.}~\bibnamefont {{Dragoun}}}, \bibinfo {author}
  {\bibfnamefont {M.}~\bibnamefont {{Drewes}}}, \bibinfo {author}
  {\bibfnamefont {G.}~\bibnamefont {{Drexlin}}}, \bibinfo {author}
  {\bibfnamefont {C.~E.}\ \bibnamefont {{D{\"u}llmann}}}, \bibinfo {author}
  {\bibfnamefont {K.}~\bibnamefont {{Eberhardt}}}, \bibinfo {author}
  {\bibfnamefont {S.}~\bibnamefont {{Eliseev}}}, \bibinfo {author}
  {\bibfnamefont {C.}~\bibnamefont {{Enss}}}, \bibinfo {author} {\bibfnamefont
  {N.~W.}\ \bibnamefont {{Evans}}}, \bibinfo {author} {\bibfnamefont
  {A.}~\bibnamefont {{Faessler}}}, \bibinfo {author} {\bibfnamefont
  {P.}~\bibnamefont {{Filianin}}}, \bibinfo {author} {\bibfnamefont
  {V.}~\bibnamefont {{Fischer}}}, \bibinfo {author} {\bibfnamefont
  {A.}~\bibnamefont {{Fleischmann}}}, \bibinfo {author} {\bibfnamefont {J.~A.}\
  \bibnamefont {{Formaggio}}}, \bibinfo {author} {\bibfnamefont
  {J.}~\bibnamefont {{Franse}}}, \bibinfo {author} {\bibfnamefont {F.~M.}\
  \bibnamefont {{Fraenkle}}}, \bibinfo {author} {\bibfnamefont {C.~S.}\
  \bibnamefont {{Frenk}}}, \bibinfo {author} {\bibfnamefont {G.}~\bibnamefont
  {{Fuller}}}, \bibinfo {author} {\bibfnamefont {L.}~\bibnamefont
  {{Gastaldo}}}, \bibinfo {author} {\bibfnamefont {A.}~\bibnamefont
  {{Garzilli}}}, \bibinfo {author} {\bibfnamefont {C.}~\bibnamefont
  {{Giunti}}}, \bibinfo {author} {\bibfnamefont {F.}~\bibnamefont
  {{Gl{\"u}ck}}}, \bibinfo {author} {\bibfnamefont {M.~C.}\ \bibnamefont
  {{Goodman}}}, \bibinfo {author} {\bibfnamefont {M.~C.}\ \bibnamefont
  {{Gonzalez-Garcia}}}, \bibinfo {author} {\bibfnamefont {D.}~\bibnamefont
  {{Gorbunov}}}, \bibinfo {author} {\bibfnamefont {J.}~\bibnamefont
  {{Hamann}}}, \bibinfo {author} {\bibfnamefont {V.}~\bibnamefont {{Hannen}}},
  \bibinfo {author} {\bibfnamefont {S.}~\bibnamefont {{Hannestad}}}, \bibinfo
  {author} {\bibfnamefont {S.~H.}\ \bibnamefont {{Hansen}}}, \bibinfo {author}
  {\bibfnamefont {C.}~\bibnamefont {{Hassel}}}, \bibinfo {author}
  {\bibfnamefont {J.}~\bibnamefont {{Heeck}}}, \bibinfo {author} {\bibfnamefont
  {F.}~\bibnamefont {{Hofmann}}}, \bibinfo {author} {\bibfnamefont
  {T.}~\bibnamefont {{Houdy}}}, \bibinfo {author} {\bibfnamefont
  {A.}~\bibnamefont {{Huber}}}, \bibinfo {author} {\bibfnamefont
  {D.}~\bibnamefont {{Iakubovskyi}}}, \bibinfo {author} {\bibfnamefont
  {A.}~\bibnamefont {{Ianni}}}, \bibinfo {author} {\bibfnamefont
  {A.}~\bibnamefont {{Ibarra}}}, \bibinfo {author} {\bibfnamefont
  {R.}~\bibnamefont {{Jacobsson}}}, \bibinfo {author} {\bibfnamefont
  {T.}~\bibnamefont {{Jeltema}}}, \bibinfo {author} {\bibfnamefont
  {J.}~\bibnamefont {{Jochum}}}, \bibinfo {author} {\bibfnamefont
  {S.}~\bibnamefont {{Kempf}}}, \bibinfo {author} {\bibfnamefont
  {T.}~\bibnamefont {{Kieck}}}, \bibinfo {author} {\bibfnamefont
  {M.}~\bibnamefont {{Korzeczek}}}, \bibinfo {author} {\bibfnamefont
  {V.}~\bibnamefont {{Kornoukhov}}}, \bibinfo {author} {\bibfnamefont
  {T.}~\bibnamefont {{Lachenmaier}}}, \bibinfo {author} {\bibfnamefont
  {M.}~\bibnamefont {{Laine}}}, \bibinfo {author} {\bibfnamefont
  {P.}~\bibnamefont {{Langacker}}}, \bibinfo {author} {\bibfnamefont
  {T.}~\bibnamefont {{Lasserre}}}, \bibinfo {author} {\bibfnamefont
  {J.}~\bibnamefont {{Lesgourgues}}}, \bibinfo {author} {\bibfnamefont
  {D.}~\bibnamefont {{Lhuillier}}}, \bibinfo {author} {\bibfnamefont {Y.~F.}\
  \bibnamefont {{Li}}}, \bibinfo {author} {\bibfnamefont {W.}~\bibnamefont
  {{Liao}}}, \bibinfo {author} {\bibfnamefont {A.~W.}\ \bibnamefont {{Long}}},
  \bibinfo {author} {\bibfnamefont {M.}~\bibnamefont {{Maltoni}}}, \bibinfo
  {author} {\bibfnamefont {G.}~\bibnamefont {{Mangano}}}, \bibinfo {author}
  {\bibfnamefont {N.~E.}\ \bibnamefont {{Mavromatos}}}, \bibinfo {author}
  {\bibfnamefont {N.}~\bibnamefont {{Menci}}}, \bibinfo {author} {\bibfnamefont
  {A.}~\bibnamefont {{Merle}}}, \bibinfo {author} {\bibfnamefont
  {S.}~\bibnamefont {{Mertens}}}, \bibinfo {author} {\bibfnamefont
  {A.}~\bibnamefont {{Mirizzi}}}, \bibinfo {author} {\bibfnamefont
  {B.}~\bibnamefont {{Monreal}}}, \bibinfo {author} {\bibfnamefont
  {A.}~\bibnamefont {{Nozik}}}, \bibinfo {author} {\bibfnamefont
  {A.}~\bibnamefont {{Neronov}}}, \bibinfo {author} {\bibfnamefont
  {V.}~\bibnamefont {{Niro}}}, \bibinfo {author} {\bibfnamefont
  {Y.}~\bibnamefont {{Novikov}}}, \bibinfo {author} {\bibfnamefont
  {L.}~\bibnamefont {{Oberauer}}}, \bibinfo {author} {\bibfnamefont
  {E.}~\bibnamefont {{Otten}}}, \bibinfo {author} {\bibfnamefont
  {N.}~\bibnamefont {{Palanque-Delabrouille}}}, \bibinfo {author}
  {\bibfnamefont {M.}~\bibnamefont {{Pallavicini}}}, \bibinfo {author}
  {\bibfnamefont {V.~S.}\ \bibnamefont {{Pantuev}}}, \bibinfo {author}
  {\bibfnamefont {E.}~\bibnamefont {{Papastergis}}}, \bibinfo {author}
  {\bibfnamefont {S.}~\bibnamefont {{Parke}}}, \bibinfo {author} {\bibfnamefont
  {S.}~\bibnamefont {{Pascoli}}}, \bibinfo {author} {\bibfnamefont
  {S.}~\bibnamefont {{Pastor}}}, \bibinfo {author} {\bibfnamefont
  {A.}~\bibnamefont {{Patwardhan}}}, \bibinfo {author} {\bibfnamefont
  {A.}~\bibnamefont {{Pilaftsis}}}, \bibinfo {author} {\bibfnamefont {D.~C.}\
  \bibnamefont {{Radford}}}, \bibinfo {author} {\bibfnamefont {P.~C.-O.}\
  \bibnamefont {{Ranitzsch}}}, \bibinfo {author} {\bibfnamefont
  {O.}~\bibnamefont {{Rest}}}, \bibinfo {author} {\bibfnamefont {D.~J.}\
  \bibnamefont {{Robinson}}}, \bibinfo {author} {\bibfnamefont {P.~S.}\
  \bibnamefont {{Rodrigues da Silva}}}, \bibinfo {author} {\bibfnamefont
  {O.}~\bibnamefont {{Ruchayskiy}}}, \bibinfo {author} {\bibfnamefont {N.~G.}\
  \bibnamefont {{Sanchez}}}, \bibinfo {author} {\bibfnamefont {M.}~\bibnamefont
  {{Sasaki}}}, \bibinfo {author} {\bibfnamefont {N.}~\bibnamefont {{Saviano}}},
  \bibinfo {author} {\bibfnamefont {A.}~\bibnamefont {{Schneider}}}, \bibinfo
  {author} {\bibfnamefont {F.}~\bibnamefont {{Schneider}}}, \bibinfo {author}
  {\bibfnamefont {T.}~\bibnamefont {{Schwetz}}}, \bibinfo {author}
  {\bibfnamefont {S.}~\bibnamefont {{Sch{\"o}nert}}}, \bibinfo {author}
  {\bibfnamefont {S.}~\bibnamefont {{Scholl}}}, \bibinfo {author}
  {\bibfnamefont {F.}~\bibnamefont {{Shankar}}}, \bibinfo {author}
  {\bibfnamefont {R.}~\bibnamefont {{Shrock}}}, \bibinfo {author}
  {\bibfnamefont {N.}~\bibnamefont {{Steinbrink}}}, \bibinfo {author}
  {\bibfnamefont {L.}~\bibnamefont {{Strigari}}}, \bibinfo {author}
  {\bibfnamefont {F.}~\bibnamefont {{Suekane}}}, \bibinfo {author}
  {\bibfnamefont {B.}~\bibnamefont {{Suerfu}}}, \bibinfo {author}
  {\bibfnamefont {R.}~\bibnamefont {{Takahashi}}}, \bibinfo {author}
  {\bibfnamefont {N.~T.~H.}\ \bibnamefont {{Van}}}, \bibinfo {author}
  {\bibfnamefont {I.}~\bibnamefont {{Tkachev}}}, \bibinfo {author}
  {\bibfnamefont {M.}~\bibnamefont {{Totzauer}}}, \bibinfo {author}
  {\bibfnamefont {Y.}~\bibnamefont {{Tsai}}}, \bibinfo {author} {\bibfnamefont
  {C.~G.}\ \bibnamefont {{Tully}}}, \bibinfo {author} {\bibfnamefont
  {K.}~\bibnamefont {{Valerius}}}, \bibinfo {author} {\bibfnamefont {J.~W.~F.}\
  \bibnamefont {{Valle}}}, \bibinfo {author} {\bibfnamefont {D.}~\bibnamefont
  {{Venos}}}, \bibinfo {author} {\bibfnamefont {M.}~\bibnamefont {{Viel}}},
  \bibinfo {author} {\bibfnamefont {M.}~\bibnamefont {{Vivier}}}, \bibinfo
  {author} {\bibfnamefont {M.~Y.}\ \bibnamefont {{Wang}}}, \bibinfo {author}
  {\bibfnamefont {C.}~\bibnamefont {{Weinheimer}}}, \bibinfo {author}
  {\bibfnamefont {K.}~\bibnamefont {{Wendt}}}, \bibinfo {author} {\bibfnamefont
  {L.}~\bibnamefont {{Winslow}}}, \bibinfo {author} {\bibfnamefont
  {J.}~\bibnamefont {{Wolf}}}, \bibinfo {author} {\bibfnamefont
  {M.}~\bibnamefont {{Wurm}}}, \bibinfo {author} {\bibfnamefont
  {Z.}~\bibnamefont {{Xing}}}, \bibinfo {author} {\bibfnamefont
  {S.}~\bibnamefont {{Zhou}}}, \ and\ \bibinfo {author} {\bibfnamefont
  {K.}~\bibnamefont {{Zuber}}},\ }\href {\doibase
  10.1088/1475-7516/2017/01/025} {\bibfield  {journal} {\bibinfo  {journal}
  {JCAP}\ }\textbf {\bibinfo {volume} {1}},\ \bibinfo {eid} {025} (\bibinfo
  {year} {2017})},\ \Eprint {http://arxiv.org/abs/1602.04816} {arXiv:1602.04816
  [hep-ph]} \BibitemShut {NoStop}%
\bibitem [{\citenamefont {{Schneider}}(2016)}]{Schneider:2016aa}%
  \BibitemOpen
  \bibfield  {author} {\bibinfo {author} {\bibfnamefont {A.}~\bibnamefont
  {{Schneider}}},\ }\href {\doibase 10.1088/1475-7516/2016/04/059} {\bibfield
  {journal} {\bibinfo  {journal} {JCAP}\ }\textbf {\bibinfo {volume} {4}},\
  \bibinfo {eid} {059} (\bibinfo {year} {2016})},\ \Eprint
  {http://arxiv.org/abs/1601.07553} {arXiv:1601.07553} \BibitemShut {NoStop}%
\bibitem [{\citenamefont {{Bulbul}}\ \emph {et~al.}(2014)\citenamefont
  {{Bulbul}}, \citenamefont {{Markevitch}}, \citenamefont {{Foster}},
  \citenamefont {{Smith}}, \citenamefont {{Loewenstein}},\ and\ \citenamefont
  {{Randall}}}]{Bulbul:2014aa}%
  \BibitemOpen
  \bibfield  {author} {\bibinfo {author} {\bibfnamefont {E.}~\bibnamefont
  {{Bulbul}}}, \bibinfo {author} {\bibfnamefont {M.}~\bibnamefont
  {{Markevitch}}}, \bibinfo {author} {\bibfnamefont {A.}~\bibnamefont
  {{Foster}}}, \bibinfo {author} {\bibfnamefont {R.~K.}\ \bibnamefont
  {{Smith}}}, \bibinfo {author} {\bibfnamefont {M.}~\bibnamefont
  {{Loewenstein}}}, \ and\ \bibinfo {author} {\bibfnamefont {S.~W.}\
  \bibnamefont {{Randall}}},\ }\href {\doibase 10.1088/0004-637X/789/1/13}
  {\bibfield  {journal} {\bibinfo  {journal} {\apj}\ }\textbf {\bibinfo
  {volume} {789}},\ \bibinfo {eid} {13} (\bibinfo {year} {2014})},\ \Eprint
  {http://arxiv.org/abs/1402.2301} {arXiv:1402.2301} \BibitemShut {NoStop}%
\bibitem [{\citenamefont {{Boyarsky}}\ \emph {et~al.}(2014)\citenamefont
  {{Boyarsky}}, \citenamefont {{Ruchayskiy}}, \citenamefont {{Iakubovskyi}},\
  and\ \citenamefont {{Franse}}}]{Boyarsky:2014aa}%
  \BibitemOpen
  \bibfield  {author} {\bibinfo {author} {\bibfnamefont {A.}~\bibnamefont
  {{Boyarsky}}}, \bibinfo {author} {\bibfnamefont {O.}~\bibnamefont
  {{Ruchayskiy}}}, \bibinfo {author} {\bibfnamefont {D.}~\bibnamefont
  {{Iakubovskyi}}}, \ and\ \bibinfo {author} {\bibfnamefont {J.}~\bibnamefont
  {{Franse}}},\ }\href {\doibase 10.1103/PhysRevLett.113.251301} {\bibfield
  {journal} {\bibinfo  {journal} {Physical Review Letters}\ }\textbf {\bibinfo
  {volume} {113}},\ \bibinfo {eid} {251301} (\bibinfo {year} {2014})},\ \Eprint
  {http://arxiv.org/abs/1402.4119} {arXiv:1402.4119} \BibitemShut {NoStop}%
\bibitem [{\citenamefont {{Urban}}\ \emph {et~al.}(2015)\citenamefont
  {{Urban}}, \citenamefont {{Werner}}, \citenamefont {{Allen}}, \citenamefont
  {{Simionescu}}, \citenamefont {{Kaastra}},\ and\ \citenamefont
  {{Strigari}}}]{Urban:2015aa}%
  \BibitemOpen
  \bibfield  {author} {\bibinfo {author} {\bibfnamefont {O.}~\bibnamefont
  {{Urban}}}, \bibinfo {author} {\bibfnamefont {N.}~\bibnamefont {{Werner}}},
  \bibinfo {author} {\bibfnamefont {S.~W.}\ \bibnamefont {{Allen}}}, \bibinfo
  {author} {\bibfnamefont {A.}~\bibnamefont {{Simionescu}}}, \bibinfo {author}
  {\bibfnamefont {J.~S.}\ \bibnamefont {{Kaastra}}}, \ and\ \bibinfo {author}
  {\bibfnamefont {L.~E.}\ \bibnamefont {{Strigari}}},\ }\href {\doibase
  10.1093/mnras/stv1142} {\bibfield  {journal} {\bibinfo  {journal} {MNRAS}\
  }\textbf {\bibinfo {volume} {451}},\ \bibinfo {pages} {2447} (\bibinfo {year}
  {2015})},\ \Eprint {http://arxiv.org/abs/1411.0050} {arXiv:1411.0050}
  \BibitemShut {NoStop}%
\bibitem [{\citenamefont {{Boyarsky}}\ \emph {et~al.}(2015)\citenamefont
  {{Boyarsky}}, \citenamefont {{Franse}}, \citenamefont {{Iakubovskyi}},\ and\
  \citenamefont {{Ruchayskiy}}}]{Boyarsky:2015aa}%
  \BibitemOpen
  \bibfield  {author} {\bibinfo {author} {\bibfnamefont {A.}~\bibnamefont
  {{Boyarsky}}}, \bibinfo {author} {\bibfnamefont {J.}~\bibnamefont
  {{Franse}}}, \bibinfo {author} {\bibfnamefont {D.}~\bibnamefont
  {{Iakubovskyi}}}, \ and\ \bibinfo {author} {\bibfnamefont {O.}~\bibnamefont
  {{Ruchayskiy}}},\ }\href {\doibase 10.1103/PhysRevLett.115.161301} {\bibfield
   {journal} {\bibinfo  {journal} {Physical Review Letters}\ }\textbf {\bibinfo
  {volume} {115}},\ \bibinfo {eid} {161301} (\bibinfo {year} {2015})},\ \Eprint
  {http://arxiv.org/abs/1408.2503} {arXiv:1408.2503} \BibitemShut {NoStop}%
\bibitem [{\citenamefont {{Iakubovskyi}}(2016)}]{Iakubovskyi:2016aa}%
  \BibitemOpen
  \bibfield  {author} {\bibinfo {author} {\bibfnamefont {D.~A.}\ \bibnamefont
  {{Iakubovskyi}}},\ }\href {\doibase 10.17721/2227-1481.6.3-15} {\bibfield
  {journal} {\bibinfo  {journal} {Advances in Astronomy and Space Physics}\
  }\textbf {\bibinfo {volume} {6}},\ \bibinfo {pages} {3} (\bibinfo {year}
  {2016})},\ \Eprint {http://arxiv.org/abs/1510.00358} {arXiv:1510.00358
  [astro-ph.HE]} \BibitemShut {NoStop}%
\bibitem [{\citenamefont {{Franse}}\ \emph {et~al.}(2016)\citenamefont
  {{Franse}}, \citenamefont {{Bulbul}}, \citenamefont {{Foster}}, \citenamefont
  {{Boyarsky}}, \citenamefont {{Markevitch}}, \citenamefont {{Bautz}},
  \citenamefont {{Iakubovskyi}}, \citenamefont {{Loewenstein}}, \citenamefont
  {{McDonald}}, \citenamefont {{Miller}}, \citenamefont {{Randall}},
  \citenamefont {{Ruchayskiy}},\ and\ \citenamefont {{Smith}}}]{Franse:2016aa}%
  \BibitemOpen
  \bibfield  {author} {\bibinfo {author} {\bibfnamefont {J.}~\bibnamefont
  {{Franse}}}, \bibinfo {author} {\bibfnamefont {E.}~\bibnamefont {{Bulbul}}},
  \bibinfo {author} {\bibfnamefont {A.}~\bibnamefont {{Foster}}}, \bibinfo
  {author} {\bibfnamefont {A.}~\bibnamefont {{Boyarsky}}}, \bibinfo {author}
  {\bibfnamefont {M.}~\bibnamefont {{Markevitch}}}, \bibinfo {author}
  {\bibfnamefont {M.}~\bibnamefont {{Bautz}}}, \bibinfo {author} {\bibfnamefont
  {D.}~\bibnamefont {{Iakubovskyi}}}, \bibinfo {author} {\bibfnamefont
  {M.}~\bibnamefont {{Loewenstein}}}, \bibinfo {author} {\bibfnamefont
  {M.}~\bibnamefont {{McDonald}}}, \bibinfo {author} {\bibfnamefont
  {E.}~\bibnamefont {{Miller}}}, \bibinfo {author} {\bibfnamefont {S.~W.}\
  \bibnamefont {{Randall}}}, \bibinfo {author} {\bibfnamefont {O.}~\bibnamefont
  {{Ruchayskiy}}}, \ and\ \bibinfo {author} {\bibfnamefont {R.~K.}\
  \bibnamefont {{Smith}}},\ }\href {\doibase 10.3847/0004-637X/829/2/124}
  {\bibfield  {journal} {\bibinfo  {journal} {\apj}\ }\textbf {\bibinfo
  {volume} {829}},\ \bibinfo {eid} {124} (\bibinfo {year} {2016})},\ \Eprint
  {http://arxiv.org/abs/1604.01759} {arXiv:1604.01759} \BibitemShut {NoStop}%
\bibitem [{\citenamefont {{Anderson}}\ \emph {et~al.}(2015)\citenamefont
  {{Anderson}}, \citenamefont {{Churazov}},\ and\ \citenamefont
  {{Bregman}}}]{Anderson:2015aa}%
  \BibitemOpen
  \bibfield  {author} {\bibinfo {author} {\bibfnamefont {M.~E.}\ \bibnamefont
  {{Anderson}}}, \bibinfo {author} {\bibfnamefont {E.}~\bibnamefont
  {{Churazov}}}, \ and\ \bibinfo {author} {\bibfnamefont {J.~N.}\ \bibnamefont
  {{Bregman}}},\ }\href {\doibase 10.1093/mnras/stv1559} {\bibfield  {journal}
  {\bibinfo  {journal} {MNRAS}\ }\textbf {\bibinfo {volume} {452}},\ \bibinfo
  {pages} {3905} (\bibinfo {year} {2015})},\ \Eprint
  {http://arxiv.org/abs/1408.4115} {arXiv:1408.4115 [astro-ph.HE]} \BibitemShut
  {NoStop}%
\bibitem [{\citenamefont {{Carlson}}\ \emph {et~al.}(2015)\citenamefont
  {{Carlson}}, \citenamefont {{Jeltema}},\ and\ \citenamefont
  {{Profumo}}}]{Carlson:2015aa}%
  \BibitemOpen
  \bibfield  {author} {\bibinfo {author} {\bibfnamefont {E.}~\bibnamefont
  {{Carlson}}}, \bibinfo {author} {\bibfnamefont {T.}~\bibnamefont
  {{Jeltema}}}, \ and\ \bibinfo {author} {\bibfnamefont {S.}~\bibnamefont
  {{Profumo}}},\ }\href {\doibase 10.1088/1475-7516/2015/02/009} {\bibfield
  {journal} {\bibinfo  {journal} {JCAP}\ }\textbf {\bibinfo {volume} {2}},\
  \bibinfo {eid} {009} (\bibinfo {year} {2015})},\ \Eprint
  {http://arxiv.org/abs/1411.1758} {arXiv:1411.1758 [astro-ph.HE]} \BibitemShut
  {NoStop}%
\bibitem [{\citenamefont {{Horiuchi}}\ \emph {et~al.}(2014)\citenamefont
  {{Horiuchi}}, \citenamefont {{Humphrey}}, \citenamefont {{O{\~n}orbe}},
  \citenamefont {{Abazajian}}, \citenamefont {{Kaplinghat}},\ and\
  \citenamefont {{Garrison-Kimmel}}}]{Horiuchi:2014aa}%
  \BibitemOpen
  \bibfield  {author} {\bibinfo {author} {\bibfnamefont {S.}~\bibnamefont
  {{Horiuchi}}}, \bibinfo {author} {\bibfnamefont {P.~J.}\ \bibnamefont
  {{Humphrey}}}, \bibinfo {author} {\bibfnamefont {J.}~\bibnamefont
  {{O{\~n}orbe}}}, \bibinfo {author} {\bibfnamefont {K.~N.}\ \bibnamefont
  {{Abazajian}}}, \bibinfo {author} {\bibfnamefont {M.}~\bibnamefont
  {{Kaplinghat}}}, \ and\ \bibinfo {author} {\bibfnamefont {S.}~\bibnamefont
  {{Garrison-Kimmel}}},\ }\href {\doibase 10.1103/PhysRevD.89.025017}
  {\bibfield  {journal} {\bibinfo  {journal} {\prd}\ }\textbf {\bibinfo
  {volume} {89}},\ \bibinfo {eid} {025017} (\bibinfo {year} {2014})},\ \Eprint
  {http://arxiv.org/abs/1311.0282} {arXiv:1311.0282} \BibitemShut {NoStop}%
\bibitem [{\citenamefont {{Tamura}}\ \emph {et~al.}(2015)\citenamefont
  {{Tamura}}, \citenamefont {{Iizuka}}, \citenamefont {{Maeda}}, \citenamefont
  {{Mitsuda}},\ and\ \citenamefont {{Yamasaki}}}]{Tamura:2015aa}%
  \BibitemOpen
  \bibfield  {author} {\bibinfo {author} {\bibfnamefont {T.}~\bibnamefont
  {{Tamura}}}, \bibinfo {author} {\bibfnamefont {R.}~\bibnamefont {{Iizuka}}},
  \bibinfo {author} {\bibfnamefont {Y.}~\bibnamefont {{Maeda}}}, \bibinfo
  {author} {\bibfnamefont {K.}~\bibnamefont {{Mitsuda}}}, \ and\ \bibinfo
  {author} {\bibfnamefont {N.~Y.}\ \bibnamefont {{Yamasaki}}},\ }\href
  {\doibase 10.1093/pasj/psu156} {\bibfield  {journal} {\bibinfo  {journal}
  {Publ. Aston. Soc. Jap.}\ }\textbf {\bibinfo {volume} {67}},\ \bibinfo {eid}
  {23} (\bibinfo {year} {2015})},\ \Eprint {http://arxiv.org/abs/1412.1869}
  {arXiv:1412.1869 [astro-ph.HE]} \BibitemShut {NoStop}%
\bibitem [{\citenamefont {{Jeltema}}\ and\ \citenamefont
  {{Profumo}}(2016)}]{Jeltema:2016aa}%
  \BibitemOpen
  \bibfield  {author} {\bibinfo {author} {\bibfnamefont {T.}~\bibnamefont
  {{Jeltema}}}\ and\ \bibinfo {author} {\bibfnamefont {S.}~\bibnamefont
  {{Profumo}}},\ }\href {\doibase 10.1093/mnras/stw578} {\bibfield  {journal}
  {\bibinfo  {journal} {MNRAS}\ }\textbf {\bibinfo {volume} {458}},\ \bibinfo
  {pages} {3592} (\bibinfo {year} {2016})},\ \Eprint
  {http://arxiv.org/abs/1512.01239} {arXiv:1512.01239 [astro-ph.HE]}
  \BibitemShut {NoStop}%
\bibitem [{\citenamefont {{Ruchayskiy}}\ \emph {et~al.}(2016)\citenamefont
  {{Ruchayskiy}}, \citenamefont {{Boyarsky}}, \citenamefont {{Iakubovskyi}},
  \citenamefont {{Bulbul}}, \citenamefont {{Eckert}}, \citenamefont {{Franse}},
  \citenamefont {{Malyshev}}, \citenamefont {{Markevitch}},\ and\ \citenamefont
  {{Neronov}}}]{Ruchayskiy:2016aa}%
  \BibitemOpen
  \bibfield  {author} {\bibinfo {author} {\bibfnamefont {O.}~\bibnamefont
  {{Ruchayskiy}}}, \bibinfo {author} {\bibfnamefont {A.}~\bibnamefont
  {{Boyarsky}}}, \bibinfo {author} {\bibfnamefont {D.}~\bibnamefont
  {{Iakubovskyi}}}, \bibinfo {author} {\bibfnamefont {E.}~\bibnamefont
  {{Bulbul}}}, \bibinfo {author} {\bibfnamefont {D.}~\bibnamefont {{Eckert}}},
  \bibinfo {author} {\bibfnamefont {J.}~\bibnamefont {{Franse}}}, \bibinfo
  {author} {\bibfnamefont {D.}~\bibnamefont {{Malyshev}}}, \bibinfo {author}
  {\bibfnamefont {M.}~\bibnamefont {{Markevitch}}}, \ and\ \bibinfo {author}
  {\bibfnamefont {A.}~\bibnamefont {{Neronov}}},\ }\href {\doibase
  10.1093/mnras/stw1026} {\bibfield  {journal} {\bibinfo  {journal} {MNRAS}\
  }\textbf {\bibinfo {volume} {460}},\ \bibinfo {pages} {1390} (\bibinfo {year}
  {2016})},\ \Eprint {http://arxiv.org/abs/1512.07217} {arXiv:1512.07217
  [astro-ph.HE]} \BibitemShut {NoStop}%
\bibitem [{\citenamefont {{Sekiya}}\ \emph {et~al.}(2016)\citenamefont
  {{Sekiya}}, \citenamefont {{Yamasaki}},\ and\ \citenamefont
  {{Mitsuda}}}]{Sekiya:2016aa}%
  \BibitemOpen
  \bibfield  {author} {\bibinfo {author} {\bibfnamefont {N.}~\bibnamefont
  {{Sekiya}}}, \bibinfo {author} {\bibfnamefont {N.~Y.}\ \bibnamefont
  {{Yamasaki}}}, \ and\ \bibinfo {author} {\bibfnamefont {K.}~\bibnamefont
  {{Mitsuda}}},\ }\href {\doibase 10.1093/pasj/psv081} {\bibfield  {journal}
  {\bibinfo  {journal} {Publ. Aston. Soc. Jap.}\ }\textbf {\bibinfo {volume}
  {68}},\ \bibinfo {eid} {S31} (\bibinfo {year} {2016})},\ \Eprint
  {http://arxiv.org/abs/1504.02826} {arXiv:1504.02826 [astro-ph.HE]}
  \BibitemShut {NoStop}%
\bibitem [{\citenamefont {{Stodolsky}}(1987)}]{stodolsky:1987QK}%
  \BibitemOpen
  \bibfield  {author} {\bibinfo {author} {\bibfnamefont {L.}~\bibnamefont
  {{Stodolsky}}},\ }\href@noop {} {\bibfield  {journal} {\bibinfo  {journal}
  {Phys. Rev. D}\ }\textbf {\bibinfo {volume} {36}} (\bibinfo {year}
  {1987})}\BibitemShut {NoStop}%
\bibitem [{\citenamefont {Barbieri}\ and\ \citenamefont
  {Dolgov}(1991)}]{Barbieri:1991aa}%
  \BibitemOpen
  \bibfield  {author} {\bibinfo {author} {\bibfnamefont {R.}~\bibnamefont
  {Barbieri}}\ and\ \bibinfo {author} {\bibfnamefont {A.}~\bibnamefont
  {Dolgov}},\ }\href@noop {} {\bibfield  {journal} {\bibinfo  {journal} {Nuc.
  Phys. B}\ }\textbf {\bibinfo {volume} {349}},\ \bibinfo {pages} {743}
  (\bibinfo {year} {1991})}\BibitemShut {NoStop}%
\bibitem [{\citenamefont {Enqvist}\ \emph {et~al.}(1991)\citenamefont
  {Enqvist}, \citenamefont {Kainulainen},\ and\ \citenamefont
  {Maalampi}}]{Enqvist:1991aa}%
  \BibitemOpen
  \bibfield  {author} {\bibinfo {author} {\bibfnamefont {K.}~\bibnamefont
  {Enqvist}}, \bibinfo {author} {\bibfnamefont {K.}~\bibnamefont
  {Kainulainen}}, \ and\ \bibinfo {author} {\bibfnamefont {J.}~\bibnamefont
  {Maalampi}},\ }\href@noop {} {\bibfield  {journal} {\bibinfo  {journal} {Nuc.
  Phys. B}\ }\textbf {\bibinfo {volume} {349}} (\bibinfo {year}
  {1991})}\BibitemShut {NoStop}%
\bibitem [{\citenamefont {{Venumadhav}}\ \emph {et~al.}(2016)\citenamefont
  {{Venumadhav}}, \citenamefont {{Cyr-Racine}}, \citenamefont {{Abazajian}},\
  and\ \citenamefont {{Hirata}}}]{Venumadhav:2015aa}%
  \BibitemOpen
  \bibfield  {author} {\bibinfo {author} {\bibfnamefont {T.}~\bibnamefont
  {{Venumadhav}}}, \bibinfo {author} {\bibfnamefont {F.-Y.}\ \bibnamefont
  {{Cyr-Racine}}}, \bibinfo {author} {\bibfnamefont {K.~N.}\ \bibnamefont
  {{Abazajian}}}, \ and\ \bibinfo {author} {\bibfnamefont {C.~M.}\ \bibnamefont
  {{Hirata}}},\ }\href@noop {} {\bibfield  {journal} {\bibinfo  {journal}
  {Phys. Rev. D}\ }\textbf {\bibinfo {volume} {94}} (\bibinfo {year} {2016})},\
  \Eprint {http://arxiv.org/abs/1507.06655} {arXiv:1507.06655} \BibitemShut
  {NoStop}%
\bibitem [{\citenamefont {Dolgov}\ \emph {et~al.}(2002)\citenamefont {Dolgov},
  \citenamefont {Hansen}, \citenamefont {Pastor}, \citenamefont {Petcov},
  \citenamefont {Raffelt},\ and\ \citenamefont {D.V.}}]{Dolgov:2002bb}%
  \BibitemOpen
  \bibfield  {author} {\bibinfo {author} {\bibfnamefont {A.}~\bibnamefont
  {Dolgov}}, \bibinfo {author} {\bibfnamefont {S.}~\bibnamefont {Hansen}},
  \bibinfo {author} {\bibfnamefont {S.}~\bibnamefont {Pastor}}, \bibinfo
  {author} {\bibfnamefont {S.}~\bibnamefont {Petcov}}, \bibinfo {author}
  {\bibfnamefont {G.}~\bibnamefont {Raffelt}}, \ and\ \bibinfo {author}
  {\bibfnamefont {S.}~\bibnamefont {D.V.}},\ }\href@noop {} {\bibfield
  {journal} {\bibinfo  {journal} {Nuc. Phys. B}\ }\textbf {\bibinfo {volume}
  {632}},\ \bibinfo {pages} {363} (\bibinfo {year} {2002})}\BibitemShut
  {NoStop}%
\bibitem [{\citenamefont {{Abazajian}}\ \emph {et~al.}(2002)\citenamefont
  {{Abazajian}}, \citenamefont {{Beacom}},\ and\ \citenamefont
  {{Bell}}}]{Abazajian:2002lr}%
  \BibitemOpen
  \bibfield  {author} {\bibinfo {author} {\bibfnamefont {K.~N.}\ \bibnamefont
  {{Abazajian}}}, \bibinfo {author} {\bibfnamefont {J.~F.}\ \bibnamefont
  {{Beacom}}}, \ and\ \bibinfo {author} {\bibfnamefont {N.~F.}\ \bibnamefont
  {{Bell}}},\ }\href {\doibase 10.1103/PhysRevD.66.013008} {\bibfield
  {journal} {\bibinfo  {journal} {Phys. Rev. D}\ }\textbf {\bibinfo {volume}
  {66}},\ \bibinfo {pages} {013008} (\bibinfo {year} {2002})},\ \Eprint
  {http://arxiv.org/abs/arXiv:astro-ph/0203442} {arXiv:astro-ph/0203442}
  \BibitemShut {NoStop}%
\bibitem [{\citenamefont {{Serpico}}\ and\ \citenamefont
  {{Raffelt}}(2005)}]{Serpico:2005aa}%
  \BibitemOpen
  \bibfield  {author} {\bibinfo {author} {\bibfnamefont {P.~D.}\ \bibnamefont
  {{Serpico}}}\ and\ \bibinfo {author} {\bibfnamefont {G.~G.}\ \bibnamefont
  {{Raffelt}}},\ }\href {\doibase 10.1103/PhysRevD.71.127301} {\bibfield
  {journal} {\bibinfo  {journal} {\prd}\ }\textbf {\bibinfo {volume} {71}},\
  \bibinfo {eid} {127301} (\bibinfo {year} {2005})},\ \Eprint
  {http://arxiv.org/abs/astro-ph/0506162} {astro-ph/0506162} \BibitemShut
  {NoStop}%
\bibitem [{\citenamefont {{Mangano}}\ \emph {et~al.}(2012)\citenamefont
  {{Mangano}}, \citenamefont {{Miele}}, \citenamefont {{Pastor}}, \citenamefont
  {{Pisanti}},\ and\ \citenamefont {{Sarikas}}}]{Mangano:2012aa}%
  \BibitemOpen
  \bibfield  {author} {\bibinfo {author} {\bibfnamefont {G.}~\bibnamefont
  {{Mangano}}}, \bibinfo {author} {\bibfnamefont {G.}~\bibnamefont {{Miele}}},
  \bibinfo {author} {\bibfnamefont {S.}~\bibnamefont {{Pastor}}}, \bibinfo
  {author} {\bibfnamefont {O.}~\bibnamefont {{Pisanti}}}, \ and\ \bibinfo
  {author} {\bibfnamefont {S.}~\bibnamefont {{Sarikas}}},\ }\href {\doibase
  10.1016/j.physletb.2012.01.015} {\bibfield  {journal} {\bibinfo  {journal}
  {Physics Letters B}\ }\textbf {\bibinfo {volume} {708}},\ \bibinfo {pages}
  {1} (\bibinfo {year} {2012})},\ \Eprint {http://arxiv.org/abs/1110.4335}
  {arXiv:1110.4335 [hep-ph]} \BibitemShut {NoStop}%
\bibitem [{\citenamefont {{Castorina}}\ \emph {et~al.}(2012)\citenamefont
  {{Castorina}}, \citenamefont {{Fran{\c c}a}}, \citenamefont {{Lattanzi}},
  \citenamefont {{Lesgourgues}}, \citenamefont {{Mangano}}, \citenamefont
  {{Melchiorri}},\ and\ \citenamefont {{Pastor}}}]{Castorina:2012aa}%
  \BibitemOpen
  \bibfield  {author} {\bibinfo {author} {\bibfnamefont {E.}~\bibnamefont
  {{Castorina}}}, \bibinfo {author} {\bibfnamefont {U.}~\bibnamefont {{Fran{\c
  c}a}}}, \bibinfo {author} {\bibfnamefont {M.}~\bibnamefont {{Lattanzi}}},
  \bibinfo {author} {\bibfnamefont {J.}~\bibnamefont {{Lesgourgues}}}, \bibinfo
  {author} {\bibfnamefont {G.}~\bibnamefont {{Mangano}}}, \bibinfo {author}
  {\bibfnamefont {A.}~\bibnamefont {{Melchiorri}}}, \ and\ \bibinfo {author}
  {\bibfnamefont {S.}~\bibnamefont {{Pastor}}},\ }\href {\doibase
  10.1103/PhysRevD.86.023517} {\bibfield  {journal} {\bibinfo  {journal}
  {\prd}\ }\textbf {\bibinfo {volume} {86}},\ \bibinfo {eid} {023517} (\bibinfo
  {year} {2012})},\ \Eprint {http://arxiv.org/abs/1204.2510} {arXiv:1204.2510
  [astro-ph.CO]} \BibitemShut {NoStop}%
\bibitem [{\citenamefont {{Steigman}}(2012)}]{Steigman:2012sp}%
  \BibitemOpen
  \bibfield  {author} {\bibinfo {author} {\bibfnamefont {G.}~\bibnamefont
  {{Steigman}}},\ }\href@noop {} {\bibfield  {journal} {\bibinfo  {journal}
  {ArXiv e-prints}\ } (\bibinfo {year} {2012})},\ \Eprint
  {http://arxiv.org/abs/1208.0032} {arXiv:1208.0032 [hep-ph]} \BibitemShut
  {NoStop}%
\bibitem [{\citenamefont {{Abazajian}}\ \emph
  {et~al.}(2001{\natexlab{a}})\citenamefont {{Abazajian}}, \citenamefont
  {{Fuller}},\ and\ \citenamefont {{Patel}}}]{Abazajian:2001ab}%
  \BibitemOpen
  \bibfield  {author} {\bibinfo {author} {\bibfnamefont {K.}~\bibnamefont
  {{Abazajian}}}, \bibinfo {author} {\bibfnamefont {G.~M.}\ \bibnamefont
  {{Fuller}}}, \ and\ \bibinfo {author} {\bibfnamefont {M.}~\bibnamefont
  {{Patel}}},\ }\href {\doibase 10.1103/PhysRevD.64.023501} {\bibfield
  {journal} {\bibinfo  {journal} {\prd}\ }\textbf {\bibinfo {volume} {64}},\
  \bibinfo {eid} {023501} (\bibinfo {year} {2001}{\natexlab{a}})},\ \Eprint
  {http://arxiv.org/abs/astro-ph/0101524} {astro-ph/0101524} \BibitemShut
  {NoStop}%
\bibitem [{\citenamefont {{Planck Collaboration}}\ \emph
  {et~al.}(2016)\citenamefont {{Planck Collaboration}}, \citenamefont {{Ade}},
  \citenamefont {{Aghanim}}, \citenamefont {{Arnaud}}, \citenamefont
  {{Ashdown}}, \citenamefont {{Aumont}}, \citenamefont {{Baccigalupi}},
  \citenamefont {{Banday}}, \citenamefont {{Barreiro}}, \citenamefont
  {{Bartlett}},\ and\ \citenamefont {et~al.}}]{Planck-Collaboration:2016aa}%
  \BibitemOpen
  \bibfield  {author} {\bibinfo {author} {\bibnamefont {{Planck
  Collaboration}}}, \bibinfo {author} {\bibfnamefont {P.~A.~R.}\ \bibnamefont
  {{Ade}}}, \bibinfo {author} {\bibfnamefont {N.}~\bibnamefont {{Aghanim}}},
  \bibinfo {author} {\bibfnamefont {M.}~\bibnamefont {{Arnaud}}}, \bibinfo
  {author} {\bibfnamefont {M.}~\bibnamefont {{Ashdown}}}, \bibinfo {author}
  {\bibfnamefont {J.}~\bibnamefont {{Aumont}}}, \bibinfo {author}
  {\bibfnamefont {C.}~\bibnamefont {{Baccigalupi}}}, \bibinfo {author}
  {\bibfnamefont {A.~J.}\ \bibnamefont {{Banday}}}, \bibinfo {author}
  {\bibfnamefont {R.~B.}\ \bibnamefont {{Barreiro}}}, \bibinfo {author}
  {\bibfnamefont {J.~G.}\ \bibnamefont {{Bartlett}}}, \ and\ \bibinfo {author}
  {\bibnamefont {et~al.}},\ }\href {\doibase 10.1051/0004-6361/201525830}
  {\bibfield  {journal} {\bibinfo  {journal} {Astron. Astrophys.}\ }\textbf
  {\bibinfo {volume} {594}},\ \bibinfo {eid} {A13} (\bibinfo {year} {2016})},\
  \Eprint {http://arxiv.org/abs/1502.01589} {arXiv:1502.01589} \BibitemShut
  {NoStop}%
\bibitem [{\citenamefont {{Tremaine}}\ and\ \citenamefont
  {{Gunn}}(1979)}]{Tremaine:1979aa}%
  \BibitemOpen
  \bibfield  {author} {\bibinfo {author} {\bibfnamefont {S.}~\bibnamefont
  {{Tremaine}}}\ and\ \bibinfo {author} {\bibfnamefont {J.~E.}\ \bibnamefont
  {{Gunn}}},\ }\href {\doibase 10.1103/PhysRevLett.42.407} {\bibfield
  {journal} {\bibinfo  {journal} {Physical Review Letters}\ }\textbf {\bibinfo
  {volume} {42}},\ \bibinfo {pages} {407} (\bibinfo {year} {1979})}\BibitemShut
  {NoStop}%
\bibitem [{\citenamefont {{Gorbunov}}\ \emph {et~al.}(2008)\citenamefont
  {{Gorbunov}}, \citenamefont {{Khmelnitsky}},\ and\ \citenamefont
  {{Rubakov}}}]{Gorbunov:2008aa}%
  \BibitemOpen
  \bibfield  {author} {\bibinfo {author} {\bibfnamefont {D.}~\bibnamefont
  {{Gorbunov}}}, \bibinfo {author} {\bibfnamefont {A.}~\bibnamefont
  {{Khmelnitsky}}}, \ and\ \bibinfo {author} {\bibfnamefont {V.}~\bibnamefont
  {{Rubakov}}},\ }\href {\doibase 10.1088/1475-7516/2008/10/041} {\bibfield
  {journal} {\bibinfo  {journal} {J. Cosmol. Astropart. Phys.}\ }\textbf
  {\bibinfo {volume} {10}},\ \bibinfo {eid} {041} (\bibinfo {year} {2008})},\
  \Eprint {http://arxiv.org/abs/0808.3910} {arXiv:0808.3910 [hep-ph]}
  \BibitemShut {NoStop}%
\bibitem [{\citenamefont {{Boyarsky}}\ \emph
  {et~al.}(2009{\natexlab{b}})\citenamefont {{Boyarsky}}, \citenamefont
  {{Ruchayskiy}},\ and\ \citenamefont {{Iakubovskyi}}}]{Boyarsky:2009aa}%
  \BibitemOpen
  \bibfield  {author} {\bibinfo {author} {\bibfnamefont {A.}~\bibnamefont
  {{Boyarsky}}}, \bibinfo {author} {\bibfnamefont {O.}~\bibnamefont
  {{Ruchayskiy}}}, \ and\ \bibinfo {author} {\bibfnamefont {D.}~\bibnamefont
  {{Iakubovskyi}}},\ }\href {\doibase 10.1088/1475-7516/2009/03/005} {\bibfield
   {journal} {\bibinfo  {journal} {JCAP}\ }\textbf {\bibinfo {volume} {3}},\
  \bibinfo {eid} {005} (\bibinfo {year} {2009}{\natexlab{b}})},\ \Eprint
  {http://arxiv.org/abs/0808.3902} {arXiv:0808.3902 [hep-ph]} \BibitemShut
  {NoStop}%
\bibitem [{\citenamefont {{Lesgourgues}}(2011)}]{Lesgourgues:2011aa}%
  \BibitemOpen
  \bibfield  {author} {\bibinfo {author} {\bibfnamefont {J.}~\bibnamefont
  {{Lesgourgues}}},\ }\href@noop {} {\bibfield  {journal} {\bibinfo  {journal}
  {ArXiv e-prints}\ } (\bibinfo {year} {2011})},\ \Eprint
  {http://arxiv.org/abs/1104.2932} {arXiv:1104.2932 [astro-ph.IM]} \BibitemShut
  {NoStop}%
\bibitem [{\citenamefont {{Silk}}(1968)}]{Silk:1968aa}%
  \BibitemOpen
  \bibfield  {author} {\bibinfo {author} {\bibfnamefont {J.}~\bibnamefont
  {{Silk}}},\ }\href {\doibase 10.1086/149449} {\bibfield  {journal} {\bibinfo
  {journal} {\apj}\ }\textbf {\bibinfo {volume} {151}},\ \bibinfo {pages} {459}
  (\bibinfo {year} {1968})}\BibitemShut {NoStop}%
\bibitem [{\citenamefont {{Polisensky}}\ and\ \citenamefont
  {{Ricotti}}(2011)}]{Polisensky:2011aa}%
  \BibitemOpen
  \bibfield  {author} {\bibinfo {author} {\bibfnamefont {E.}~\bibnamefont
  {{Polisensky}}}\ and\ \bibinfo {author} {\bibfnamefont {M.}~\bibnamefont
  {{Ricotti}}},\ }\href {\doibase 10.1103/PhysRevD.83.043506} {\bibfield
  {journal} {\bibinfo  {journal} {\prd}\ }\textbf {\bibinfo {volume} {83}},\
  \bibinfo {eid} {043506} (\bibinfo {year} {2011})},\ \Eprint
  {http://arxiv.org/abs/1004.1459} {arXiv:1004.1459 [astro-ph.CO]} \BibitemShut
  {NoStop}%
\bibitem [{\citenamefont {{Abazajian}}\ and\ \citenamefont
  {{Koushiappas}}(2006)}]{Abazajian:2006aa}%
  \BibitemOpen
  \bibfield  {author} {\bibinfo {author} {\bibfnamefont {K.}~\bibnamefont
  {{Abazajian}}}\ and\ \bibinfo {author} {\bibfnamefont {S.~M.}\ \bibnamefont
  {{Koushiappas}}},\ }\href@noop {} {\bibfield  {journal} {\bibinfo  {journal}
  {Phys. Rev. D}\ }\textbf {\bibinfo {volume} {47}} (\bibinfo {year} {2006})},\
  \Eprint {http://arxiv.org/abs/astro-ph/0605271v2} {arXiv:astro-ph/0605271v2
  [astro-ph]} \BibitemShut {NoStop}%
\bibitem [{\citenamefont {{Viel}}\ \emph {et~al.}(2006)\citenamefont {{Viel}},
  \citenamefont {{Haehnelt}},\ and\ \citenamefont {{Lewis}}}]{Viel:2006aa}%
  \BibitemOpen
  \bibfield  {author} {\bibinfo {author} {\bibfnamefont {M.}~\bibnamefont
  {{Viel}}}, \bibinfo {author} {\bibfnamefont {M.~G.}\ \bibnamefont
  {{Haehnelt}}}, \ and\ \bibinfo {author} {\bibfnamefont {A.}~\bibnamefont
  {{Lewis}}},\ }\href@noop {} {\bibfield  {journal} {\bibinfo  {journal} {Mon.
  Not. Roy. Astron. Soc.}\ }\textbf {\bibinfo {volume} {370}},\ \bibinfo
  {pages} {51} (\bibinfo {year} {2006})},\ \Eprint
  {http://arxiv.org/abs/astro-ph/0604310v2} {arXiv:astro-ph/0604310v2
  [astro-ph]} \BibitemShut {NoStop}%
\bibitem [{\citenamefont {{Seljak}}\ \emph {et~al.}(2006)\citenamefont
  {{Seljak}}, \citenamefont {{Slosar}},\ and\ \citenamefont
  {{McDonald}}}]{Seljak:2006aa}%
  \BibitemOpen
  \bibfield  {author} {\bibinfo {author} {\bibfnamefont {U.}~\bibnamefont
  {{Seljak}}}, \bibinfo {author} {\bibfnamefont {A.}~\bibnamefont {{Slosar}}},
  \ and\ \bibinfo {author} {\bibfnamefont {P.}~\bibnamefont {{McDonald}}},\
  }\href@noop {} {\bibfield  {journal} {\bibinfo  {journal} {J. Cosmol.
  Astropart. Phys.}\ }\textbf {\bibinfo {volume} {2006}} (\bibinfo {year}
  {2006})},\ \Eprint {http://arxiv.org/abs/astro-ph/0604335v4}
  {arXiv:astro-ph/0604335v4 [astro-ph]} \BibitemShut {NoStop}%
\bibitem [{\citenamefont {{Boyarsky}}\ \emph
  {et~al.}(2009{\natexlab{c}})\citenamefont {{Boyarsky}}, \citenamefont
  {{Lesgourgues}}, \citenamefont {{Ruchayskiy}},\ and\ \citenamefont
  {{Viel}}}]{Boyarsky:2009ac}%
  \BibitemOpen
  \bibfield  {author} {\bibinfo {author} {\bibfnamefont {A.}~\bibnamefont
  {{Boyarsky}}}, \bibinfo {author} {\bibfnamefont {J.}~\bibnamefont
  {{Lesgourgues}}}, \bibinfo {author} {\bibfnamefont {O.}~\bibnamefont
  {{Ruchayskiy}}}, \ and\ \bibinfo {author} {\bibfnamefont {M.}~\bibnamefont
  {{Viel}}},\ }\href@noop {} {\bibfield  {journal} {\bibinfo  {journal} {J.
  Cosmol. Astropart. Phys.}\ }\textbf {\bibinfo {volume} {2009}} (\bibinfo
  {year} {2009}{\natexlab{c}})},\ \Eprint {http://arxiv.org/abs/0812.0010v2}
  {arXiv:0812.0010v2 [astro-ph]} \BibitemShut {NoStop}%
\bibitem [{\citenamefont {{Viel}}\ \emph {et~al.}(2013)\citenamefont {{Viel}},
  \citenamefont {{Becker}}, \citenamefont {{Bolton}},\ and\ \citenamefont
  {{Haehnelt}}}]{Viel:2013aa}%
  \BibitemOpen
  \bibfield  {author} {\bibinfo {author} {\bibfnamefont {M.}~\bibnamefont
  {{Viel}}}, \bibinfo {author} {\bibfnamefont {G.~D.}\ \bibnamefont
  {{Becker}}}, \bibinfo {author} {\bibfnamefont {J.~S.}\ \bibnamefont
  {{Bolton}}}, \ and\ \bibinfo {author} {\bibfnamefont {M.~G.}\ \bibnamefont
  {{Haehnelt}}},\ }\href {\doibase 10.1103/PhysRevD.88.043502} {\bibfield
  {journal} {\bibinfo  {journal} {\prd}\ }\textbf {\bibinfo {volume} {88}},\
  \bibinfo {eid} {043502} (\bibinfo {year} {2013})},\ \Eprint
  {http://arxiv.org/abs/1306.2314} {arXiv:1306.2314} \BibitemShut {NoStop}%
\bibitem [{\citenamefont {{Garzilli}}\ \emph {et~al.}(2015)\citenamefont
  {{Garzilli}}, \citenamefont {{Boyarsky}},\ and\ \citenamefont
  {{Ruchayskiy}}}]{Garzilli:2015aa}%
  \BibitemOpen
  \bibfield  {author} {\bibinfo {author} {\bibfnamefont {A.}~\bibnamefont
  {{Garzilli}}}, \bibinfo {author} {\bibfnamefont {A.}~\bibnamefont
  {{Boyarsky}}}, \ and\ \bibinfo {author} {\bibfnamefont {O.}~\bibnamefont
  {{Ruchayskiy}}},\ }\href@noop {} {\bibfield  {journal} {\bibinfo  {journal}
  {ArXiv e-prints}\ } (\bibinfo {year} {2015})},\ \Eprint
  {http://arxiv.org/abs/1510.07006} {arXiv:1510.07006 [astro-ph.CO]}
  \BibitemShut {NoStop}%
\bibitem [{\citenamefont {{Bullock}}\ \emph {et~al.}(2000)\citenamefont
  {{Bullock}}, \citenamefont {{Kravtsov}},\ and\ \citenamefont
  {{Weinberg}}}]{Bullock:2000aa}%
  \BibitemOpen
  \bibfield  {author} {\bibinfo {author} {\bibfnamefont {J.~S.}\ \bibnamefont
  {{Bullock}}}, \bibinfo {author} {\bibfnamefont {A.~V.}\ \bibnamefont
  {{Kravtsov}}}, \ and\ \bibinfo {author} {\bibfnamefont {D.~H.}\ \bibnamefont
  {{Weinberg}}},\ }\href {\doibase 10.1086/309279} {\bibfield  {journal}
  {\bibinfo  {journal} {\apj}\ }\textbf {\bibinfo {volume} {539}},\ \bibinfo
  {pages} {517} (\bibinfo {year} {2000})},\ \Eprint
  {http://arxiv.org/abs/astro-ph/0002214} {astro-ph/0002214} \BibitemShut
  {NoStop}%
\bibitem [{\citenamefont {Governato}\ \emph {et~al.}(2012)\citenamefont
  {Governato}, \citenamefont {Zolotov}, \citenamefont {Pontzen}, \citenamefont
  {Christensen}, \citenamefont {Oh}, \citenamefont {Brooks}, \citenamefont
  {Quinn}, \citenamefont {Shen},\ and\ \citenamefont
  {Wadsley}}]{Governato:2012fa}%
  \BibitemOpen
  \bibfield  {author} {\bibinfo {author} {\bibfnamefont {F.}~\bibnamefont
  {Governato}}, \bibinfo {author} {\bibfnamefont {A.}~\bibnamefont {Zolotov}},
  \bibinfo {author} {\bibfnamefont {A.}~\bibnamefont {Pontzen}}, \bibinfo
  {author} {\bibfnamefont {C.}~\bibnamefont {Christensen}}, \bibinfo {author}
  {\bibfnamefont {S.~H.}\ \bibnamefont {Oh}}, \bibinfo {author} {\bibfnamefont
  {A.~M.}\ \bibnamefont {Brooks}}, \bibinfo {author} {\bibfnamefont
  {T.}~\bibnamefont {Quinn}}, \bibinfo {author} {\bibfnamefont
  {S.}~\bibnamefont {Shen}}, \ and\ \bibinfo {author} {\bibfnamefont
  {J.}~\bibnamefont {Wadsley}},\ }\href {\doibase
  10.1111/j.1365-2966.2012.20696.x} {\bibfield  {journal} {\bibinfo  {journal}
  {Mon. Not. Roy. Astron. Soc.}\ }\textbf {\bibinfo {volume} {422}},\ \bibinfo
  {pages} {1231} (\bibinfo {year} {2012})},\ \Eprint
  {http://arxiv.org/abs/1202.0554} {arXiv:1202.0554 [astro-ph.CO]} \BibitemShut
  {NoStop}%
\bibitem [{\citenamefont {{Font}}\ \emph {et~al.}(2011)\citenamefont {{Font}},
  \citenamefont {{Benson}}, \citenamefont {{Bower}}, \citenamefont {{Frenk}},
  \citenamefont {{Cooper}}, \citenamefont {{Lucia}}, \citenamefont {{Helly}},
  \citenamefont {{Helmi}}, \citenamefont {{Li}}, \citenamefont {{McCarthy}},
  \citenamefont {{Navarro}}, \citenamefont {{Springel}}, \citenamefont
  {{Starkenburg}},\ and\ \citenamefont {{Wang}}}]{Font:2011aa}%
  \BibitemOpen
  \bibfield  {author} {\bibinfo {author} {\bibfnamefont {A.~S.}\ \bibnamefont
  {{Font}}}, \bibinfo {author} {\bibfnamefont {A.~J.}\ \bibnamefont
  {{Benson}}}, \bibinfo {author} {\bibfnamefont {R.~G.}\ \bibnamefont
  {{Bower}}}, \bibinfo {author} {\bibfnamefont {C.~F.}\ \bibnamefont
  {{Frenk}}}, \bibinfo {author} {\bibfnamefont {A.~P.}\ \bibnamefont
  {{Cooper}}}, \bibinfo {author} {\bibfnamefont {G.~D.}\ \bibnamefont
  {{Lucia}}}, \bibinfo {author} {\bibfnamefont {J.~C.}\ \bibnamefont
  {{Helly}}}, \bibinfo {author} {\bibfnamefont {A.}~\bibnamefont {{Helmi}}},
  \bibinfo {author} {\bibfnamefont {Y.-S.}\ \bibnamefont {{Li}}}, \bibinfo
  {author} {\bibfnamefont {I.~G.}\ \bibnamefont {{McCarthy}}}, \bibinfo
  {author} {\bibfnamefont {J.~F.}\ \bibnamefont {{Navarro}}}, \bibinfo {author}
  {\bibfnamefont {V.}~\bibnamefont {{Springel}}}, \bibinfo {author}
  {\bibfnamefont {E.}~\bibnamefont {{Starkenburg}}}, \ and\ \bibinfo {author}
  {\bibfnamefont {J.}~\bibnamefont {{Wang}}},\ }\href@noop {} {\bibfield
  {journal} {\bibinfo  {journal} {Mon. Not. Roy. Astron. Soc.}\ }\textbf
  {\bibinfo {volume} {417}},\ \bibinfo {pages} {1260} (\bibinfo {year}
  {2011})},\ \Eprint {http://arxiv.org/abs/1103.0024v2} {arXiv:1103.0024v2
  [astro-ph.GA]} \BibitemShut {NoStop}%
\bibitem [{\citenamefont {{Lovell}}\ \emph
  {et~al.}(2016{\natexlab{b}})\citenamefont {{Lovell}}, \citenamefont {{Bose}},
  \citenamefont {{Boyarsky}}, \citenamefont {{Cole}}, \citenamefont {{Frenk}},
  \citenamefont {{Gonzalez-Perez}}, \citenamefont {{Kennedy}}, \citenamefont
  {{Ruchayskiy}},\ and\ \citenamefont {{Smith}}}]{Lovell:2016aa}%
  \BibitemOpen
  \bibfield  {author} {\bibinfo {author} {\bibfnamefont {M.~R.}\ \bibnamefont
  {{Lovell}}}, \bibinfo {author} {\bibfnamefont {S.}~\bibnamefont {{Bose}}},
  \bibinfo {author} {\bibfnamefont {A.}~\bibnamefont {{Boyarsky}}}, \bibinfo
  {author} {\bibfnamefont {S.}~\bibnamefont {{Cole}}}, \bibinfo {author}
  {\bibfnamefont {C.~S.}\ \bibnamefont {{Frenk}}}, \bibinfo {author}
  {\bibfnamefont {V.}~\bibnamefont {{Gonzalez-Perez}}}, \bibinfo {author}
  {\bibfnamefont {R.}~\bibnamefont {{Kennedy}}}, \bibinfo {author}
  {\bibfnamefont {O.}~\bibnamefont {{Ruchayskiy}}}, \ and\ \bibinfo {author}
  {\bibfnamefont {A.}~\bibnamefont {{Smith}}},\ }\href {\doibase
  10.1093/mnras/stw1317} {\bibfield  {journal} {\bibinfo  {journal} {Mon. Not.
  Roy. Astron. Soc.}\ }\textbf {\bibinfo {volume} {461}},\ \bibinfo {pages}
  {60} (\bibinfo {year} {2016}{\natexlab{b}})},\ \Eprint
  {http://arxiv.org/abs/1511.04078} {arXiv:1511.04078} \BibitemShut {NoStop}%
\bibitem [{\citenamefont {{Schneider}}(2015)}]{Schneider:2015aa}%
  \BibitemOpen
  \bibfield  {author} {\bibinfo {author} {\bibfnamefont {A.}~\bibnamefont
  {{Schneider}}},\ }\href {\doibase 10.1093/mnras/stv1169} {\bibfield
  {journal} {\bibinfo  {journal} {MNRAS}\ }\textbf {\bibinfo {volume} {451}},\
  \bibinfo {pages} {3117} (\bibinfo {year} {2015})},\ \Eprint
  {http://arxiv.org/abs/1412.2133} {arXiv:1412.2133} \BibitemShut {NoStop}%
\bibitem [{\citenamefont {{Dunstan}}\ \emph {et~al.}(2011)\citenamefont
  {{Dunstan}}, \citenamefont {{Abazajian}}, \citenamefont {{Polisensky}},\ and\
  \citenamefont {{Ricotti}}}]{Dunstan:2011aa}%
  \BibitemOpen
  \bibfield  {author} {\bibinfo {author} {\bibfnamefont {R.~M.}\ \bibnamefont
  {{Dunstan}}}, \bibinfo {author} {\bibfnamefont {K.~N.}\ \bibnamefont
  {{Abazajian}}}, \bibinfo {author} {\bibfnamefont {E.}~\bibnamefont
  {{Polisensky}}}, \ and\ \bibinfo {author} {\bibfnamefont {M.}~\bibnamefont
  {{Ricotti}}},\ }\href@noop {} {\bibfield  {journal} {\bibinfo  {journal}
  {ArXiv e-prints}\ } (\bibinfo {year} {2011})},\ \Eprint
  {http://arxiv.org/abs/1109.6291} {arXiv:1109.6291 [astro-ph.CO]} \BibitemShut
  {NoStop}%
\bibitem [{\citenamefont {{Press}}\ and\ \citenamefont
  {{Schechter}}(1974)}]{Press:1974aa}%
  \BibitemOpen
  \bibfield  {author} {\bibinfo {author} {\bibfnamefont {W.~H.}\ \bibnamefont
  {{Press}}}\ and\ \bibinfo {author} {\bibfnamefont {P.}~\bibnamefont
  {{Schechter}}},\ }\href {\doibase 10.1086/152650} {\bibfield  {journal}
  {\bibinfo  {journal} {\apj}\ }\textbf {\bibinfo {volume} {187}},\ \bibinfo
  {pages} {425} (\bibinfo {year} {1974})}\BibitemShut {NoStop}%
\bibitem [{\citenamefont {{Bond}}\ \emph {et~al.}(1991)\citenamefont {{Bond}},
  \citenamefont {{Cole}}, \citenamefont {{Efstathiou}},\ and\ \citenamefont
  {{Kaiser}}}]{Bond:1991aa}%
  \BibitemOpen
  \bibfield  {author} {\bibinfo {author} {\bibfnamefont {J.~R.}\ \bibnamefont
  {{Bond}}}, \bibinfo {author} {\bibfnamefont {S.}~\bibnamefont {{Cole}}},
  \bibinfo {author} {\bibfnamefont {G.}~\bibnamefont {{Efstathiou}}}, \ and\
  \bibinfo {author} {\bibfnamefont {N.}~\bibnamefont {{Kaiser}}},\ }\href
  {\doibase 10.1086/170520} {\bibfield  {journal} {\bibinfo  {journal} {\apj}\
  }\textbf {\bibinfo {volume} {379}},\ \bibinfo {pages} {440} (\bibinfo {year}
  {1991})}\BibitemShut {NoStop}%
\bibitem [{\citenamefont {{Komatsu}}\ \emph {et~al.}(2011)\citenamefont
  {{Komatsu}}, \citenamefont {{Smith}}, \citenamefont {{Dunkley}},
  \citenamefont {{Bennett}}, \citenamefont {{Gold}}, \citenamefont {{Hinshaw}},
  \citenamefont {{Jarosik}}, \citenamefont {{Larson}}, \citenamefont {{Nolta}},
  \citenamefont {{Page}}, \citenamefont {{Spergel}}, \citenamefont {{Halpern}},
  \citenamefont {{Hill}}, \citenamefont {{Kogut}}, \citenamefont {{Limon}},
  \citenamefont {{Meyer}}, \citenamefont {{Odegard}}, \citenamefont {{Tucker}},
  \citenamefont {{Weiland}}, \citenamefont {{Wollack}},\ and\ \citenamefont
  {{Wright}}}]{Komatsu:2011aa}%
  \BibitemOpen
  \bibfield  {author} {\bibinfo {author} {\bibfnamefont {E.}~\bibnamefont
  {{Komatsu}}}, \bibinfo {author} {\bibfnamefont {K.~M.}\ \bibnamefont
  {{Smith}}}, \bibinfo {author} {\bibfnamefont {J.}~\bibnamefont {{Dunkley}}},
  \bibinfo {author} {\bibfnamefont {C.~L.}\ \bibnamefont {{Bennett}}}, \bibinfo
  {author} {\bibfnamefont {B.}~\bibnamefont {{Gold}}}, \bibinfo {author}
  {\bibfnamefont {G.}~\bibnamefont {{Hinshaw}}}, \bibinfo {author}
  {\bibfnamefont {N.}~\bibnamefont {{Jarosik}}}, \bibinfo {author}
  {\bibfnamefont {D.}~\bibnamefont {{Larson}}}, \bibinfo {author}
  {\bibfnamefont {M.~R.}\ \bibnamefont {{Nolta}}}, \bibinfo {author}
  {\bibfnamefont {L.}~\bibnamefont {{Page}}}, \bibinfo {author} {\bibfnamefont
  {D.~N.}\ \bibnamefont {{Spergel}}}, \bibinfo {author} {\bibfnamefont
  {M.}~\bibnamefont {{Halpern}}}, \bibinfo {author} {\bibfnamefont {R.~S.}\
  \bibnamefont {{Hill}}}, \bibinfo {author} {\bibfnamefont {A.}~\bibnamefont
  {{Kogut}}}, \bibinfo {author} {\bibfnamefont {M.}~\bibnamefont {{Limon}}},
  \bibinfo {author} {\bibfnamefont {S.~S.}\ \bibnamefont {{Meyer}}}, \bibinfo
  {author} {\bibfnamefont {N.}~\bibnamefont {{Odegard}}}, \bibinfo {author}
  {\bibfnamefont {G.~S.}\ \bibnamefont {{Tucker}}}, \bibinfo {author}
  {\bibfnamefont {J.~L.}\ \bibnamefont {{Weiland}}}, \bibinfo {author}
  {\bibfnamefont {E.}~\bibnamefont {{Wollack}}}, \ and\ \bibinfo {author}
  {\bibfnamefont {E.~L.}\ \bibnamefont {{Wright}}},\ }\href {\doibase
  10.1088/0067-0049/192/2/18} {\bibfield  {journal} {\bibinfo  {journal} {The
  Astrophysical Journal Supplement}\ }\textbf {\bibinfo {volume} {192}},\
  \bibinfo {eid} {18} (\bibinfo {year} {2011})},\ \Eprint
  {http://arxiv.org/abs/1001.4538} {arXiv:1001.4538 [astro-ph.CO]} \BibitemShut
  {NoStop}%
\bibitem [{\citenamefont {{Planck Collaboration}}\ \emph
  {et~al.}(2014)\citenamefont {{Planck Collaboration}}, \citenamefont {{Ade}},
  \citenamefont {{Aghanim}}, \citenamefont {{Armitage-Caplan}}, \citenamefont
  {{Arnaud}}, \citenamefont {{Ashdown}}, \citenamefont {{Atrio-Barandela}},
  \citenamefont {{Aumont}}, \citenamefont {{Baccigalupi}}, \citenamefont
  {{Banday}},\ and\ \citenamefont {et~al.}}]{Planck-Collaboration:2014aa}%
  \BibitemOpen
  \bibfield  {author} {\bibinfo {author} {\bibnamefont {{Planck
  Collaboration}}}, \bibinfo {author} {\bibfnamefont {P.~A.~R.}\ \bibnamefont
  {{Ade}}}, \bibinfo {author} {\bibfnamefont {N.}~\bibnamefont {{Aghanim}}},
  \bibinfo {author} {\bibfnamefont {C.}~\bibnamefont {{Armitage-Caplan}}},
  \bibinfo {author} {\bibfnamefont {M.}~\bibnamefont {{Arnaud}}}, \bibinfo
  {author} {\bibfnamefont {M.}~\bibnamefont {{Ashdown}}}, \bibinfo {author}
  {\bibfnamefont {F.}~\bibnamefont {{Atrio-Barandela}}}, \bibinfo {author}
  {\bibfnamefont {J.}~\bibnamefont {{Aumont}}}, \bibinfo {author}
  {\bibfnamefont {C.}~\bibnamefont {{Baccigalupi}}}, \bibinfo {author}
  {\bibfnamefont {A.~J.}\ \bibnamefont {{Banday}}}, \ and\ \bibinfo {author}
  {\bibnamefont {et~al.}},\ }\href {\doibase 10.1051/0004-6361/201321591}
  {\bibfield  {journal} {\bibinfo  {journal} {Astron. Astrophys.}\ }\textbf
  {\bibinfo {volume} {571}},\ \bibinfo {eid} {A16} (\bibinfo {year} {2014})},\
  \Eprint {http://arxiv.org/abs/1303.5076} {arXiv:1303.5076} \BibitemShut
  {NoStop}%
\bibitem [{\citenamefont {{Wang}}\ \emph {et~al.}(2015)\citenamefont {{Wang}},
  \citenamefont {{Han}}, \citenamefont {{Cooper}}, \citenamefont {{Cole}},
  \citenamefont {{Frenk}},\ and\ \citenamefont {{Lowing}}}]{Wang:2015aa}%
  \BibitemOpen
  \bibfield  {author} {\bibinfo {author} {\bibfnamefont {W.}~\bibnamefont
  {{Wang}}}, \bibinfo {author} {\bibfnamefont {J.}~\bibnamefont {{Han}}},
  \bibinfo {author} {\bibfnamefont {A.~P.}\ \bibnamefont {{Cooper}}}, \bibinfo
  {author} {\bibfnamefont {S.}~\bibnamefont {{Cole}}}, \bibinfo {author}
  {\bibfnamefont {C.}~\bibnamefont {{Frenk}}}, \ and\ \bibinfo {author}
  {\bibfnamefont {B.}~\bibnamefont {{Lowing}}},\ }\href@noop {} {\bibfield
  {journal} {\bibinfo  {journal} {Mon. Not. Roy. Astron. Soc.}\ }\textbf
  {\bibinfo {volume} {453}},\ \bibinfo {pages} {377} (\bibinfo {year}
  {2015})},\ \Eprint {http://arxiv.org/abs/1502.03477v3} {arXiv:1502.03477v3
  [astro-ph.GA]} \BibitemShut {NoStop}%
\bibitem [{\citenamefont {{Brooks}}\ and\ \citenamefont
  {{Zolotov}}(2014)}]{Brooks:2014aa}%
  \BibitemOpen
  \bibfield  {author} {\bibinfo {author} {\bibfnamefont {A.~M.}\ \bibnamefont
  {{Brooks}}}\ and\ \bibinfo {author} {\bibfnamefont {A.}~\bibnamefont
  {{Zolotov}}},\ }\href {\doibase 10.1088/0004-637X/786/2/87} {\bibfield
  {journal} {\bibinfo  {journal} {\apj}\ }\textbf {\bibinfo {volume} {786}},\
  \bibinfo {eid} {87} (\bibinfo {year} {2014})},\ \Eprint
  {http://arxiv.org/abs/1207.2468} {arXiv:1207.2468} \BibitemShut {NoStop}%
\bibitem [{\citenamefont {{Bullock}}\ \emph {et~al.}(2001)\citenamefont
  {{Bullock}}, \citenamefont {{Kolatt}}, \citenamefont {{Sigad}}, \citenamefont
  {{Somerville}}, \citenamefont {{Kravtsov}}, \citenamefont {{Klypin}},
  \citenamefont {{Primack}},\ and\ \citenamefont {{Dekel}}}]{Bullock:2001aa}%
  \BibitemOpen
  \bibfield  {author} {\bibinfo {author} {\bibfnamefont {J.~S.}\ \bibnamefont
  {{Bullock}}}, \bibinfo {author} {\bibfnamefont {T.~S.}\ \bibnamefont
  {{Kolatt}}}, \bibinfo {author} {\bibfnamefont {Y.}~\bibnamefont {{Sigad}}},
  \bibinfo {author} {\bibfnamefont {R.~S.}\ \bibnamefont {{Somerville}}},
  \bibinfo {author} {\bibfnamefont {A.~V.}\ \bibnamefont {{Kravtsov}}},
  \bibinfo {author} {\bibfnamefont {A.~A.}\ \bibnamefont {{Klypin}}}, \bibinfo
  {author} {\bibfnamefont {J.~R.}\ \bibnamefont {{Primack}}}, \ and\ \bibinfo
  {author} {\bibfnamefont {A.}~\bibnamefont {{Dekel}}},\ }\href {\doibase
  10.1046/j.1365-8711.2001.04068.x} {\bibfield  {journal} {\bibinfo  {journal}
  {MNRAS}\ }\textbf {\bibinfo {volume} {321}},\ \bibinfo {pages} {559}
  (\bibinfo {year} {2001})},\ \Eprint {http://arxiv.org/abs/astro-ph/9908159}
  {astro-ph/9908159} \BibitemShut {NoStop}%
\bibitem [{\citenamefont {{Strigari}}\ \emph {et~al.}(2008)\citenamefont
  {{Strigari}}, \citenamefont {{Bullock}}, \citenamefont {{Kaplinghat}},
  \citenamefont {{Simon}}, \citenamefont {{Geha}}, \citenamefont {{Willman}},\
  and\ \citenamefont {{Walker}}}]{Strigari:2008aa}%
  \BibitemOpen
  \bibfield  {author} {\bibinfo {author} {\bibfnamefont {L.~E.}\ \bibnamefont
  {{Strigari}}}, \bibinfo {author} {\bibfnamefont {J.~S.}\ \bibnamefont
  {{Bullock}}}, \bibinfo {author} {\bibfnamefont {M.}~\bibnamefont
  {{Kaplinghat}}}, \bibinfo {author} {\bibfnamefont {J.~D.}\ \bibnamefont
  {{Simon}}}, \bibinfo {author} {\bibfnamefont {M.}~\bibnamefont {{Geha}}},
  \bibinfo {author} {\bibfnamefont {B.}~\bibnamefont {{Willman}}}, \ and\
  \bibinfo {author} {\bibfnamefont {M.~G.}\ \bibnamefont {{Walker}}},\ }\href
  {\doibase 10.1038/nature07222} {\bibfield  {journal} {\bibinfo  {journal}
  {Nature}\ }\textbf {\bibinfo {volume} {454}},\ \bibinfo {pages} {1096}
  (\bibinfo {year} {2008})},\ \Eprint {http://arxiv.org/abs/0808.3772}
  {arXiv:0808.3772} \BibitemShut {NoStop}%
\bibitem [{\citenamefont {{Abbott}}\ \emph {et~al.}(2016)\citenamefont
  {{Abbott}} \emph {et~al.}}]{DES-Collaboration:2016aa}%
  \BibitemOpen
  \bibfield  {author} {\bibinfo {author} {\bibfnamefont {T.}~\bibnamefont
  {{Abbott}}} \emph {et~al.},\ }\href {\doibase 10.1093/mnras/stw641}
  {\bibfield  {journal} {\bibinfo  {journal} {MNRAS}\ }\textbf {\bibinfo
  {volume} {460}},\ \bibinfo {pages} {1270} (\bibinfo {year} {2016})},\ \Eprint
  {http://arxiv.org/abs/1601.00329} {arXiv:1601.00329} \BibitemShut {NoStop}%
\bibitem [{\citenamefont {{Beringer}}\ and\ \citenamefont {{(Particle Data
  Group)}}(2012)}]{PDG:2013xk}%
  \BibitemOpen
  \bibfield  {author} {\bibinfo {author} {\bibfnamefont {J.}~\bibnamefont
  {{Beringer}}}\ and\ \bibinfo {author} {\bibfnamefont {e.}~\bibnamefont
  {{(Particle Data Group)}}},\ }\href@noop {} {\bibfield  {journal} {\bibinfo
  {journal} {Phys. Rev. D}\ }\textbf {\bibinfo {volume} {86}},\ \bibinfo
  {pages} {010001} (\bibinfo {year} {2012})}\BibitemShut {NoStop}%
\bibitem [{\citenamefont {{Wilks}}(1938)}]{Wilks:1938uq}%
  \BibitemOpen
  \bibfield  {author} {\bibinfo {author} {\bibfnamefont {S.~S.}\ \bibnamefont
  {{Wilks}}},\ }\href@noop {} {\bibfield  {journal} {\bibinfo  {journal} {The
  Annals of Mathematical Statistics}\ }\textbf {\bibinfo {volume} {9}},\
  \bibinfo {pages} {60} (\bibinfo {year} {1938})}\BibitemShut {NoStop}%
\bibitem [{\citenamefont {{McConnachie}}(2012)}]{McConnachie:2012aa}%
  \BibitemOpen
  \bibfield  {author} {\bibinfo {author} {\bibfnamefont {A.~W.}\ \bibnamefont
  {{McConnachie}}},\ }\href {\doibase 10.1088/0004-6256/144/1/4} {\bibfield
  {journal} {\bibinfo  {journal} {The Astronomical Journal}\ }\textbf {\bibinfo
  {volume} {144}},\ \bibinfo {eid} {4} (\bibinfo {year} {2012})},\ \Eprint
  {http://arxiv.org/abs/1204.1562} {arXiv:1204.1562} \BibitemShut {NoStop}%
\bibitem [{\citenamefont {{SDSS Collaboration}}\ \emph
  {et~al.}(2016)\citenamefont {{SDSS Collaboration}}, \citenamefont
  {{Albareti}}, \citenamefont {{Allende Prieto}}, \citenamefont {{Almeida}},
  \citenamefont {{Anders}}, \citenamefont {{Anderson}}, \citenamefont
  {{Andrews}}, \citenamefont {{Aragon-Salamanca}}, \citenamefont
  {{Argudo-Fernandez}}, \citenamefont {{Armengaud}},\ and\ \citenamefont
  {et~al.}}]{SDSS-Collaboration:2016aa}%
  \BibitemOpen
  \bibfield  {author} {\bibinfo {author} {\bibnamefont {{SDSS Collaboration}}},
  \bibinfo {author} {\bibfnamefont {F.~D.}\ \bibnamefont {{Albareti}}},
  \bibinfo {author} {\bibfnamefont {C.}~\bibnamefont {{Allende Prieto}}},
  \bibinfo {author} {\bibfnamefont {A.}~\bibnamefont {{Almeida}}}, \bibinfo
  {author} {\bibfnamefont {F.}~\bibnamefont {{Anders}}}, \bibinfo {author}
  {\bibfnamefont {S.}~\bibnamefont {{Anderson}}}, \bibinfo {author}
  {\bibfnamefont {B.~H.}\ \bibnamefont {{Andrews}}}, \bibinfo {author}
  {\bibfnamefont {A.}~\bibnamefont {{Aragon-Salamanca}}}, \bibinfo {author}
  {\bibfnamefont {M.}~\bibnamefont {{Argudo-Fernandez}}}, \bibinfo {author}
  {\bibfnamefont {E.}~\bibnamefont {{Armengaud}}}, \ and\ \bibinfo {author}
  {\bibnamefont {et~al.}},\ }\href@noop {} {\bibfield  {journal} {\bibinfo
  {journal} {ArXiv e-prints}\ } (\bibinfo {year} {2016})},\ \Eprint
  {http://arxiv.org/abs/1608.02013} {arXiv:1608.02013} \BibitemShut {NoStop}%
\bibitem [{\citenamefont {{Koposov}}\ \emph
  {et~al.}(2015{\natexlab{a}})\citenamefont {{Koposov}}, \citenamefont
  {{Belokurov}}, \citenamefont {{Torrealba}},\ and\ \citenamefont
  {{Evans}}}]{Koposov:2015aa}%
  \BibitemOpen
  \bibfield  {author} {\bibinfo {author} {\bibfnamefont {S.}~\bibnamefont
  {{Koposov}}}, \bibinfo {author} {\bibfnamefont {V.}~\bibnamefont
  {{Belokurov}}}, \bibinfo {author} {\bibfnamefont {G.}~\bibnamefont
  {{Torrealba}}}, \ and\ \bibinfo {author} {\bibfnamefont {W.}~\bibnamefont
  {{Evans}}},\ }\href@noop {} {\bibfield  {journal} {\bibinfo  {journal} {IAU
  General Assembly}\ }\textbf {\bibinfo {volume} {22}},\ \bibinfo {eid}
  {2256759} (\bibinfo {year} {2015}{\natexlab{a}})}\BibitemShut {NoStop}%
\bibitem [{\citenamefont {{Bechtol}}\ \emph {et~al.}(2015)\citenamefont
  {{Bechtol}} \emph {et~al.}}]{Bechtol:2015aa}%
  \BibitemOpen
  \bibfield  {author} {\bibinfo {author} {\bibfnamefont {K.}~\bibnamefont
  {{Bechtol}}} \emph {et~al.},\ }\href {\doibase 10.1088/0004-637X/807/1/50}
  {\bibfield  {journal} {\bibinfo  {journal} {\apj}\ }\textbf {\bibinfo
  {volume} {807}},\ \bibinfo {eid} {50} (\bibinfo {year} {2015})},\ \Eprint
  {http://arxiv.org/abs/1503.02584} {arXiv:1503.02584} \BibitemShut {NoStop}%
\bibitem [{\citenamefont {{Drlica-Wagner}}\ \emph {et~al.}(2015)\citenamefont
  {{Drlica-Wagner}} \emph {et~al.}}]{Drlica-Wagner:2015aa}%
  \BibitemOpen
  \bibfield  {author} {\bibinfo {author} {\bibfnamefont {A.}~\bibnamefont
  {{Drlica-Wagner}}} \emph {et~al.},\ }\href {\doibase
  10.1088/0004-637X/813/2/109} {\bibfield  {journal} {\bibinfo  {journal}
  {\apj}\ }\textbf {\bibinfo {volume} {813}},\ \bibinfo {eid} {109} (\bibinfo
  {year} {2015})},\ \Eprint {http://arxiv.org/abs/1508.03622}
  {arXiv:1508.03622} \BibitemShut {NoStop}%
\bibitem [{\citenamefont {{Perez}}\ \emph {et~al.}(2016)\citenamefont
  {{Perez}}, \citenamefont {{Ng}}, \citenamefont {{Beacom}}, \citenamefont
  {{Hersh}}, \citenamefont {{Horiuchi}},\ and\ \citenamefont
  {{Krivonos}}}]{Perez:2016aa}%
  \BibitemOpen
  \bibfield  {author} {\bibinfo {author} {\bibfnamefont {K.}~\bibnamefont
  {{Perez}}}, \bibinfo {author} {\bibfnamefont {K.~C.~Y.}\ \bibnamefont
  {{Ng}}}, \bibinfo {author} {\bibfnamefont {J.~F.}\ \bibnamefont {{Beacom}}},
  \bibinfo {author} {\bibfnamefont {C.}~\bibnamefont {{Hersh}}}, \bibinfo
  {author} {\bibfnamefont {S.}~\bibnamefont {{Horiuchi}}}, \ and\ \bibinfo
  {author} {\bibfnamefont {R.}~\bibnamefont {{Krivonos}}},\ }\href@noop {}
  {\bibfield  {journal} {\bibinfo  {journal} {ArXiv e-prints}\ } (\bibinfo
  {year} {2016})},\ \Eprint {http://arxiv.org/abs/1609.00667} {arXiv:1609.00667
  [astro-ph.HE]} \BibitemShut {NoStop}%
\bibitem [{\citenamefont {{Pal}}\ and\ \citenamefont
  {{Wolfenstein}}(1982)}]{Pal:1982aa}%
  \BibitemOpen
  \bibfield  {author} {\bibinfo {author} {\bibfnamefont {P.~B.}\ \bibnamefont
  {{Pal}}}\ and\ \bibinfo {author} {\bibfnamefont {L.}~\bibnamefont
  {{Wolfenstein}}},\ }\href {\doibase 10.1103/PhysRevD.25.766} {\bibfield
  {journal} {\bibinfo  {journal} {\prd}\ }\textbf {\bibinfo {volume} {25}},\
  \bibinfo {pages} {766} (\bibinfo {year} {1982})}\BibitemShut {NoStop}%
\bibitem [{\citenamefont {{Abazajian}}\ \emph
  {et~al.}(2001{\natexlab{b}})\citenamefont {{Abazajian}}, \citenamefont
  {{Fuller}},\ and\ \citenamefont {{Tucker}}}]{Abazajian:2001aa}%
  \BibitemOpen
  \bibfield  {author} {\bibinfo {author} {\bibfnamefont {K.}~\bibnamefont
  {{Abazajian}}}, \bibinfo {author} {\bibfnamefont {G.~M.}\ \bibnamefont
  {{Fuller}}}, \ and\ \bibinfo {author} {\bibfnamefont {W.~H.}\ \bibnamefont
  {{Tucker}}},\ }\href {\doibase 10.1086/323867} {\bibfield  {journal}
  {\bibinfo  {journal} {\apj}\ }\textbf {\bibinfo {volume} {562}},\ \bibinfo
  {pages} {593} (\bibinfo {year} {2001}{\natexlab{b}})},\ \Eprint
  {http://arxiv.org/abs/astro-ph/0106002} {astro-ph/0106002} \BibitemShut
  {NoStop}%
\bibitem [{\citenamefont {Dolgov}\ and\ \citenamefont
  {Hansen}(2002)}]{Dolgov:2002aa}%
  \BibitemOpen
  \bibfield  {author} {\bibinfo {author} {\bibfnamefont {A.}~\bibnamefont
  {Dolgov}}\ and\ \bibinfo {author} {\bibfnamefont {S.}~\bibnamefont
  {Hansen}},\ }\href {\doibase 10.1016/S0927-6505(01)00115-3} {\bibfield
  {journal} {\bibinfo  {journal} {Astroparticle Physics}\ }\textbf {\bibinfo
  {volume} {16}},\ \bibinfo {pages} {339} (\bibinfo {year} {2002})},\ \Eprint
  {http://arxiv.org/abs/hep-ph/0009083} {arXiv:hep-ph/0009083 [hep-ph]}
  \BibitemShut {NoStop}%
\bibitem [{\citenamefont {{Watson}}\ \emph {et~al.}(2006)\citenamefont
  {{Watson}}, \citenamefont {{Beacom}}, \citenamefont {{Y{\"u}ksel}},\ and\
  \citenamefont {{Walker}}}]{Watson:2006aa}%
  \BibitemOpen
  \bibfield  {author} {\bibinfo {author} {\bibfnamefont {C.~R.}\ \bibnamefont
  {{Watson}}}, \bibinfo {author} {\bibfnamefont {J.~F.}\ \bibnamefont
  {{Beacom}}}, \bibinfo {author} {\bibfnamefont {H.}~\bibnamefont
  {{Y{\"u}ksel}}}, \ and\ \bibinfo {author} {\bibfnamefont {T.~P.}\
  \bibnamefont {{Walker}}},\ }\href {\doibase 10.1103/PhysRevD.74.033009}
  {\bibfield  {journal} {\bibinfo  {journal} {\prd}\ }\textbf {\bibinfo
  {volume} {74}},\ \bibinfo {eid} {033009} (\bibinfo {year} {2006})},\ \Eprint
  {http://arxiv.org/abs/astro-ph/0605424} {astro-ph/0605424} \BibitemShut
  {NoStop}%
\bibitem [{\citenamefont {{Y{\"u}ksel}}\ \emph {et~al.}(2008)\citenamefont
  {{Y{\"u}ksel}}, \citenamefont {{Beacom}},\ and\ \citenamefont
  {{Watson}}}]{Yuksel:2008aa}%
  \BibitemOpen
  \bibfield  {author} {\bibinfo {author} {\bibfnamefont {H.}~\bibnamefont
  {{Y{\"u}ksel}}}, \bibinfo {author} {\bibfnamefont {J.~F.}\ \bibnamefont
  {{Beacom}}}, \ and\ \bibinfo {author} {\bibfnamefont {C.~R.}\ \bibnamefont
  {{Watson}}},\ }\href {\doibase 10.1103/PhysRevLett.101.121301} {\bibfield
  {journal} {\bibinfo  {journal} {Physical Review Letters}\ }\textbf {\bibinfo
  {volume} {101}},\ \bibinfo {eid} {121301} (\bibinfo {year} {2008})},\ \Eprint
  {http://arxiv.org/abs/0706.4084} {arXiv:0706.4084} \BibitemShut {NoStop}%
\bibitem [{\citenamefont {{Baur}}\ \emph {et~al.}(2016)\citenamefont {{Baur}},
  \citenamefont {{Palanque-Delabrouille}}, \citenamefont {{Y{\`e}che}},
  \citenamefont {{Magneville}},\ and\ \citenamefont {{Viel}}}]{Baur:2016aa}%
  \BibitemOpen
  \bibfield  {author} {\bibinfo {author} {\bibfnamefont {J.}~\bibnamefont
  {{Baur}}}, \bibinfo {author} {\bibfnamefont {N.}~\bibnamefont
  {{Palanque-Delabrouille}}}, \bibinfo {author} {\bibfnamefont
  {C.}~\bibnamefont {{Y{\`e}che}}}, \bibinfo {author} {\bibfnamefont
  {C.}~\bibnamefont {{Magneville}}}, \ and\ \bibinfo {author} {\bibfnamefont
  {M.}~\bibnamefont {{Viel}}},\ }\href {\doibase 10.1088/1475-7516/2016/08/012}
  {\bibfield  {journal} {\bibinfo  {journal} {JCAP}\ }\textbf {\bibinfo
  {volume} {8}},\ \bibinfo {eid} {012} (\bibinfo {year} {2016})},\ \Eprint
  {http://arxiv.org/abs/1512.01981} {arXiv:1512.01981} \BibitemShut {NoStop}%
\bibitem [{\citenamefont {{Kulkarni}}\ \emph {et~al.}(2015)\citenamefont
  {{Kulkarni}}, \citenamefont {{Hennawi}}, \citenamefont {{O{\~n}orbe}},
  \citenamefont {{Rorai}},\ and\ \citenamefont {{Springel}}}]{Kulkarni:2015aa}%
  \BibitemOpen
  \bibfield  {author} {\bibinfo {author} {\bibfnamefont {G.}~\bibnamefont
  {{Kulkarni}}}, \bibinfo {author} {\bibfnamefont {J.~F.}\ \bibnamefont
  {{Hennawi}}}, \bibinfo {author} {\bibfnamefont {J.}~\bibnamefont
  {{O{\~n}orbe}}}, \bibinfo {author} {\bibfnamefont {A.}~\bibnamefont
  {{Rorai}}}, \ and\ \bibinfo {author} {\bibfnamefont {V.}~\bibnamefont
  {{Springel}}},\ }\href {\doibase 10.1088/0004-637X/812/1/30} {\bibfield
  {journal} {\bibinfo  {journal} {\apj}\ }\textbf {\bibinfo {volume} {812}},\
  \bibinfo {eid} {30} (\bibinfo {year} {2015})},\ \Eprint
  {http://arxiv.org/abs/1504.00366} {arXiv:1504.00366} \BibitemShut {NoStop}%
\bibitem [{\citenamefont {Asaka}\ \emph {et~al.}(2005)\citenamefont {Asaka},
  \citenamefont {Blanchet},\ and\ \citenamefont {Shaposhnikov}}]{Asaka:2005an}%
  \BibitemOpen
  \bibfield  {author} {\bibinfo {author} {\bibfnamefont {T.}~\bibnamefont
  {Asaka}}, \bibinfo {author} {\bibfnamefont {S.}~\bibnamefont {Blanchet}}, \
  and\ \bibinfo {author} {\bibfnamefont {M.}~\bibnamefont {Shaposhnikov}},\
  }\href {\doibase 10.1016/j.physletb.2005.09.070} {\bibfield  {journal}
  {\bibinfo  {journal} {Phys. Lett.}\ }\textbf {\bibinfo {volume} {B631}},\
  \bibinfo {pages} {151} (\bibinfo {year} {2005})},\ \Eprint
  {http://arxiv.org/abs/hep-ph/0503065} {arXiv:hep-ph/0503065 [hep-ph]}
  \BibitemShut {NoStop}%
\bibitem [{\citenamefont {Boyarsky}\ \emph {et~al.}(2006)\citenamefont
  {Boyarsky}, \citenamefont {Neronov}, \citenamefont {Ruchayskiy},\ and\
  \citenamefont {Shaposhnikov}}]{Boyarsky:2006jm}%
  \BibitemOpen
  \bibfield  {author} {\bibinfo {author} {\bibfnamefont {A.}~\bibnamefont
  {Boyarsky}}, \bibinfo {author} {\bibfnamefont {A.}~\bibnamefont {Neronov}},
  \bibinfo {author} {\bibfnamefont {O.}~\bibnamefont {Ruchayskiy}}, \ and\
  \bibinfo {author} {\bibfnamefont {M.}~\bibnamefont {Shaposhnikov}},\ }\href
  {\doibase 10.1134/S0021364006040011} {\bibfield  {journal} {\bibinfo
  {journal} {JETP Lett.}\ }\textbf {\bibinfo {volume} {83}},\ \bibinfo {pages}
  {133} (\bibinfo {year} {2006})},\ \Eprint
  {http://arxiv.org/abs/hep-ph/0601098} {arXiv:hep-ph/0601098 [hep-ph]}
  \BibitemShut {NoStop}%
\bibitem [{\citenamefont {Asaka}\ \emph {et~al.}(2007)\citenamefont {Asaka},
  \citenamefont {Laine},\ and\ \citenamefont {Shaposhnikov}}]{Asaka:2006nq}%
  \BibitemOpen
  \bibfield  {author} {\bibinfo {author} {\bibfnamefont {T.}~\bibnamefont
  {Asaka}}, \bibinfo {author} {\bibfnamefont {M.}~\bibnamefont {Laine}}, \ and\
  \bibinfo {author} {\bibfnamefont {M.}~\bibnamefont {Shaposhnikov}},\ }\href
  {\doibase 10.1088/1126-6708/2007/01/091, 10.1007/JHEP02(2015)028} {\bibfield
  {journal} {\bibinfo  {journal} {JHEP}\ }\textbf {\bibinfo {volume} {01}},\
  \bibinfo {pages} {091} (\bibinfo {year} {2007})},\ \bibinfo {note} {[Erratum:
  JHEP02,028(2015)]},\ \Eprint {http://arxiv.org/abs/hep-ph/0612182}
  {arXiv:hep-ph/0612182 [hep-ph]} \BibitemShut {NoStop}%
\bibitem [{\citenamefont {Canetti}\ \emph
  {et~al.}(2013{\natexlab{a}})\citenamefont {Canetti}, \citenamefont {Drewes},\
  and\ \citenamefont {Shaposhnikov}}]{Canetti:2012vf}%
  \BibitemOpen
  \bibfield  {author} {\bibinfo {author} {\bibfnamefont {L.}~\bibnamefont
  {Canetti}}, \bibinfo {author} {\bibfnamefont {M.}~\bibnamefont {Drewes}}, \
  and\ \bibinfo {author} {\bibfnamefont {M.}~\bibnamefont {Shaposhnikov}},\
  }\href {\doibase 10.1103/PhysRevLett.110.061801} {\bibfield  {journal}
  {\bibinfo  {journal} {Phys. Rev. Lett.}\ }\textbf {\bibinfo {volume} {110}},\
  \bibinfo {pages} {061801} (\bibinfo {year} {2013}{\natexlab{a}})},\ \Eprint
  {http://arxiv.org/abs/1204.3902} {arXiv:1204.3902 [hep-ph]} \BibitemShut
  {NoStop}%
\bibitem [{\citenamefont {Canetti}\ \emph
  {et~al.}(2013{\natexlab{b}})\citenamefont {Canetti}, \citenamefont {Drewes},
  \citenamefont {Frossard},\ and\ \citenamefont
  {Shaposhnikov}}]{Canetti:2012kh}%
  \BibitemOpen
  \bibfield  {author} {\bibinfo {author} {\bibfnamefont {L.}~\bibnamefont
  {Canetti}}, \bibinfo {author} {\bibfnamefont {M.}~\bibnamefont {Drewes}},
  \bibinfo {author} {\bibfnamefont {T.}~\bibnamefont {Frossard}}, \ and\
  \bibinfo {author} {\bibfnamefont {M.}~\bibnamefont {Shaposhnikov}},\ }\href
  {\doibase 10.1103/PhysRevD.87.093006} {\bibfield  {journal} {\bibinfo
  {journal} {Phys. Rev.}\ }\textbf {\bibinfo {volume} {D87}},\ \bibinfo {pages}
  {093006} (\bibinfo {year} {2013}{\natexlab{b}})},\ \Eprint
  {http://arxiv.org/abs/1208.4607} {arXiv:1208.4607 [hep-ph]} \BibitemShut
  {NoStop}%
\bibitem [{\citenamefont {{Koposov}}\ \emph
  {et~al.}(2015{\natexlab{b}})\citenamefont {{Koposov}}, \citenamefont
  {{Casey}}, \citenamefont {{Belokurov}}, \citenamefont {{Lewis}},
  \citenamefont {{Gilmore}}, \citenamefont {{Worley}}, \citenamefont
  {{Hourihane}}, \citenamefont {{Randich}}, \citenamefont {{Bensby}},
  \citenamefont {{Bragaglia}}, \citenamefont {{Bergemann}}, \citenamefont
  {{Carraro}}, \citenamefont {{Costado}}, \citenamefont {{Flaccomio}},
  \citenamefont {{Francois}}, \citenamefont {{Heiter}}, \citenamefont {{Hill}},
  \citenamefont {{Jofre}}, \citenamefont {{Lando}}, \citenamefont
  {{Lanzafame}}, \citenamefont {{de Laverny}}, \citenamefont {{Monaco}},
  \citenamefont {{Morbidelli}}, \citenamefont {{Sbordone}}, \citenamefont
  {{Mikolaitis}},\ and\ \citenamefont {{Ryde}}}]{Koposov:2015ab}%
  \BibitemOpen
  \bibfield  {author} {\bibinfo {author} {\bibfnamefont {S.~E.}\ \bibnamefont
  {{Koposov}}}, \bibinfo {author} {\bibfnamefont {A.~R.}\ \bibnamefont
  {{Casey}}}, \bibinfo {author} {\bibfnamefont {V.}~\bibnamefont
  {{Belokurov}}}, \bibinfo {author} {\bibfnamefont {J.~R.}\ \bibnamefont
  {{Lewis}}}, \bibinfo {author} {\bibfnamefont {G.}~\bibnamefont {{Gilmore}}},
  \bibinfo {author} {\bibfnamefont {C.}~\bibnamefont {{Worley}}}, \bibinfo
  {author} {\bibfnamefont {A.}~\bibnamefont {{Hourihane}}}, \bibinfo {author}
  {\bibfnamefont {S.}~\bibnamefont {{Randich}}}, \bibinfo {author}
  {\bibfnamefont {T.}~\bibnamefont {{Bensby}}}, \bibinfo {author}
  {\bibfnamefont {A.}~\bibnamefont {{Bragaglia}}}, \bibinfo {author}
  {\bibfnamefont {M.}~\bibnamefont {{Bergemann}}}, \bibinfo {author}
  {\bibfnamefont {G.}~\bibnamefont {{Carraro}}}, \bibinfo {author}
  {\bibfnamefont {M.~T.}\ \bibnamefont {{Costado}}}, \bibinfo {author}
  {\bibfnamefont {E.}~\bibnamefont {{Flaccomio}}}, \bibinfo {author}
  {\bibfnamefont {P.}~\bibnamefont {{Francois}}}, \bibinfo {author}
  {\bibfnamefont {U.}~\bibnamefont {{Heiter}}}, \bibinfo {author}
  {\bibfnamefont {V.}~\bibnamefont {{Hill}}}, \bibinfo {author} {\bibfnamefont
  {P.}~\bibnamefont {{Jofre}}}, \bibinfo {author} {\bibfnamefont
  {C.}~\bibnamefont {{Lando}}}, \bibinfo {author} {\bibfnamefont {A.~C.}\
  \bibnamefont {{Lanzafame}}}, \bibinfo {author} {\bibfnamefont
  {P.}~\bibnamefont {{de Laverny}}}, \bibinfo {author} {\bibfnamefont
  {L.}~\bibnamefont {{Monaco}}}, \bibinfo {author} {\bibfnamefont
  {L.}~\bibnamefont {{Morbidelli}}}, \bibinfo {author} {\bibfnamefont
  {L.}~\bibnamefont {{Sbordone}}}, \bibinfo {author} {\bibfnamefont {{\v
  S}.}~\bibnamefont {{Mikolaitis}}}, \ and\ \bibinfo {author} {\bibfnamefont
  {N.}~\bibnamefont {{Ryde}}},\ }\href {\doibase 10.1088/0004-637X/811/1/62}
  {\bibfield  {journal} {\bibinfo  {journal} {\apj}\ }\textbf {\bibinfo
  {volume} {811}},\ \bibinfo {eid} {62} (\bibinfo {year}
  {2015}{\natexlab{b}})},\ \Eprint {http://arxiv.org/abs/1504.07916}
  {arXiv:1504.07916} \BibitemShut {NoStop}%
\bibitem [{\citenamefont {{Simon}}\ \emph {et~al.}(2015)\citenamefont
  {{Simon}}, \citenamefont {{Drlica-Wagner}}, \citenamefont {{Li}},
  \citenamefont {{Nord}}, \citenamefont {{Geha}}, \citenamefont {{Bechtol}},
  \citenamefont {{Balbinot}}, \citenamefont {{Buckley-Geer}}, \citenamefont
  {{Lin}}, \citenamefont {{Marshall}}, \citenamefont {{Santiago}},
  \citenamefont {{Strigari}}, \citenamefont {{Wang}}, \citenamefont
  {{Wechsler}}, \citenamefont {{Yanny}}, \citenamefont {{Abbott}},
  \citenamefont {{Bauer}}, \citenamefont {{Bernstein}}, \citenamefont
  {{Bertin}}, \citenamefont {{Brooks}}, \citenamefont {{Burke}}, \citenamefont
  {{Capozzi}}, \citenamefont {{Carnero Rosell}}, \citenamefont {{Carrasco
  Kind}}, \citenamefont {{D'Andrea}}, \citenamefont {{da Costa}}, \citenamefont
  {{DePoy}}, \citenamefont {{Desai}}, \citenamefont {{Diehl}}, \citenamefont
  {{Dodelson}}, \citenamefont {{Cunha}}, \citenamefont {{Estrada}},
  \citenamefont {{Evrard}}, \citenamefont {{Fausti Neto}}, \citenamefont
  {{Fernandez}}, \citenamefont {{Finley}}, \citenamefont {{Flaugher}},
  \citenamefont {{Frieman}}, \citenamefont {{Gaztanaga}}, \citenamefont
  {{Gerdes}}, \citenamefont {{Gruen}}, \citenamefont {{Gruendl}}, \citenamefont
  {{Honscheid}}, \citenamefont {{James}}, \citenamefont {{Kent}}, \citenamefont
  {{Kuehn}}, \citenamefont {{Kuropatkin}}, \citenamefont {{Lahav}},
  \citenamefont {{Maia}}, \citenamefont {{March}}, \citenamefont {{Martini}},
  \citenamefont {{Miller}}, \citenamefont {{Miquel}}, \citenamefont {{Ogando}},
  \citenamefont {{Romer}}, \citenamefont {{Roodman}}, \citenamefont {{Rykoff}},
  \citenamefont {{Sako}}, \citenamefont {{Sanchez}}, \citenamefont
  {{Schubnell}}, \citenamefont {{Sevilla}}, \citenamefont {{Smith}},
  \citenamefont {{Soares-Santos}}, \citenamefont {{Sobreira}}, \citenamefont
  {{Suchyta}}, \citenamefont {{Swanson}}, \citenamefont {{Tarle}},
  \citenamefont {{Thaler}}, \citenamefont {{Tucker}}, \citenamefont {{Vikram}},
  \citenamefont {{Walker}}, \citenamefont {{Wester}},\ and\ \citenamefont {{DES
  Collaboration}}}]{Simon:2015aa}%
  \BibitemOpen
  \bibfield  {author} {\bibinfo {author} {\bibfnamefont {J.~D.}\ \bibnamefont
  {{Simon}}}, \bibinfo {author} {\bibfnamefont {A.}~\bibnamefont
  {{Drlica-Wagner}}}, \bibinfo {author} {\bibfnamefont {T.~S.}\ \bibnamefont
  {{Li}}}, \bibinfo {author} {\bibfnamefont {B.}~\bibnamefont {{Nord}}},
  \bibinfo {author} {\bibfnamefont {M.}~\bibnamefont {{Geha}}}, \bibinfo
  {author} {\bibfnamefont {K.}~\bibnamefont {{Bechtol}}}, \bibinfo {author}
  {\bibfnamefont {E.}~\bibnamefont {{Balbinot}}}, \bibinfo {author}
  {\bibfnamefont {E.}~\bibnamefont {{Buckley-Geer}}}, \bibinfo {author}
  {\bibfnamefont {H.}~\bibnamefont {{Lin}}}, \bibinfo {author} {\bibfnamefont
  {J.}~\bibnamefont {{Marshall}}}, \bibinfo {author} {\bibfnamefont
  {B.}~\bibnamefont {{Santiago}}}, \bibinfo {author} {\bibfnamefont
  {L.}~\bibnamefont {{Strigari}}}, \bibinfo {author} {\bibfnamefont
  {M.}~\bibnamefont {{Wang}}}, \bibinfo {author} {\bibfnamefont {R.~H.}\
  \bibnamefont {{Wechsler}}}, \bibinfo {author} {\bibfnamefont
  {B.}~\bibnamefont {{Yanny}}}, \bibinfo {author} {\bibfnamefont
  {T.}~\bibnamefont {{Abbott}}}, \bibinfo {author} {\bibfnamefont {A.~H.}\
  \bibnamefont {{Bauer}}}, \bibinfo {author} {\bibfnamefont {G.~M.}\
  \bibnamefont {{Bernstein}}}, \bibinfo {author} {\bibfnamefont
  {E.}~\bibnamefont {{Bertin}}}, \bibinfo {author} {\bibfnamefont
  {D.}~\bibnamefont {{Brooks}}}, \bibinfo {author} {\bibfnamefont {D.~L.}\
  \bibnamefont {{Burke}}}, \bibinfo {author} {\bibfnamefont {D.}~\bibnamefont
  {{Capozzi}}}, \bibinfo {author} {\bibfnamefont {A.}~\bibnamefont {{Carnero
  Rosell}}}, \bibinfo {author} {\bibfnamefont {M.}~\bibnamefont {{Carrasco
  Kind}}}, \bibinfo {author} {\bibfnamefont {C.~B.}\ \bibnamefont
  {{D'Andrea}}}, \bibinfo {author} {\bibfnamefont {L.~N.}\ \bibnamefont {{da
  Costa}}}, \bibinfo {author} {\bibfnamefont {D.~L.}\ \bibnamefont {{DePoy}}},
  \bibinfo {author} {\bibfnamefont {S.}~\bibnamefont {{Desai}}}, \bibinfo
  {author} {\bibfnamefont {H.~T.}\ \bibnamefont {{Diehl}}}, \bibinfo {author}
  {\bibfnamefont {S.}~\bibnamefont {{Dodelson}}}, \bibinfo {author}
  {\bibfnamefont {C.~E.}\ \bibnamefont {{Cunha}}}, \bibinfo {author}
  {\bibfnamefont {J.}~\bibnamefont {{Estrada}}}, \bibinfo {author}
  {\bibfnamefont {A.~E.}\ \bibnamefont {{Evrard}}}, \bibinfo {author}
  {\bibfnamefont {A.}~\bibnamefont {{Fausti Neto}}}, \bibinfo {author}
  {\bibfnamefont {E.}~\bibnamefont {{Fernandez}}}, \bibinfo {author}
  {\bibfnamefont {D.~A.}\ \bibnamefont {{Finley}}}, \bibinfo {author}
  {\bibfnamefont {B.}~\bibnamefont {{Flaugher}}}, \bibinfo {author}
  {\bibfnamefont {J.}~\bibnamefont {{Frieman}}}, \bibinfo {author}
  {\bibfnamefont {E.}~\bibnamefont {{Gaztanaga}}}, \bibinfo {author}
  {\bibfnamefont {D.}~\bibnamefont {{Gerdes}}}, \bibinfo {author}
  {\bibfnamefont {D.}~\bibnamefont {{Gruen}}}, \bibinfo {author} {\bibfnamefont
  {R.~A.}\ \bibnamefont {{Gruendl}}}, \bibinfo {author} {\bibfnamefont
  {K.}~\bibnamefont {{Honscheid}}}, \bibinfo {author} {\bibfnamefont
  {D.}~\bibnamefont {{James}}}, \bibinfo {author} {\bibfnamefont
  {S.}~\bibnamefont {{Kent}}}, \bibinfo {author} {\bibfnamefont
  {K.}~\bibnamefont {{Kuehn}}}, \bibinfo {author} {\bibfnamefont
  {N.}~\bibnamefont {{Kuropatkin}}}, \bibinfo {author} {\bibfnamefont
  {O.}~\bibnamefont {{Lahav}}}, \bibinfo {author} {\bibfnamefont {M.~A.~G.}\
  \bibnamefont {{Maia}}}, \bibinfo {author} {\bibfnamefont {M.}~\bibnamefont
  {{March}}}, \bibinfo {author} {\bibfnamefont {P.}~\bibnamefont {{Martini}}},
  \bibinfo {author} {\bibfnamefont {C.~J.}\ \bibnamefont {{Miller}}}, \bibinfo
  {author} {\bibfnamefont {R.}~\bibnamefont {{Miquel}}}, \bibinfo {author}
  {\bibfnamefont {R.}~\bibnamefont {{Ogando}}}, \bibinfo {author}
  {\bibfnamefont {A.~K.}\ \bibnamefont {{Romer}}}, \bibinfo {author}
  {\bibfnamefont {A.}~\bibnamefont {{Roodman}}}, \bibinfo {author}
  {\bibfnamefont {E.~S.}\ \bibnamefont {{Rykoff}}}, \bibinfo {author}
  {\bibfnamefont {M.}~\bibnamefont {{Sako}}}, \bibinfo {author} {\bibfnamefont
  {E.}~\bibnamefont {{Sanchez}}}, \bibinfo {author} {\bibfnamefont
  {M.}~\bibnamefont {{Schubnell}}}, \bibinfo {author} {\bibfnamefont
  {I.}~\bibnamefont {{Sevilla}}}, \bibinfo {author} {\bibfnamefont {R.~C.}\
  \bibnamefont {{Smith}}}, \bibinfo {author} {\bibfnamefont {M.}~\bibnamefont
  {{Soares-Santos}}}, \bibinfo {author} {\bibfnamefont {F.}~\bibnamefont
  {{Sobreira}}}, \bibinfo {author} {\bibfnamefont {E.}~\bibnamefont
  {{Suchyta}}}, \bibinfo {author} {\bibfnamefont {M.~E.~C.}\ \bibnamefont
  {{Swanson}}}, \bibinfo {author} {\bibfnamefont {G.}~\bibnamefont {{Tarle}}},
  \bibinfo {author} {\bibfnamefont {J.}~\bibnamefont {{Thaler}}}, \bibinfo
  {author} {\bibfnamefont {D.}~\bibnamefont {{Tucker}}}, \bibinfo {author}
  {\bibfnamefont {V.}~\bibnamefont {{Vikram}}}, \bibinfo {author}
  {\bibfnamefont {A.~R.}\ \bibnamefont {{Walker}}}, \bibinfo {author}
  {\bibfnamefont {W.}~\bibnamefont {{Wester}}}, \ and\ \bibinfo {author}
  {\bibnamefont {{DES Collaboration}}},\ }\href {\doibase
  10.1088/0004-637X/808/1/95} {\bibfield  {journal} {\bibinfo  {journal}
  {\apj}\ }\textbf {\bibinfo {volume} {808}},\ \bibinfo {eid} {95} (\bibinfo
  {year} {2015})},\ \Eprint {http://arxiv.org/abs/1504.02889}
  {arXiv:1504.02889} \BibitemShut {NoStop}%
\bibitem [{\citenamefont {{Walker}}\ \emph {et~al.}(2015)\citenamefont
  {{Walker}}, \citenamefont {{Mateo}}, \citenamefont {{Olszewski}},
  \citenamefont {{Bailey}}, \citenamefont {{Koposov}}, \citenamefont
  {{Belokurov}},\ and\ \citenamefont {{Evans}}}]{Walker:2015aa}%
  \BibitemOpen
  \bibfield  {author} {\bibinfo {author} {\bibfnamefont {M.~G.}\ \bibnamefont
  {{Walker}}}, \bibinfo {author} {\bibfnamefont {M.}~\bibnamefont {{Mateo}}},
  \bibinfo {author} {\bibfnamefont {E.~W.}\ \bibnamefont {{Olszewski}}},
  \bibinfo {author} {\bibfnamefont {J.~I.}\ \bibnamefont {{Bailey}},
  \bibfnamefont {III}}, \bibinfo {author} {\bibfnamefont {S.~E.}\ \bibnamefont
  {{Koposov}}}, \bibinfo {author} {\bibfnamefont {V.}~\bibnamefont
  {{Belokurov}}}, \ and\ \bibinfo {author} {\bibfnamefont {N.~W.}\ \bibnamefont
  {{Evans}}},\ }\href {\doibase 10.1088/0004-637X/808/2/108} {\bibfield
  {journal} {\bibinfo  {journal} {\apj}\ }\textbf {\bibinfo {volume} {808}},\
  \bibinfo {eid} {108} (\bibinfo {year} {2015})},\ \Eprint
  {http://arxiv.org/abs/1504.03060} {arXiv:1504.03060} \BibitemShut {NoStop}%
\bibitem [{\citenamefont {{Walker}}\ \emph {et~al.}(2016)\citenamefont
  {{Walker}}, \citenamefont {{Mateo}}, \citenamefont {{Olszewski}},
  \citenamefont {{Koposov}}, \citenamefont {{Belokurov}}, \citenamefont
  {{Jethwa}}, \citenamefont {{Nidever}}, \citenamefont {{Bonnivard}},
  \citenamefont {{Bailey}}, \citenamefont {{Bell}},\ and\ \citenamefont
  {{Loebman}}}]{Walker:2016aa}%
  \BibitemOpen
  \bibfield  {author} {\bibinfo {author} {\bibfnamefont {M.~G.}\ \bibnamefont
  {{Walker}}}, \bibinfo {author} {\bibfnamefont {M.}~\bibnamefont {{Mateo}}},
  \bibinfo {author} {\bibfnamefont {E.~W.}\ \bibnamefont {{Olszewski}}},
  \bibinfo {author} {\bibfnamefont {S.}~\bibnamefont {{Koposov}}}, \bibinfo
  {author} {\bibfnamefont {V.}~\bibnamefont {{Belokurov}}}, \bibinfo {author}
  {\bibfnamefont {P.}~\bibnamefont {{Jethwa}}}, \bibinfo {author}
  {\bibfnamefont {D.~L.}\ \bibnamefont {{Nidever}}}, \bibinfo {author}
  {\bibfnamefont {V.}~\bibnamefont {{Bonnivard}}}, \bibinfo {author}
  {\bibfnamefont {J.~I.}\ \bibnamefont {{Bailey}}, \bibfnamefont {III}},
  \bibinfo {author} {\bibfnamefont {E.~F.}\ \bibnamefont {{Bell}}}, \ and\
  \bibinfo {author} {\bibfnamefont {S.~R.}\ \bibnamefont {{Loebman}}},\ }\href
  {\doibase 10.3847/0004-637X/819/1/53} {\bibfield  {journal} {\bibinfo
  {journal} {\apj}\ }\textbf {\bibinfo {volume} {819}},\ \bibinfo {eid} {53}
  (\bibinfo {year} {2016})},\ \Eprint {http://arxiv.org/abs/1511.06296}
  {arXiv:1511.06296} \BibitemShut {NoStop}%
\bibitem [{\citenamefont {Speckhard}\ \emph {et~al.}(2016)\citenamefont
  {Speckhard}, \citenamefont {Ng}, \citenamefont {Beacom},\ and\ \citenamefont
  {Laha}}]{Speckhard:2015eva}%
  \BibitemOpen
  \bibfield  {author} {\bibinfo {author} {\bibfnamefont {E.~G.}\ \bibnamefont
  {Speckhard}}, \bibinfo {author} {\bibfnamefont {K.~C.~Y.}\ \bibnamefont
  {Ng}}, \bibinfo {author} {\bibfnamefont {J.~F.}\ \bibnamefont {Beacom}}, \
  and\ \bibinfo {author} {\bibfnamefont {R.}~\bibnamefont {Laha}},\ }\href
  {\doibase 10.1103/PhysRevLett.116.031301} {\bibfield  {journal} {\bibinfo
  {journal} {Phys. Rev. Lett.}\ }\textbf {\bibinfo {volume} {116}},\ \bibinfo
  {pages} {031301} (\bibinfo {year} {2016})},\ \Eprint
  {http://arxiv.org/abs/1507.04744} {arXiv:1507.04744 [astro-ph.CO]}
  \BibitemShut {NoStop}%
\bibitem [{\citenamefont {Powell}\ \emph {et~al.}(2016)\citenamefont {Powell},
  \citenamefont {Laha}, \citenamefont {Ng},\ and\ \citenamefont
  {Abel}}]{Powell:2016zbo}%
  \BibitemOpen
  \bibfield  {author} {\bibinfo {author} {\bibfnamefont {D.}~\bibnamefont
  {Powell}}, \bibinfo {author} {\bibfnamefont {R.}~\bibnamefont {Laha}},
  \bibinfo {author} {\bibfnamefont {K.~C.~Y.}\ \bibnamefont {Ng}}, \ and\
  \bibinfo {author} {\bibfnamefont {T.}~\bibnamefont {Abel}},\ }\href@noop {}
  {\  (\bibinfo {year} {2016})},\ \Eprint {http://arxiv.org/abs/1611.02714}
  {arXiv:1611.02714 [astro-ph.CO]} \BibitemShut {NoStop}%
\end{thebibliography}%

\end{document}